\documentclass[journal]{IEEEtran}
\usepackage{cite}
\usepackage{caption}
\usepackage{subfigure}
\usepackage{amsmath,amssymb,amsfonts}
\usepackage[linesnumbered,ruled,vlined]{algorithm2e}
\usepackage{graphicx}
\usepackage{textcomp}
\usepackage{xcolor}
\usepackage{makecell}
\usepackage{url}
\usepackage{booktabs}
\usepackage[misc]{ifsym}
\usepackage{lineno}
\usepackage{multirow}
\usepackage[colorlinks,linkcolor=blue]{hyperref}
\usepackage{algpseudocode}

\begin{document}
\title{CentiTrack: Towards Centimeter-Level Passive Gesture Tracking with Commodity WiFi}

\author{Zijun Han, Zhaoming Lu\textsuperscript{\dag}, Xiangming Wen, Wei Zheng, Jingbo Zhao and Lingchao Guo
\thanks{The authors are with the School of Information and Communication Engineering,
Beijing University of Posts and Telecommunications, Beijing, China. The authors are in Beijing Key Laboratory of Network System Architecture and Convergence and Beijing Laboratory of Advanced Information Networks. E-mail:\{zijunhan, lzy0372, xiangmw, zhengweius, Zhjb, rita guo\}@bupt.edu.cn.}
\thanks{\textsuperscript{\dag}This is corresponding author.}
}
\maketitle

\begin{abstract}
Gesture awareness plays a crucial role in promoting human-computer interface. Previous works either depend on customized hardware or need a priori learning of wireless signal patterns, facing downsides in terms of the privacy concern, availability and reliability. In this paper, we propose CentiTrack, the first centimeter-level passive gesture tracking system that works with only three commodity WiFi devices, without any extra hardware modifications or wearable sensors. To this end, we first identify the Channel State Information (CSI) measurement error sources in the physical layer process, and then denoise CSI by the complex ratio between adjacent antennas. Principal Component Analysis (PCA) is further adopted to separate the reflected signals from noises. Benchmark experiments are conducted to verify that the phase changes of denoised CSI are proportional to the length changes of dynamic path reflected off the hand. In addition, we adopt the Multiple Signal Classification (MUSIC) algorithm to estimate the Angle-of-Arrivals (AoAs) of dynamic paths, and then locate the initial position of hands with triangulation. We also propose a novel static componnets elimination algorithm for tracking correction by eliminating the components unrelated to motion. A prototype of CentiTrack is fully realized and evaluated in various real scenarios. Extensive experiments show that CentiTrack is superior in terms of tracking accuracy, sensing range and device cost, compared with the state-of-the-arts.

\end{abstract}


\IEEEpeerreviewmaketitle

\section{Introduction}

Gesture awareness is one of the most ancient and intuitive ways for communications, which carries rich, profound and subtle information. Human  beings also desire to interact with computers or other electronic devices by gestures. In this way, gesture recognition has been a basic human-computer interaction mechanism to control not only Internet of Things (IoT) devices but also VR/AR devices \cite{witrace}. It seems to be with broad application prospect due to its simplicity and intuitiveness, e.g., users can easily raise the temperature of air conditioners or switch on the lights just by making gestures in smart home.

Thus far, diverse solutions have been proposed including acoustic signals- \cite{acoustic1, acoustic2, acoustic3, acoustic4}, cameras- \cite{camera1, camera2, camera3, camera4, camera5}, infrared- \cite{infrared1, infrared2}, radio frequency (RF)-based \cite{rf1,rf2,rf3,rf4,rf5,rf6,rf7,rf8,rf9}. Among these solutions, WiFi-based systems stand out as particularly promising due to the ubiquity, pervasive availability as well as privacy preservation. The development of WiFi Network Interface Cards (NICs), such as Intel 5300 and Atheros AR9580 series, has granted the researchers access to a fine-grained channel measurement from the physical layer \cite{wcnc2019}, i.e., Channel State Information (CSI). CSI describes the detailed propagation of signals from the transmitter (TX) to the receiver (RX) through multiple paths at the granularity of Orthogonal Frequency Division Multiplexing (OFDM) subcarriers \cite{csi}. Multiple attempts \cite{learn1, learn2} have been done for gesture recognition based on CSI. However, most of solutions go through the process of extracting features, and further perform classification by a priori learning of signal propagation patterns (e.g., machine learning), which is limited to recognize only a group of pre-defined gestures with limited categories. Moreover, their generalization abilities to diverse domains (e.g., diverse environments, users or locations) are insufficient. In light of this, \cite{wicapture} and \cite{milliback} explore learning-free solutions by tracking various hand motion traces. Nevertheless, these schemes in practice track users' hand-held devices, which belong to active tracking, and are reluctant and discomfort for users. 

Generally, passive tracking is based on the signals reflected off the objects to extract essential motion information. As far as we know, there are mainly six WiFi-based passive tracking schemes, i.e., WiTrace \cite{witrace}, WiDraw \cite{rf4}, IndoTrack \cite{indotrack}, Dynamic Music \cite{dynamic}, Tracking from One Side \cite{trackside} and Widar \cite{widar1, widar2}. In particular, WiTrace \cite{witrace} aggregates Doppler Frequency Shift (DFS) from multiple links, but it relies on specialized transceivers (USRP-N210) and leverages an external clock for synchronization. Furthermore, it estimates the initial position based on the result of two preamble gestures as the fingerprints, which is labor-intensive and inconvenient. WiDraw \cite{rf4} adopts Angle-of-Arrival (AoA) values of incident Line-of-Sight (LoS) signals for gesture tracking using commodity WiFi devices. Whenever the user's hand occludes a signal coming from a certain LoS, the signal strength of the AoA representing the same LoS will experience a drop. However, it needs calibration in advance and relies on densely deployed TXs, which leaves rich channel parameters unexplored and makes it less conducive for practical usage. Moreover, IndoTrack \cite{indotrack}, Dynamic Music \cite{dynamic}, Tracking from One Side \cite{trackside} and Widar \cite{widar1, widar2} belong to human body tracking schemes, with parameters including AoA, DFS, Angle-of-Departure (AoD) or Time-of-Flight (ToF), regarding the human body as a single cylinder with coarse-grained decimeter-level granularity. Table \ref{table1} provides a comprehensive comparison of the recent WiFi-based tracking systems.
\begin{table*}
\centering
\caption{Comparison of state-of-the-art WiFi-based tracking systems}
\label{table1}

\begin{tabular}{cccccccc}
\hline
$\textbf{Category}$       & $\textbf{System  Name}$    & $\textbf{Objective}$    & $\textbf{Granularity}$    & $\textbf{Range}$      & $\textbf{\#(TX, RX)/Link}$    & $\textbf{Parameter}$ & $\textbf{Cost}$  \\
\hline
\multirow{3}{*}{Active}        
& SpotFi \cite{spotfi}            & TX & 132 cm   & 5 m                      & 2                         & AoA       & General\\
& WiCapture \cite{wicapture}      & TX & 0.88 cm  & 5 m                      & 2                     & AoD       & General\\
& MilliBack \cite{milliback}      & Backscatter device & 0.49 cm   & 0.8 m        & 4/5                        & DFS       & General\\
\hline
\multirow{8}{*}{Passive}
& IndoTrack \cite{indotrack}    & Body         & 0.48 m         & 6 m                      & 2         & AoA, DFS       & Cheap\\
& Dynamic Music \cite{dynamic}  & Body         & 0.6 m          & 8 m                      & 2           & AoA            & Cheap\\
& Tracking from One Side \cite{trackside}& Body  & 0.55 m         & 7 m                   & at least 3   & AoA, Attenuation  & Cheap\\
& Widar \cite{widar1}           & Body         & 0.35 m         & 4 m                      & 2            & DFS       & Cheap\\
& Widar2.0 \cite{widar2}        & Body         & 0.75 m         & 8 m                      & 1            & AoA, DFS, ToF  & Cheap\\ 
& WiTrace \cite{witrace}        & Hand         & 2.09 cm         & 1.2 m                    & 2                & DFS, Fingerprint       & Expensive\\
& WiDraw \cite{rf4}             & Hand         & 5 cm              & 0.6 m                    & at least 25        & AoA       & General\\
& $\textbf{Ours}$    & $\textbf{Hand}$  & $\textbf{1.5 cm}$  & $\textbf{1.5 m}$  & $\textbf{2}$   & $\textbf{AoA, DFS}$  & $\textbf{Cheap}$ \\
\hline
\end{tabular}
\centering
\end{table*}

In this paper, we raise the question: Is it possible to build a fine-grained passive gesture tracking system with limited WiFi infrastructures in smart home where users would not like to intensively deploy devices for sensing, which: (i) accurately and automatically tracks gestures without customized hardware or modifications on devices; (ii) robustly works without influence of various parameters including weak illumination, occlusion, user diversity and scenario layout, etc.; and (iii) low cost and compatible with wireless transmissions? To this end,  we propose CentiTrack, aimed at passive centimeter-level gesture tracking with only three commodity WiFi devices. The key idea is to leverage fluctuated CSI values induced by hand motions via signal reflections for gesture tracking. However, three challenges need to be addressed: (i) How to discover and reflect the hand motion with noisy CSI; (ii) How to establish the model linking CSI with hand traces; (iii) How to eliminate the motion-unrelated static components? 

For the first challenge, the physical world WiFi channels are typically far from being clean. CSI samples have abundant distortions from both aspects of amplitude and phase. To achieve centimeter-level gesture tracking, we first identify the CSI measurement error sources in the physical layer process, and then denoise CSI by the complex ratio between two adjacent antennas. Principal Component Analysis (PCA) is further leveraged to separate the reflected signals from other measurement noises. We also carry out several benchmark experiments to verify that the phase changes of denoised CSI are proportional to the length changes of dynamic path reflected off the hand. In this way, CentiTrack can aggregate the length changes of dynamic paths from multiple links to reflect 2D hand motion.

For the second challenge, despite precise changes of dynamic paths, it is tricky to estimate the absolute hand traces directly without the initial position. To cope with it, we adopt the Multiple Signal Classification (MUSIC) algorithm to calculate AoAs of dynamic paths, and then locate the initial position with triangulation. Intuitively, the signals from dynamic paths are incoherent with those from static paths, yet the signals from static paths are coherent and thus can be merged into one static component. Therefore, a three-antenna array provided by most commodity WiFi NICs is competent to estimate dynamic AoAs. Experimental results yield that CentiTrack achieves the mean estimation error of the initial position by 5.4 cm, and whole 2D hand traces by 1.5 cm within a 1.5 m $\times$ 1.5 m tracking area.

For the third challenge, the signals of static paths composed of LoS and static reflectors are orders of magnitude stronger than those of dynamic paths. In addition, static paths may slowly change due to the hand blocking static reflectors, and erratic motions like posture changes or limb swings during tracking. These static components will translate the complex CSI traces in whole or in part, and seriously degrade the tracking accuracy. We observe that static components applied to the CSI trace are actually the translations of the trace segments to diverse degrees, so we propose a novel static components elimination algorithm for tracking correction. For CSI traces, we first translate it to the origin to remove the strong static components, and further decompose it into several independent circles by multi-peak searching algorithm. Linear interpolation between adjacent circle centers is finally adopted to eliminate slow changing static components.

We prototype CentiTrack on three WiFi devices equipped with Intel 5300 NICs. We evaluate its effectiveness and robustness by numerous experiments in three indoor scenarios. Benefits provided by its accurate tracking performance can track users' writings/drawings in practical applications, be it handwriting, intelligent household appliance controlling in smart homes or artificial intelligence enabled healthcare in hospitals where typically WiFi devices would be easily deployed, without dedicated infrastructures.

The rest of the paper is organized as follows. First related work is reviewed in Section II. We introduce the background and observations of CSI in Section III and describe CSI-based distance estimation in Section IV. Section V presents the system architecture. The implementation, evaluation and use case are provided in Section VI. We give the discussion in Section VII, and finally conclude in Section VIII.

\section{Related Work}
CentiTrack is mainly related to works in the field of gesture recognition and gesture tracking. The former focuses on recognizing a set of pre-defined gestures \cite{rf1, rf3, learn1, learn2, wisee}, while the latter intends to track hand traces \cite{rf4, rf8, wicapture, milliback}. Over the years, various gesture recognition and tracking systems have been developed including acoustic signals- \cite{acoustic1, acoustic2, acoustic3, acoustic4}, cameras- \cite{camera1, camera2, camera3, camera4, camera5}, infrared- \cite{infrared1, infrared2}, inertial sensors- \cite{inertial1, inertial2} and RF-based \cite{rf1,rf2,rf3,rf4,rf5,rf6,rf7,rf8,rf9}, which can be mainly divided into RF-based and non-RF-based.

\subsection{Non-RF-based Gesture Recognition and Tracking}
Non-RF-based gesture recognition and tracking systems mainly relies on cameras and acoustic signals. Camera-based solutions are a well-researched topic for years, which leverage depth or infrared cameras \cite{camera2} for human-computer interactions. In addition to their high computational cost, the dedicated hardware setup, lighting conditions and privacy concerns may incur users' reluctance. Besides, they cover only limited LoS areas. Acoustic signals-based solutions \cite{acoustic1, acoustic2, acoustic3, acoustic4} typically adopt the DFS for tracking tasks, but they only have a small tracking range, which is a major barrier to the large-scale deployment. Other works on inertial sensing \cite{inertial1, inertial2} or on-body systems \cite{humantenna1,humantenna2} explore sensing devices carried by or attached to the user target. For our proposed scheme, it is device-free, transparent to users and can apply to large-scale space, despite weak illumination or occlusion.

\subsection{RF-based Gesture Recognition and Tracking}
Recent years have witnessed the ever-fast development of RF-based sensing applications. Received Signal Strength (RSS) and CSI extracted in wireless signals have been widely used for gesture recognition \cite{rf1, rf2, rf3, wisee}. For instance, WiFinger \cite{rf1} leverages CSI to recognize a set of nine digits finger-grained gestures with accuracy of 90.4\%. WiFi Gestures \cite{rf2} achieves recognition accuracy by 91\% while classifying four gestures across six participants, without the need for per-participant training. WiGest \cite{rf3} explores changes in RSS to sense in-air hand gestures around the users' mobile devices, and achieves a recognition accuracy by 96\%. However, most of the systems demand a priori learning of signal patterns, which is limited to recognize only a fixed set of pre-defined gestures. With multiple antennas or devices, RF-based gesture tracking schemes make breakthroughs in estimating the trace and speed of hands \cite{rf4, rf5, rf6, rf7, rf8, rf9}. RF-IDraw \cite{rf8} is the first RF-based system that enables a virtual touch screen and adopts AoA for fine-grained tracking. WiCapture \cite{wicapture} leverages commodity WiFi radios and achieves an accuracy of 0.88 cm with higher sensing range. More recently, MilliBack \cite{milliback} proposes phase differential iterative schemes to infer the position of a writing tool, and further track various handwriting traces with a median error of 0.49 cm. While these schemes can achieve considerable tracking accuracy, they require the user to hold an RF TX or backscatter device, which are reluctant and inconvenient for users in the long run. Works \cite{wisee, wisee2} make progress in whole-home passive gesture detection, resistance to occlusion. However, they require either costly customized wireless hardware \cite{wisee2}, or specific modifications on devices \cite{wisee}. WiDraw \cite{rf4} first passively tracks users' hands  based on commodity WiFi devices with a median error of 5 cm, but it has a limited tracking range which is less than 0.6 m. Besides, it relies on at least 25 TXs to cover all incident LoS signal directions. In contrast, CentiTrack enables a fine-grained tracking accuracy by 1.5 cm over a distance of 1.5 m, with only three devices. WiTrace \cite{witrace} adopts templates to restrict random hand movements, and achieves the mean tracking error of 2.09 cm within 1.2 m sensing range. However, it is implemented by specialized USRP-N210 platforms, raising downsides in availability issue. Furthermore, it introduces an external clock for phase offsets elimination, and is cumbersome to estimate the initial location based on the result of two preamble gestures as the fingerprints. 
\begin{figure}[htbp]
\begin{center}
\includegraphics[width=7.2cm,height=3.0cm]{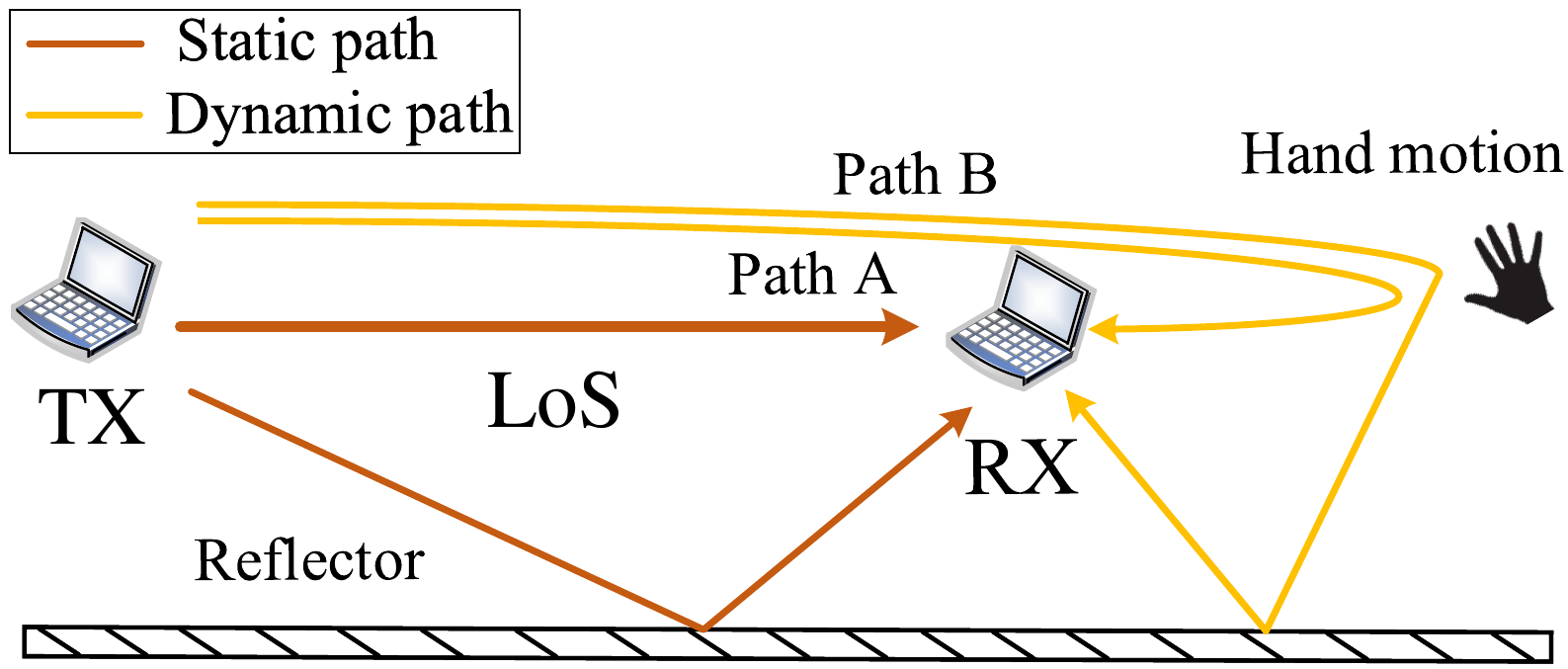}
\caption{Illustration of multiple paths.}
\label{fig1}
\end{center}
\end{figure}Based on commodity devices, CentiTrack automatically locates the initial location by AoA estimation, and corrects CSI phase offsets by signal processing.

\section{Background and Observations}
\label{sec3}
In this section, we first describe the basic concept of CSI, and then present the theoretical model of CSI regarding dynamic hand motion as a background introduction.

\subsection{Overview of CSI}

CSI estimates the properties of wireless channels at the granularity of OFDM subcarriers \cite{csi}. Specifically, the transient CSI $H(f, t)$ with carrier frequency $f$ at time $t$, can be derived as:
\begin{equation}
H(f,t)=Y(f,t)/X(f,t),
\end{equation}
where $Y(f,t)$ and $X(f,t)$ are the Fourier transformation of transmitted and received signals respectively. The channel noises are implicitly included in $H(f, t)$. Due to reflection, refraction or scattering, wireless signals propagate from TX to RX through multiple paths, resulting in multipath distortions \cite{wcnc2019}. Mathematically, CSI is the superposition of signals from all the propagation paths, which can be represented as:
\begin{equation}
H(f,t)=\sum \limits_{l=1} \limits^{L} \alpha_l e^{-j2\pi\frac{d_l(t)}{\lambda}},
\end{equation}
where $L$ is the total number of paths, $\alpha_l$ and $d_l(t)$ are the amplitude attenuation and the propagation length of signal through path $l$ respectively, $\lambda$ is the wavelength. An 802.11a/g/n RX implements an OFDM system with 56 subcarriers for each of its antennas. Theoretically, each CSI sample contains 56 matrices with dimensions of $N_{TX}$ $\times$ $N_{RX}$, where $N_{TX}$ and $N_{RX}$ are the number of transceiver antennas.

\subsection{From CSI to Geometric Distance}
\label{CSI-Based}
As shown in Fig. \ref{fig1}, all of the propagation paths can be divided into static paths or dynamic paths. Static paths consist of LoS path and the reflection paths from such static objects as walls and ceiling, which do not change with time, thus they can be considered as a constant. While for dynamic paths, they are reflected by active objects, and each path has a DFS and variable length. Thus, Eq. (2) can be expanded as:
\begin{equation}
H(f,t)=H_s(f,t) + H_d(f,t)=H_s(f,t) + \sum \limits_{l \in P_d} \alpha_le^{-j2\pi\frac{d_l(t)}{\lambda}},
\end{equation}
\begin{figure}[htbp]
\begin{center}
\includegraphics[width=8cm,height=2.9cm]{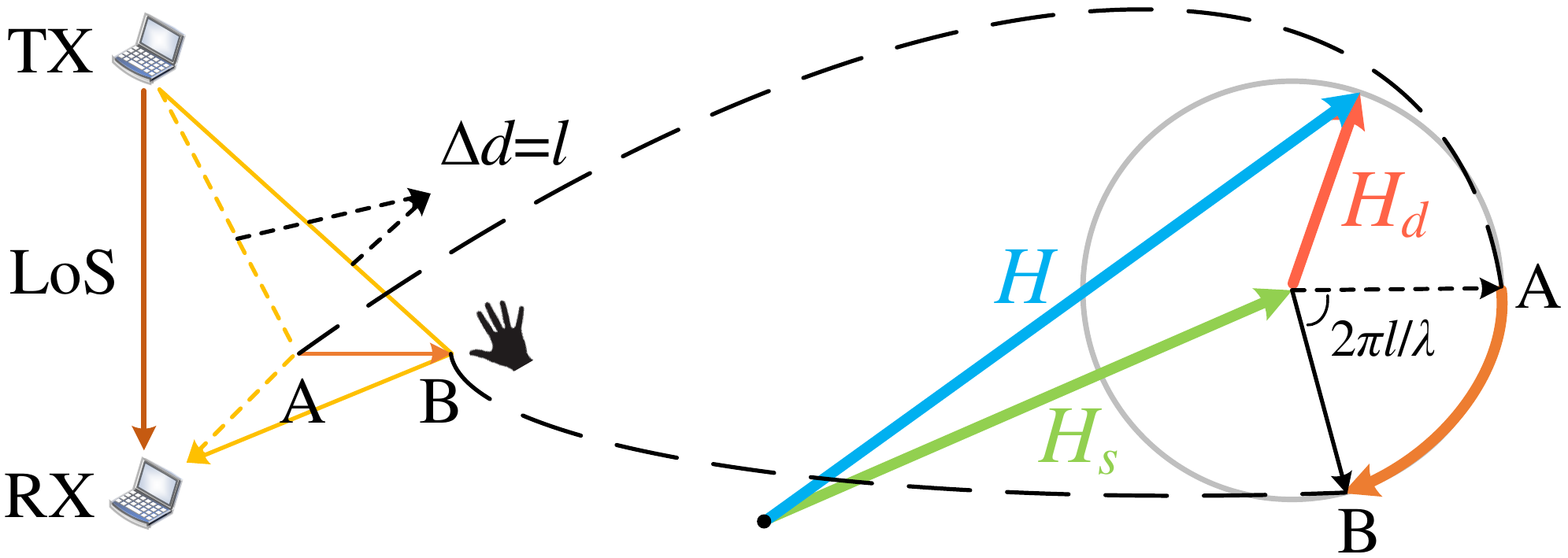}
\caption{Illustration of the CSI dynamics.}
\label{illustrate}
\end{center}
\end{figure}where $H_s(f,t)$ is the aggregate CSI of all static paths, and $P_d$ is the set of dynamic paths. When the environment is static, $P_d$ is empty.

The hand-reflected signals may transmit through multiple paths (e.g., path A and B in Fig. \ref{fig1}), which will generate multiple dynamic components. Based on the observation that the strength of signal which is reflected for more than once (e.g., path B) will be further attenuated than that of direct reflection path signal (e.g., path A). In this case, the signal of direct reflection path is dominated among all the dynamic components. Therefore, without loss of generality, we postulate that hand motion introduces only one dynamic path, i.e., $P_d=1$. Mathematically, $H_d(f,t)$ can be considered as a point $(\alpha\cos(2\pi\frac{d(t)}{\lambda}), -\alpha\sin(2\pi\frac{d(t)}{\lambda}))$ on complex plane. When hand motion results in an increase of $d(t)$ by $l$, there is a rotation of $H_d(f,t)$ around its origin, i.e., $(\alpha\cos(2\pi\frac{d(t)}{\lambda}), -\alpha\sin(2\pi\frac{d(t)}{\lambda})) \to (\alpha\cos(2\pi\frac{d(t)+l}{\lambda}), -\alpha\sin(2\pi\frac{d(t)+l}{\lambda}))$. To be specific, when $d(t)$ increases or decreases by $k\lambda$, $H_d(f,t)$ rotates $2k\pi$ clockwise or counterclockwise. As time evolves, $H_d(f,t)$ samples will draw In-phase (I)/Quadrature (Q) points to construct circular traces with radius $\alpha$, and $H(f,t)$ is a $H_s(f,t)$ translation based on $H_d(f,t)$, which is shown in Fig. \ref{illustrate}. The phase changes of CSI can be leveraged to estimate the motion distance, while the the rotation direction determines the motion direction (i.e., close to or away from transceiver). The amplitude $\alpha$ (i.e., the radius of circle) is inversely proportional to the length of reflection path, so the circles formed by the rotation are not exactly with the same radius. Our intuition is to measure the phase changes of the dynamic path caused by hand motion, and derivate the length changes for gesture tracking in 2D space.

\section{CSI-Based Accurate Distance Estimation}
In this section, we first identify the CSI measurement error sources and present the denosing process for accurate estimation, and then carry out several benchmark experiments to verify our derivation.

\subsection{CSI Error Sources and Correction}

\begin{figure}[htbp]
\begin{center}
\includegraphics[width=8.4cm,height=2.4cm]{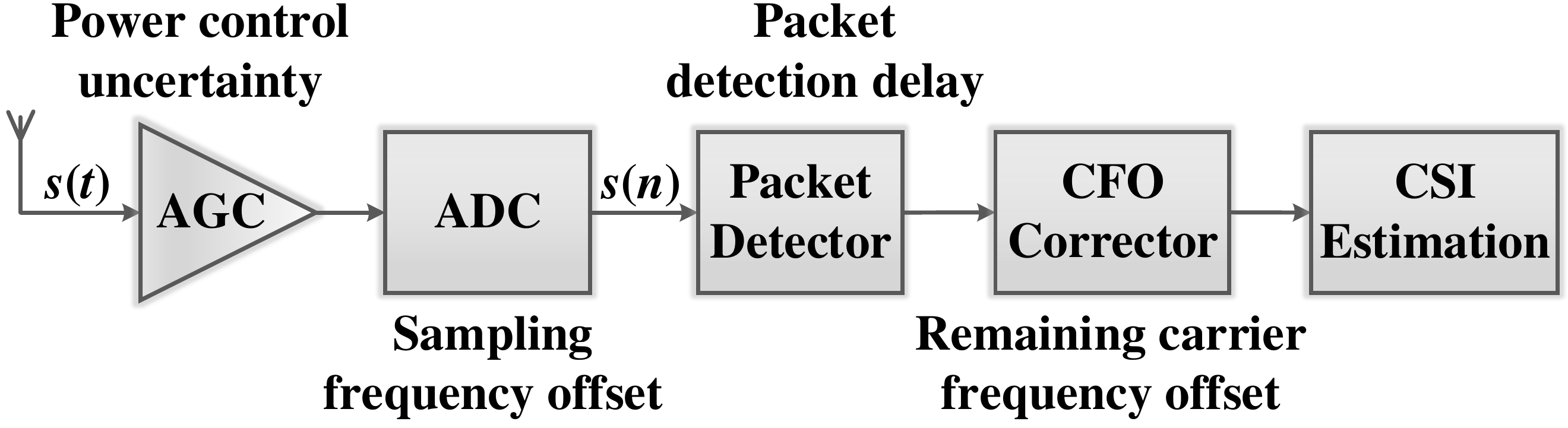}
\caption{Illustration of signal processing in WiFi NICs.}
\label{fig_signal}
\end{center}
\end{figure}

In real-world WiFi systems, CSI measurements have abundant distortions from both aspects of amplitude and phase, which cannot be directly used for tracking tasks. The main wireless signal processing in WiFi NICs is pointed out in Fig. \ref{fig_signal}. Firstly, the incident passband signal is down converted to the baseband signal $s(t)$ based on antennas, and then processed by the Automatic Gain Controller (AGC) to compensate for the amplitude of signal and sampled by Analog-to-Digital (ADC) to obtain the digital $s[n]$. The Packet Boundary Detector (PBD) executes energy detection or correlation between $s[n]$ and a predefined 802.11 preamble to identify the packet. Once a packet is confirmed, the signal central frequency will be calibrated by the Central Frequency Offset (CFO) corrector. The RX then estimates the CSI based on $s[n]$ with interpolation of 802.11 preamble. Therefore, CSI describes not only the frequency response of the wireless channel in passband, but also the errors of the internal circuit in baseband \cite{pisplicer}. 

\begin{figure*}[htbp]
\begin{center}
\subfigure[Amp. at 20$^{th}$ subcarrier on antenna 1 and 2.]{
\includegraphics[width=5.4cm,height=3.8cm]{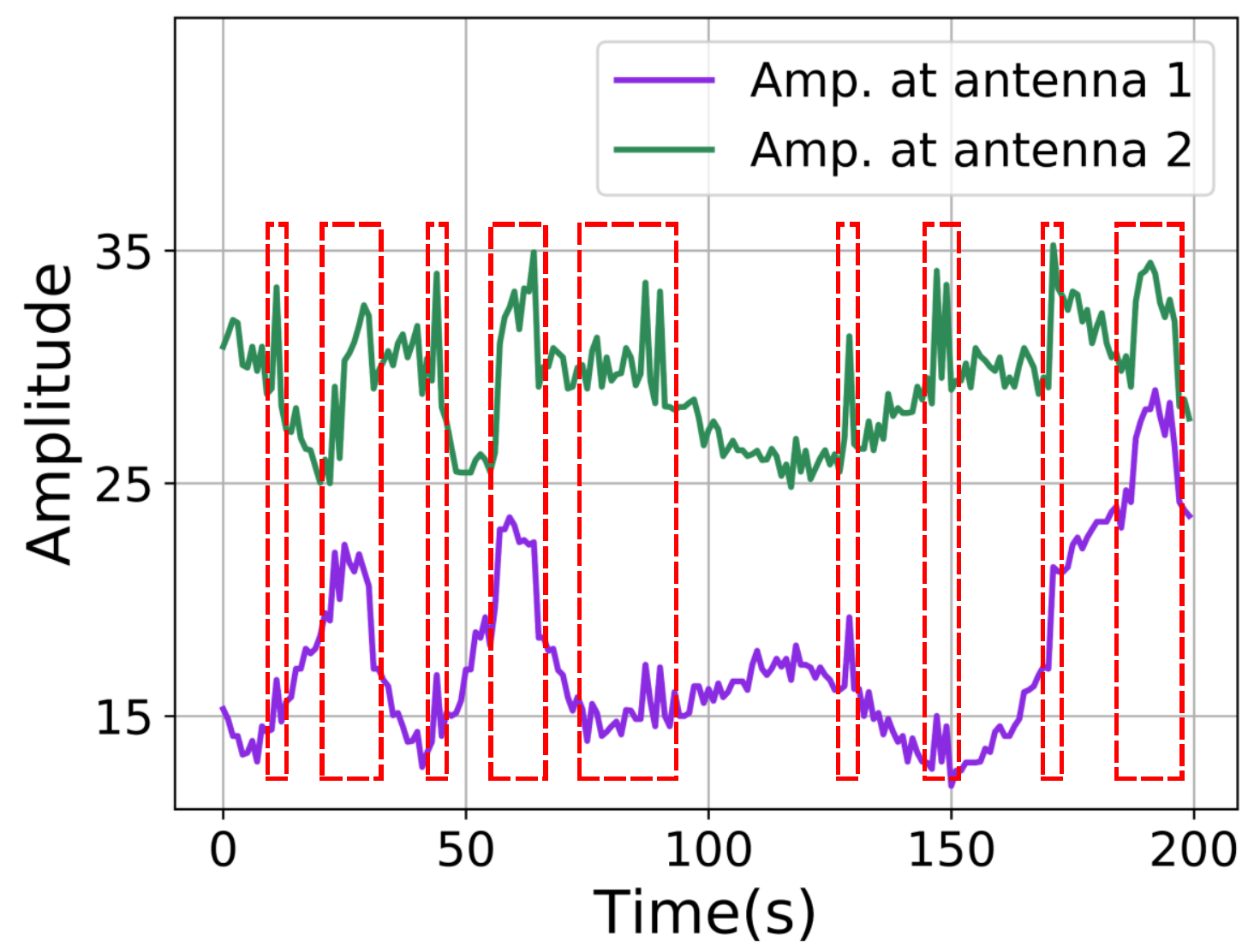}
}
\subfigure[Ratio of Amp. at 20$^{th}$ subcarrier.]{
\includegraphics[width=5.4cm,height=3.8cm]{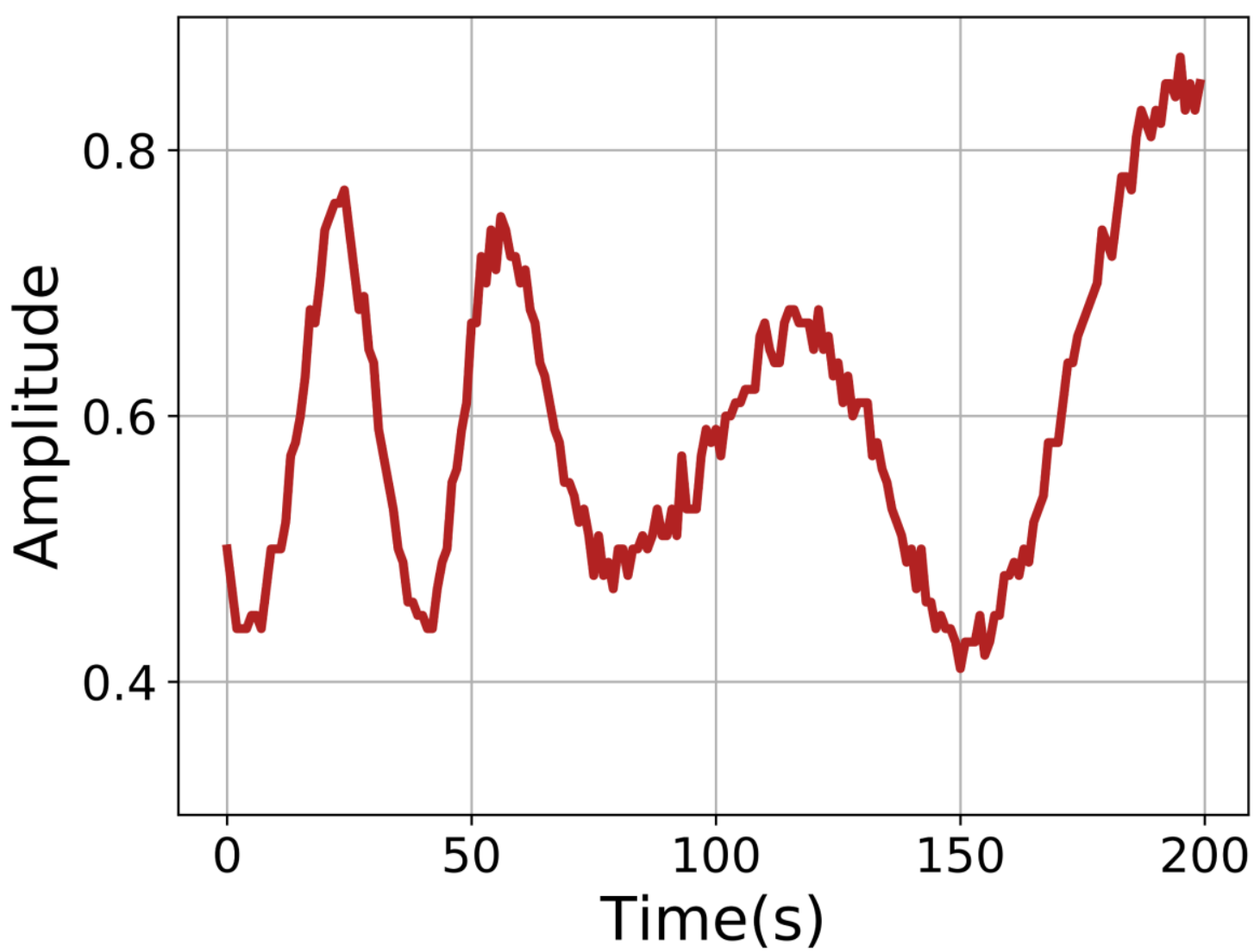}
}
\subfigure[Amp. of PCA results.]{
\includegraphics[width=5.4cm,height=3.8cm]{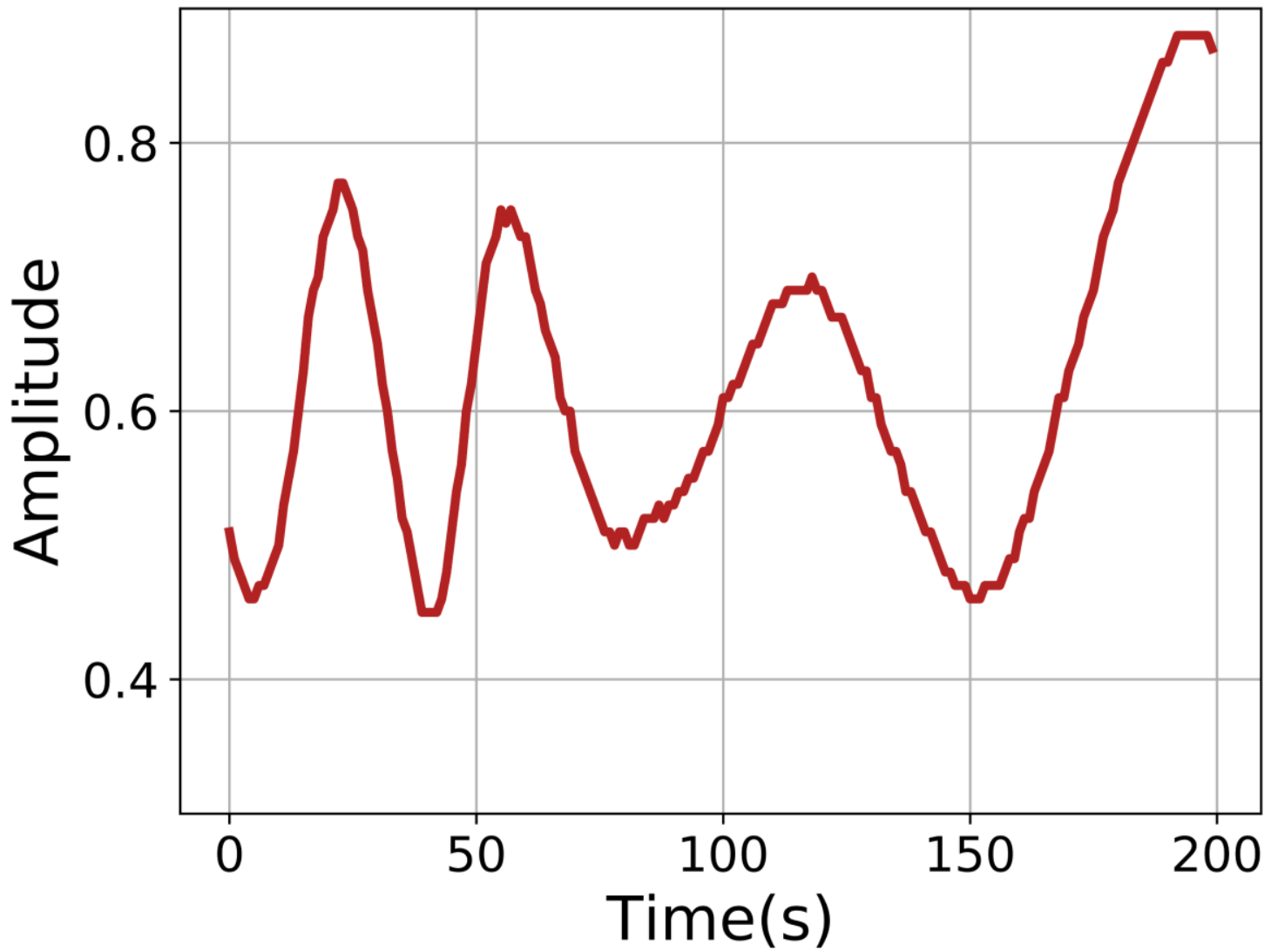}
}
\caption{Before and after Amp. denoising, where (a) is raw Amp., (b) is the Amp. ratio, and (c) is the PCA result.}
\label{fig2}
\end{center}
\end{figure*}

\begin{figure*}[htbp]
\begin{center}
\subfigure[Phase at 20$^{th}$ subcarrier on antenna 1 and 2.]{
\includegraphics[width=5.4cm,height=3.8cm]{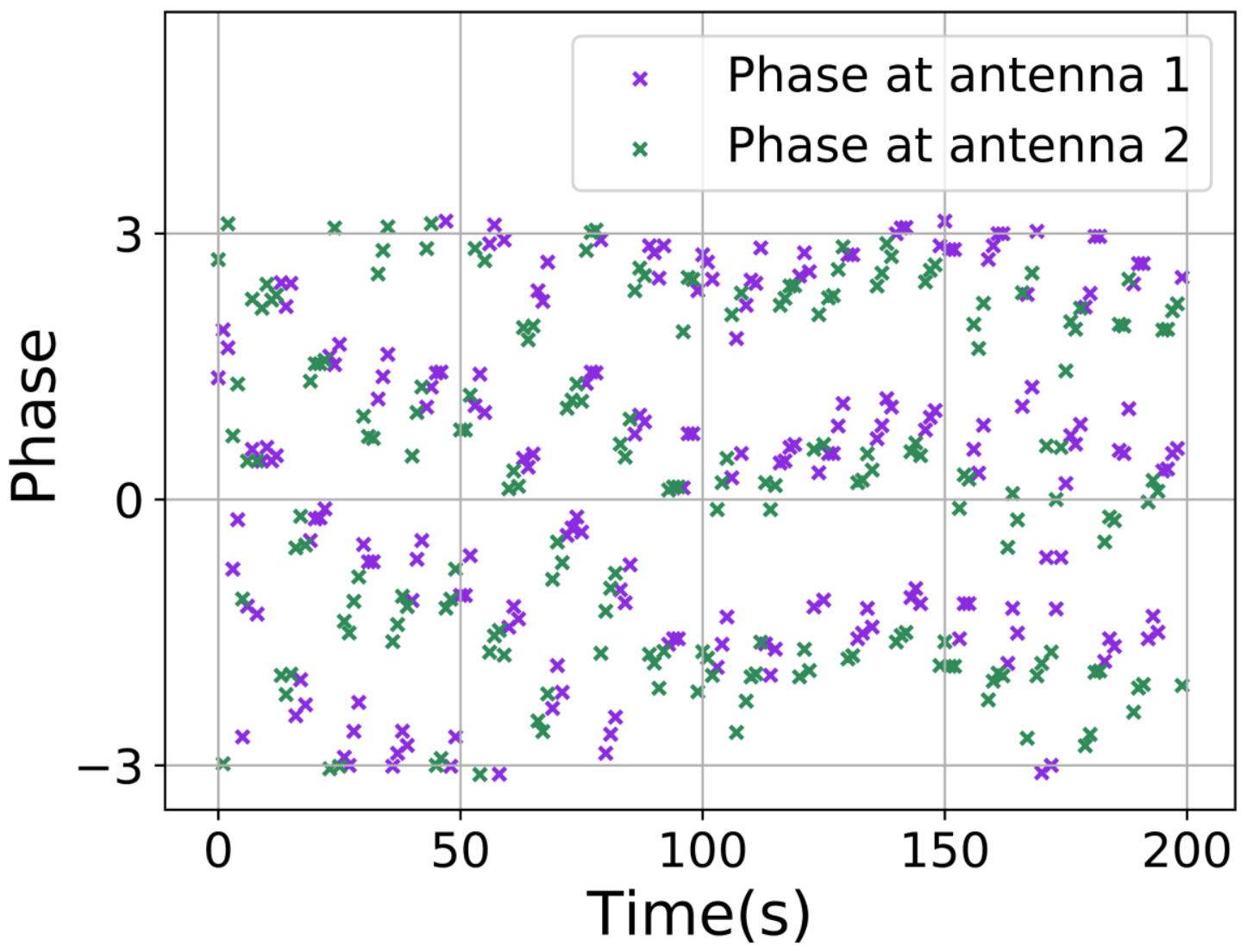}
}
\subfigure[Phase difference at 20$^{th}$ subcarrier.]{
\includegraphics[width=5.4cm,height=3.8cm]{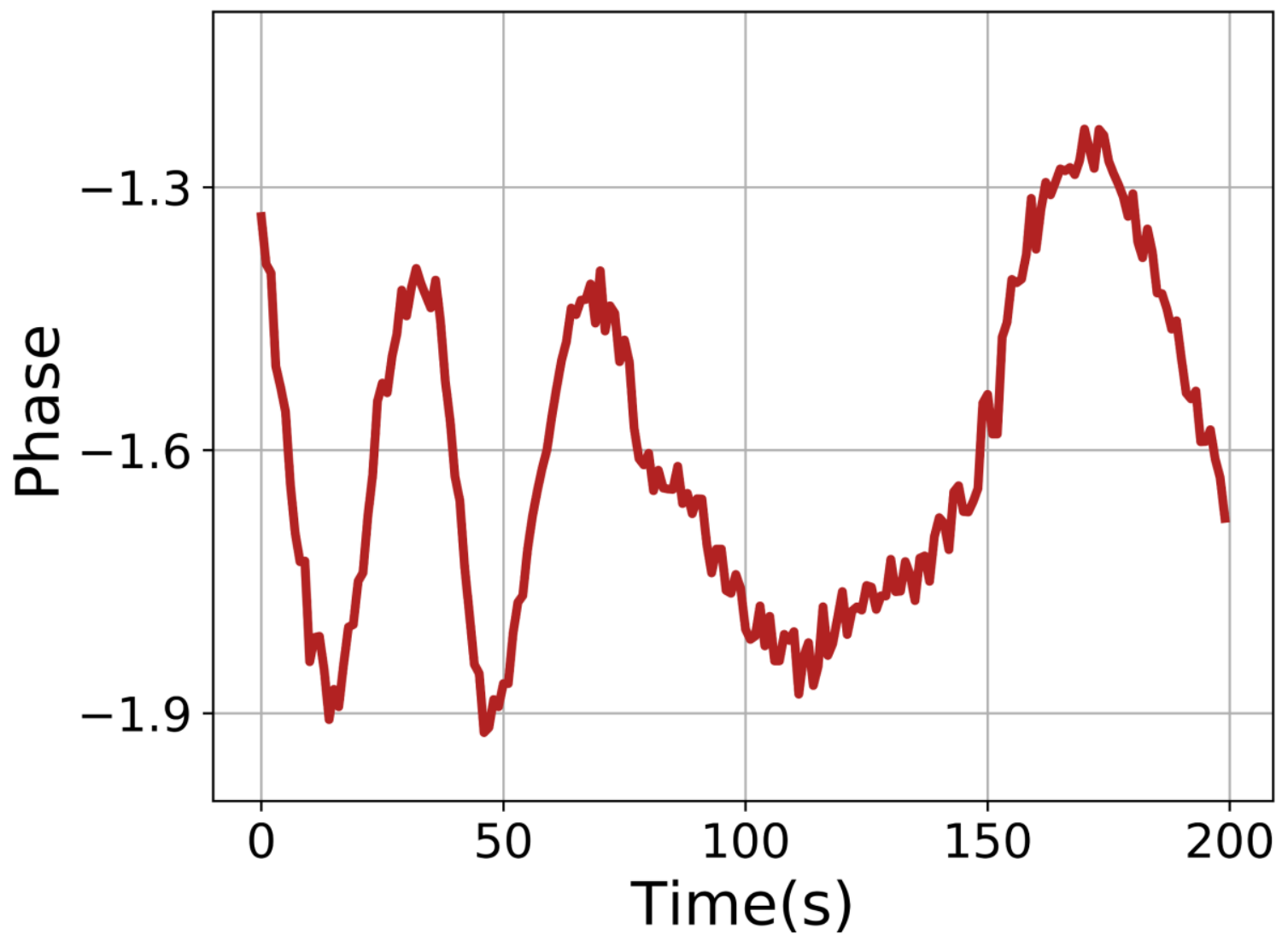}
}
\subfigure[Phase of PCA results.]{
\includegraphics[width=5.4cm,height=3.8cm]{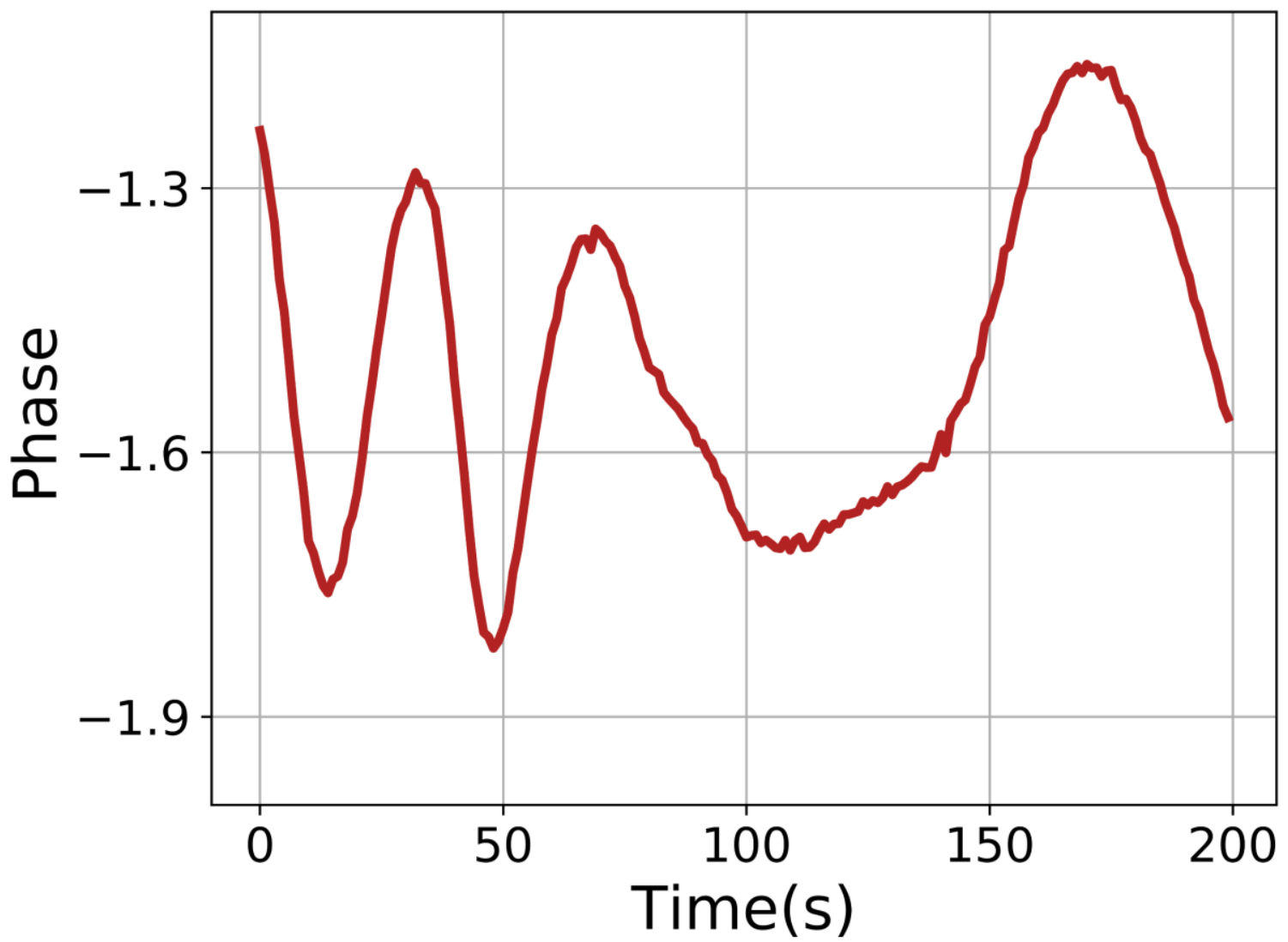}
}
\caption{Before and after phase denoising, where (a) is raw phase, (b) is the phase difference, and (c) is the PCA result.}
\label{fig3}
\end{center}
\end{figure*}

For the amplitude, it fluctuates due to the power control uncertainty errors of AGC and surrounding electromagnetic noises. Fig. \ref{fig2} (a) depicts high impulses and burst noises when hand moves back and forth around the WiFi transceiver. We observe that the moment, duration and profile of these noises are nearly identical across RX antennas. Indeed, the impulse noise is scale noise, which is consistent across the antennas of the same RX, despite the varying of power scale with time. WiFi cards are generally installed with several antennas for Multiple Input and Multiple Output (MIMO) transmission. Thus, we leverage the ratio of amplitude between two adjacent antennas to eliminate the noises. It can also be seen that the amplitude fluctuations related to motions vary with antennas due to the physical spatial spacing in between. Thus, amplitude ratio can remove noises and persist signals of interest, as illustrated in Fig. \ref{fig2} (b).

For the phase, the sampling frequencies and central frequencies of the transceiver are not perfectly synchronized, which will lead to Sampling Frequency Offset (SFO) and CFO. The SFO causes $s[n]$ a time shift $\tau_o$ with respect to $s[t]$, and thus introduces near constant errors $\lambda_o$ to the CSI phases from different subcarriers. The CFO is compensated by the CFO corrector, but due to the hardware imperfection, the compensation is incomplete \cite{csi}, leading to time-variant random phase offsets $\beta$. Besides, due to correlator sensitivity of PBD, the packet detection introduces another time shift $\tau_b$, resulting in errors $\lambda_b$ to the phases. From the above error sources, the phase measurement $\phi_{k}$ for subcarrier $k$ can be expressed:
\begin{equation}
\begin{aligned}
\begin{split}
\phi_{k}=&\theta_{k} + k\cdot(\lambda_o + \lambda_b) + \lambda_c + \beta + z,
\end{split}
\end{aligned}
\end{equation}
where $\theta_{k}$ is the true phase rotation caused by signal propagation, $\lambda_o$,  $\lambda_b$ and $\beta$ are the phase offsets caused by PBD, SFO and CFO, respectively. $\lambda_c$ is the random initial phase offset due to the jitters of Phase-Locked Loop (PLL) \cite{pisplicer}. $k$ ranges from -28 to 28 for 20 MHz bandwidth in IEEE 802.11n specification, and $z$ is the additive white Gaussian noise. As shown in Fig. \ref{fig3} (a), these
errors significantly distort CSI true phase, and thus prevent its practical usage. Note that these errors are identical across RX antennas for a particular subcarrier, as all the receiver chains on the same WiFi NIC are time synchronized. Therefore, we eliminate these errors by the phase difference \cite{indotrack, rtfall} between antennas across all subcarriers. Take antenna 1 and 2 for example:
\begin{equation}
\begin{aligned}
\begin{split}
\widetilde{\theta_k} = \phi_{1, k} - \phi_{2, k}=\theta_{1,k} - \theta_{2,k} + \Delta z.
\end{split}
\end{aligned}
\end{equation}

The phase deniosing results can be shown in Fig. \ref{fig3} (b). With the denoised amplitude $\widetilde{\alpha}$ and phase $\widetilde{\theta}$ of CSI, a new complex sample can be derived: $H_{ratio}(f,t) = \widetilde{\alpha}\cdot \cos (\widetilde{\theta}) + j\widetilde{\alpha}\cdot \sin (\widetilde{\theta})$, which is exactly the complex ratio of CSI between adjacent antennas in substance. Based on the observation \cite{farsense}, when LoS path exits, $H_{ratio}(f,t)$ will rotate clockwise as the length of reflection path increases. Otherwise, when LoS is occluded, it will rotate counterclockwise. In other words, the estimated motion direction (i.e., close to or away from transceiver) based on rotation direction of $H_{ratio}(f,t)$ is diametrically opposite when LoS exists compared to not. We refer readers to \cite{farsense} for more detailed theoretical derivation, and we will take the existence of LoS as an example for research in later part. We also note that, due to hand motion, the changes of CSI on different subcarriers are correlated. Therefore, we further separate $H_{ratio}(f,t)$ from noises by performing PCA across all the subcarriers. The first PCA component is preserved, and the results can be shown in Fig. \ref{fig2} (c) and Fig. \ref{fig3} (c).

\begin{figure}[htbp]
\begin{center}
\includegraphics[width=8.0cm,height=3.8cm]{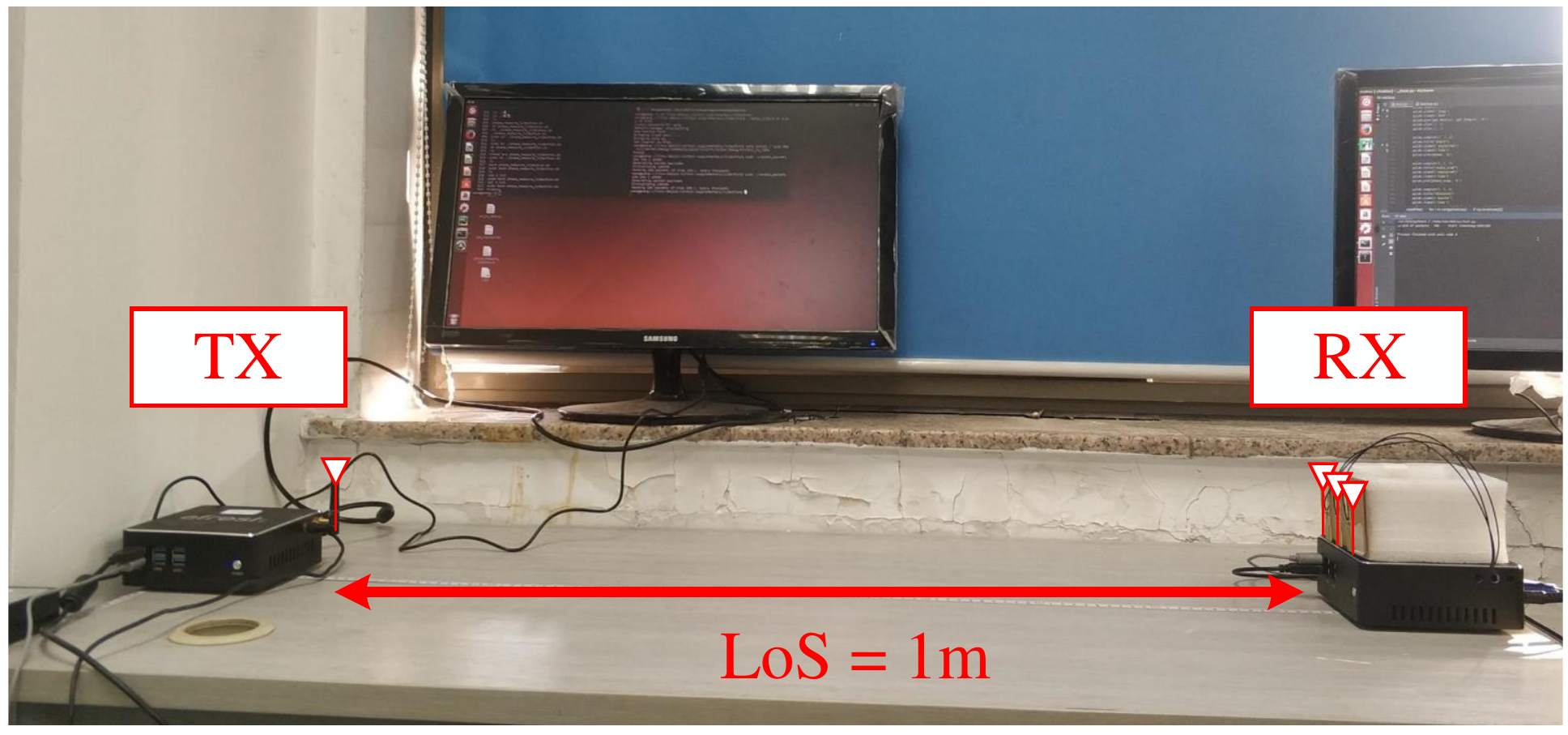}
\caption{Scenario for benchmark experiments.}
\label{fig4}
\end{center}
\end{figure}

\subsection{Verification via Benchmark Experiments}
\label{111}

In this part, we will verify the performance of $H_{ratio}(f,t)$ via the proof-of-concept experiments. As shown in Fig. \ref{fig4}, the verification experiments are carried out in a typical laboratory environment. We set a pair of WiFi transceiver 1 m apart, and they are equipped with Intel 5300 NICs and vertically-polarized omni-directional commodity patch antennas. We upgrade the device driver following \cite{csitool} and configure NICs to run at the central frequency of 5.32 GHz with 20 MHz bandwidth. Transceiver and hands are set on the same line, so the length of dynamic path changes by two times of the distance of hand motion. To precisely control the displacement of hands, the movement distance is measured by the ruler along the line. When hands move towards or away from the transceiver, the reflection path length changes accordingly. 

\begin{figure}[htbp]
\begin{center}
\subfigure[Moving away from transceiver.]{
\includegraphics[width=7.5cm,height=1.7cm]{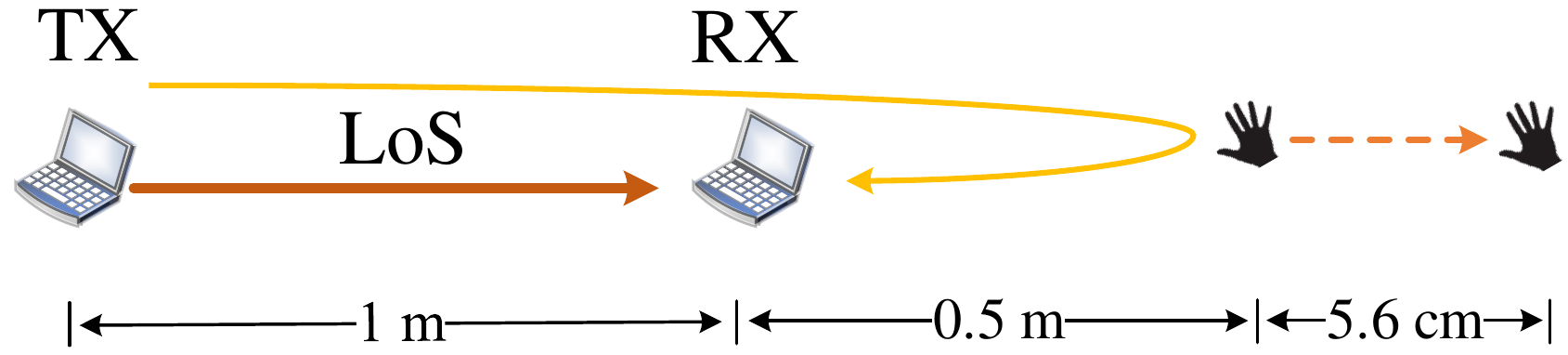}
}
\subfigure[Raw CSI trace.]{
\includegraphics[width=3.7cm,height=3.4cm]{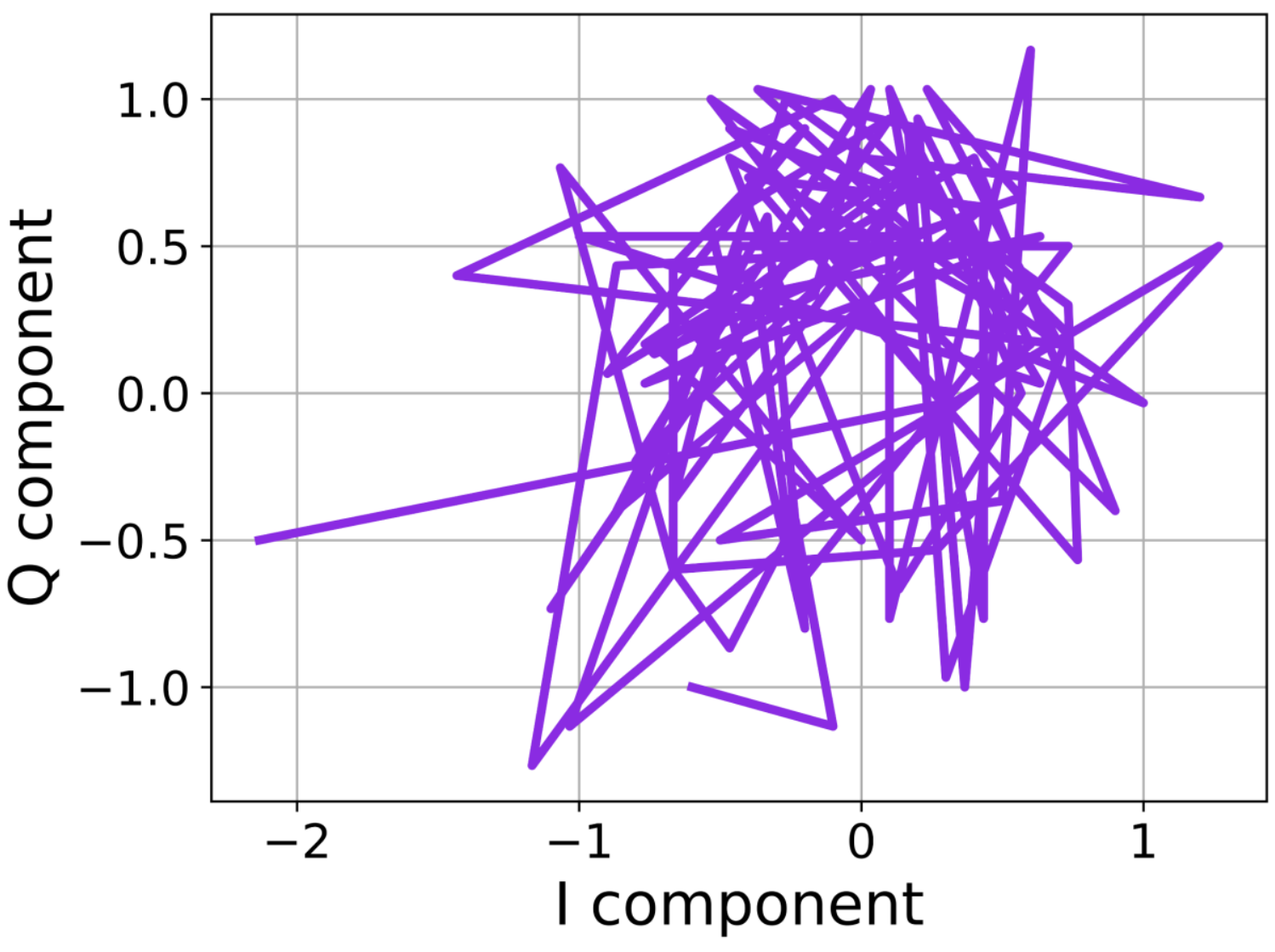}
}
\subfigure[$H_{ratio}(f,t)$ trace.]{
\includegraphics[width=3.7cm,height=3.4cm]{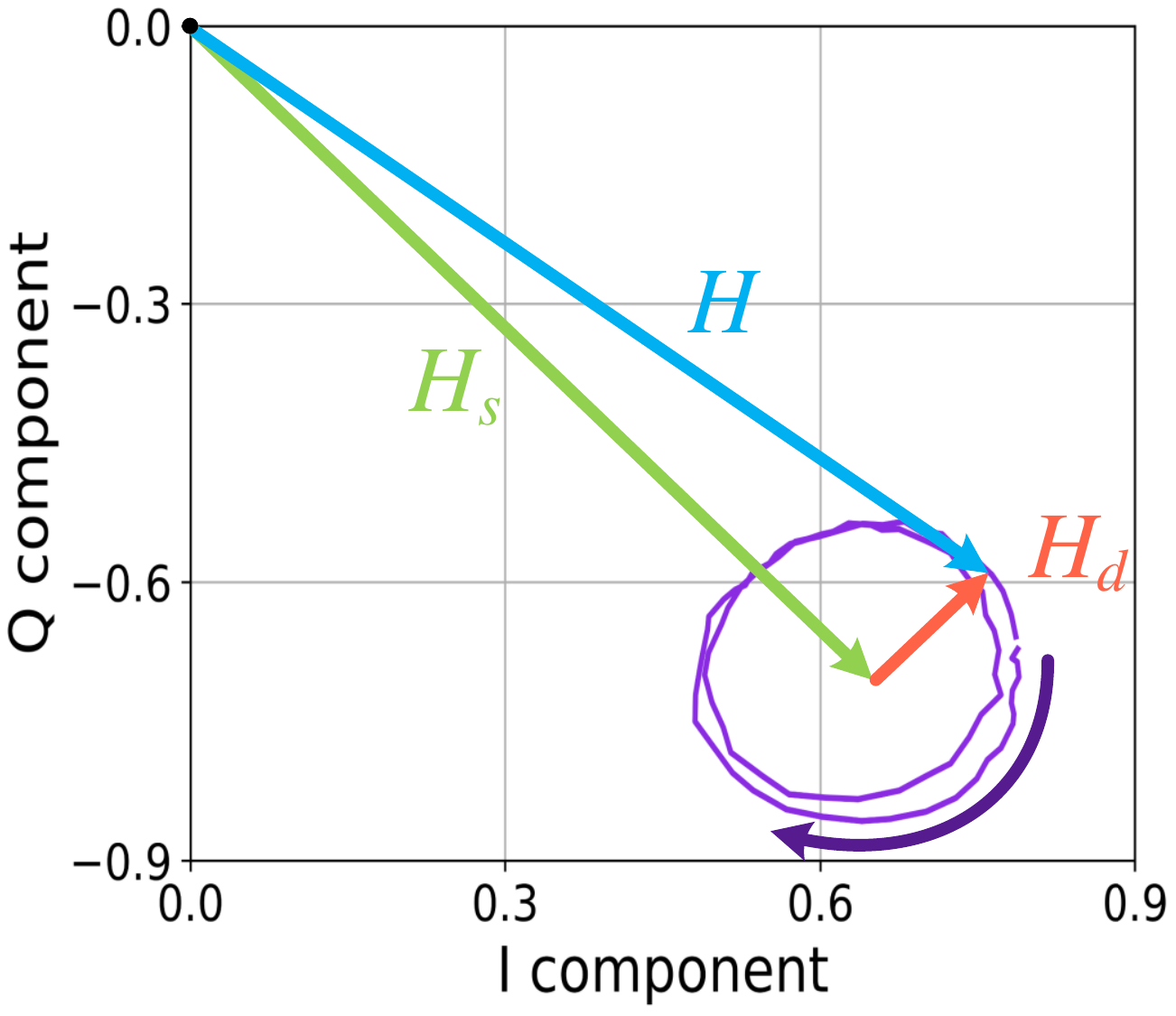}
}
\caption{Hand moves away from transceiver.}
\label{fig5}
\end{center}
\end{figure}

\begin{figure}[htbp]
\begin{center}
\subfigure[Moving towards transceiver.]{
\includegraphics[width=7.5cm,height=1.7cm]{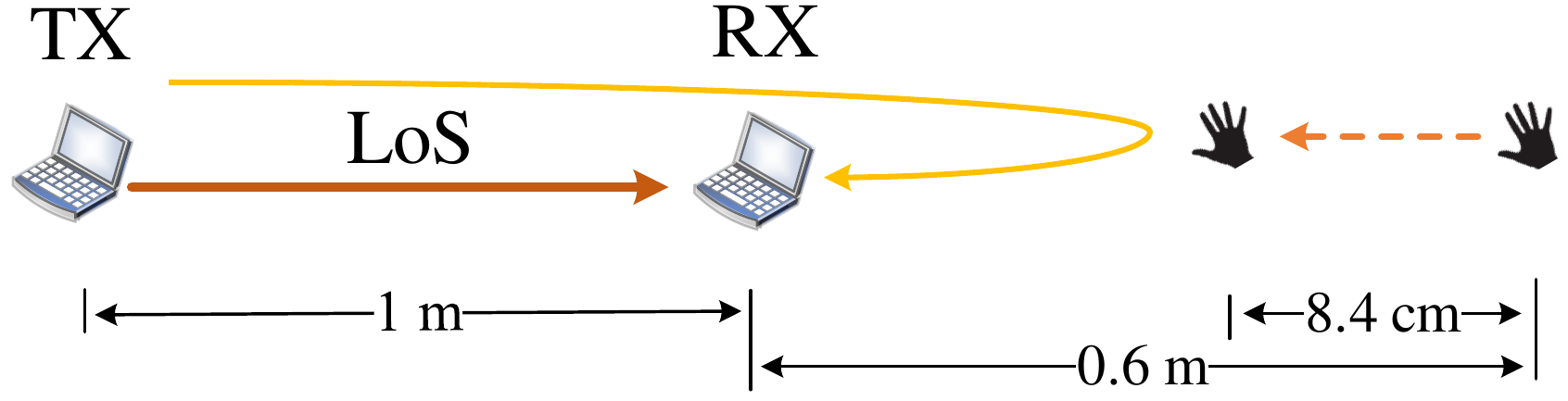}
}
\subfigure[Raw CSI trace.]{
\includegraphics[width=3.7cm,height=3.4cm]{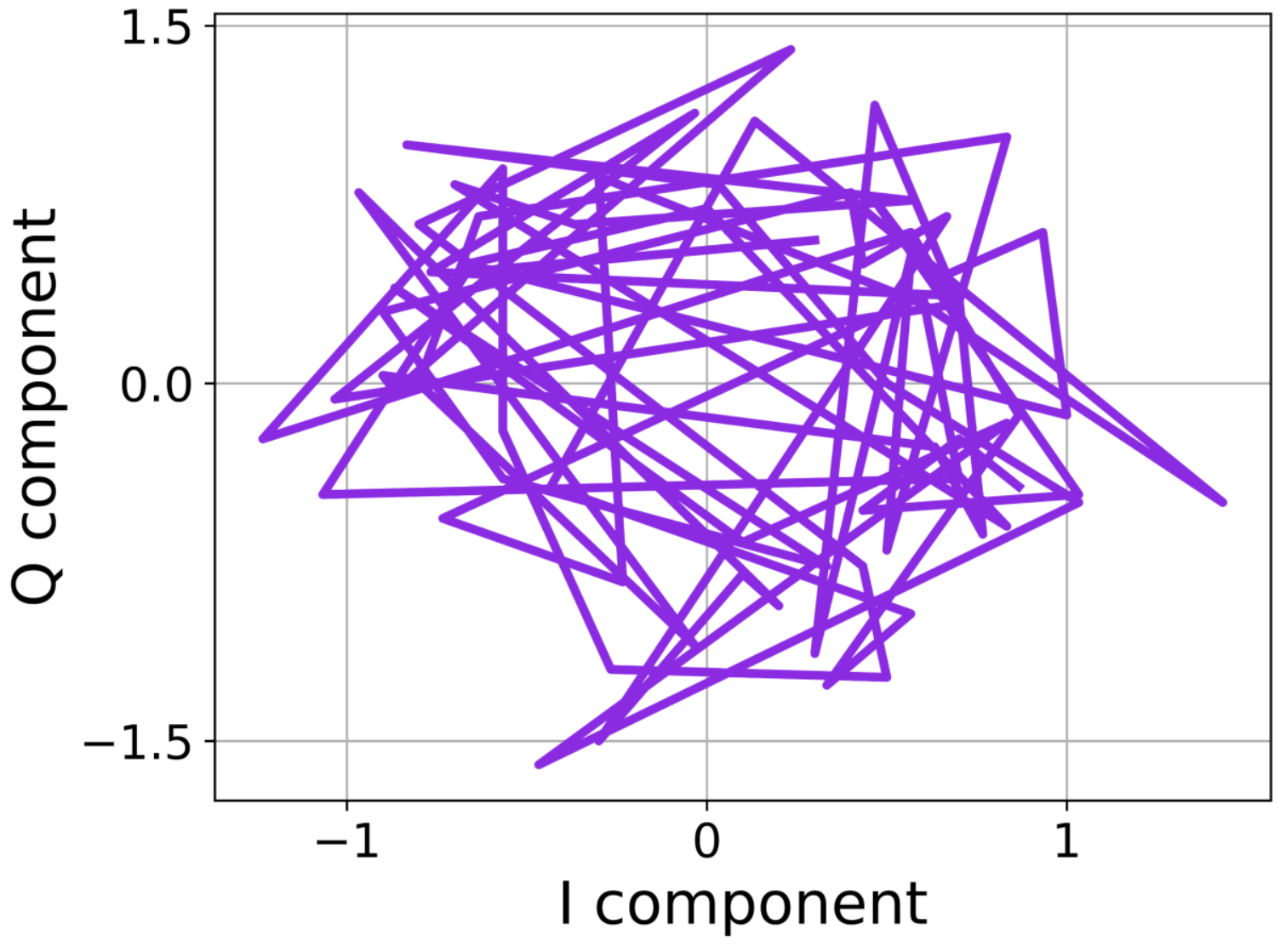}
}
\subfigure[$H_{ratio}(f,t)$ trace.]{
\includegraphics[width=3.7cm,height=3.4cm]{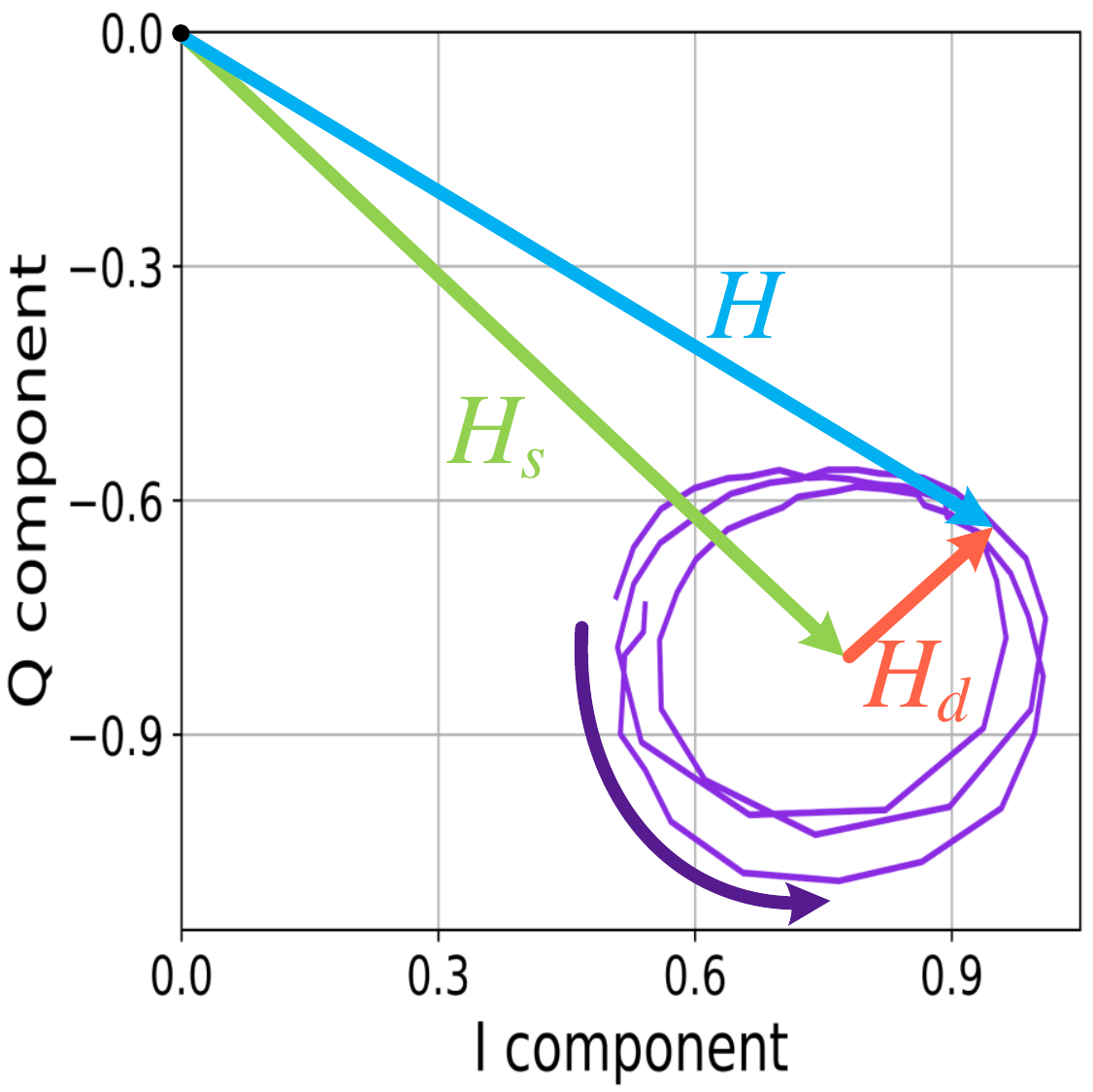}
}
\caption{Hand moves towards transceiver.}
\label{fig6}
\end{center}
\end{figure}

As shown in Fig. \ref{fig5}, hand first moves 5.6 cm (i.e., around one $\lambda$) horizontally away from the transceiver with the start position of 0.5 m from RX. The complex traces in Fig. \ref{fig5} (b) and (c) are made up of the I/Q parts of raw CSI measurements and $H_{ratio}(f,t)$, respectively. It can be seen that the raw CSI trace is chaotic, and $H_{ratio}(f,t)$ rotates clockwise in accordance with circle in complex plane, with a 4$\pi$ radian. Then, we take 8.5 cm (i.e., around 1.5 $\lambda$) towards transceiver with the start point at 0.6 m. We can get a counterclockwise rotation of $H_{ratio}(f,t)$, with a 6$\pi$ radian in Fig. \ref{fig6}. These above prove that the denoised CSI, i.e., $H_{ratio}(f,t)$, can verify our derivation, and thus competent for the gesture tracking tasks.

\section{The Proposed Scheme}
\begin{figure*}[htbp]
\begin{center}
\includegraphics[width=16cm,height=3.3cm]{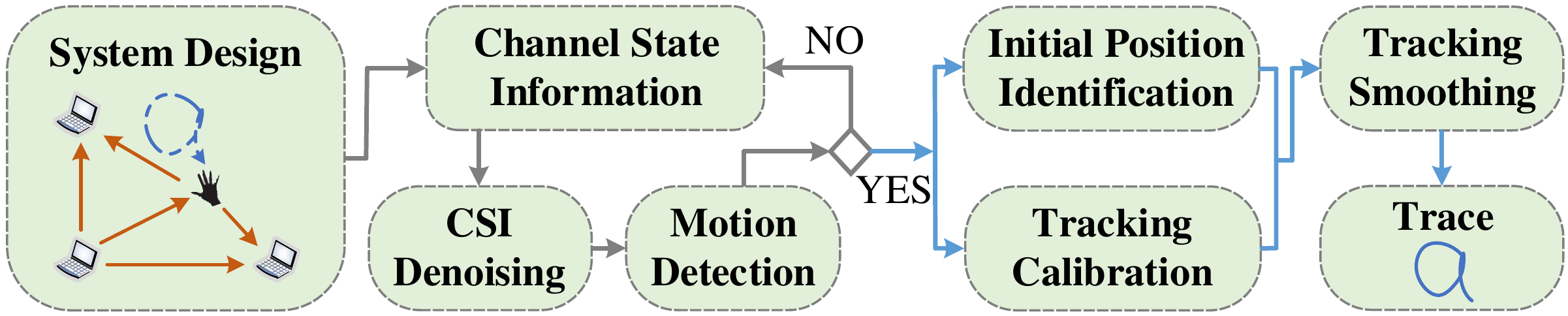}
\caption{System architecture of CentiTrack.}
\label{fig8}
\end{center}
\end{figure*}

In this section, we present our design and implementation of CentiTrack by drawing/writing in the air. The system consists of five basic modules: System Design, Motion Detection, Initial Position Identification, Tracking Calibration and Tracking Smoothing, as shown in Fig. \ref{fig8}.

\subsection{System Design}
\begin{figure}[htbp]
\begin{center}
\subfigure[One pair of transceiver.]{
\includegraphics[width=4.1cm,height=2.5cm]{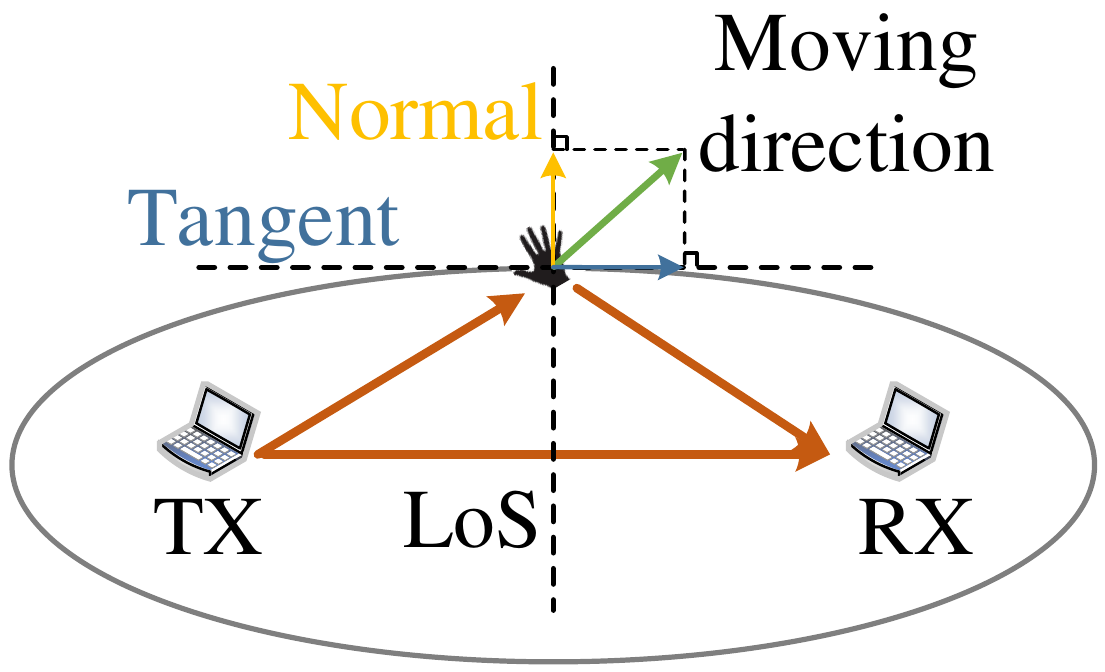}
}
\subfigure[Two pairs of vertical transceiver.]{
\includegraphics[width=4.1cm,height=3.0cm]{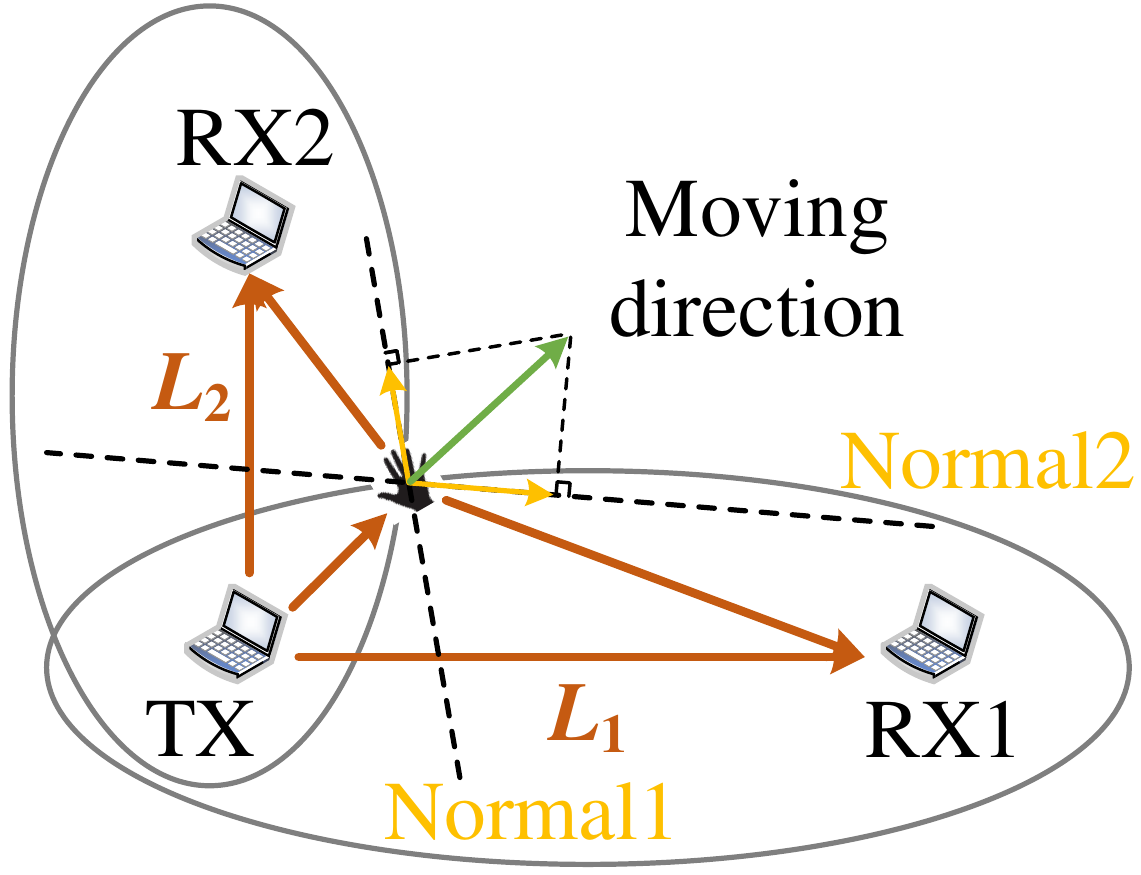}
}
\caption{Geometrical relationship between hand motion and the length change of reflection path.}
\label{fig9}
\end{center}
\end{figure}

CentiTrack leverages one TX and two RXs to track hands on a 2D plane. Consider a free-space scenario and one pair of WiFi transceiver, there are several ellipses with foci at TX and RX, where each ellipse corresponds to one reflection path. All the reflectors are on their own ellipses, and hands are no exception. According to the properties of ellipse, the length of reflection path remains unchanged when the hand moves in the tangent direction of its ellipse (i.e., equal to the major axis), as shown in Fig. \ref{fig9} (a). If the motion has a moving component in the normal direction of ellipse, the movement will change the path length and thus result in DFS. 
Therefore, another pair of transceiver is arranged perpendicular to the first transceiver pair as illustrated in Fig. \ref{fig9} (b), to aggregate movement components in the normal direction of both ellipses. Here, we take TX as the origin, transceiver pair TX-RX1 as the x-axis, and TX-RX2 as the y-axis to establish a Cartesian coordinate, where two RXs are at $(L_1, 0)$, $(0, L_2)$. If we know the reflection path length $p$ and $q$ of the two ellipses, the coordinates of hands $(x,y)$ can be solved:
\begin{equation}
\begin{aligned}
\begin{split}
&\sqrt{x^2+y^2} + \sqrt{(x-L_1)^2+y^2}=p,\\
&\sqrt{x^2+y^2} + \sqrt{x^2+(y-L_2)^2}=q,
\end{split}
\end{aligned}
\end{equation}
where $x>0, y>0$. The length changes of dynamic path can be estimated based on the radians of arcs in $H_{ratio}(f,t)$. Therefore, given the initial length of reflection path $p_0$ and $q_0$, i.e., initial position $(x_0, y_0)$, the absolute values of $p$, $q$ and the coordinates of hands $(x, y)$ can be updated thereafter.

\subsection{Motion Detection}
\begin{figure}[htbp]
\begin{center}
\includegraphics[width=5.8cm,height=4.3cm]{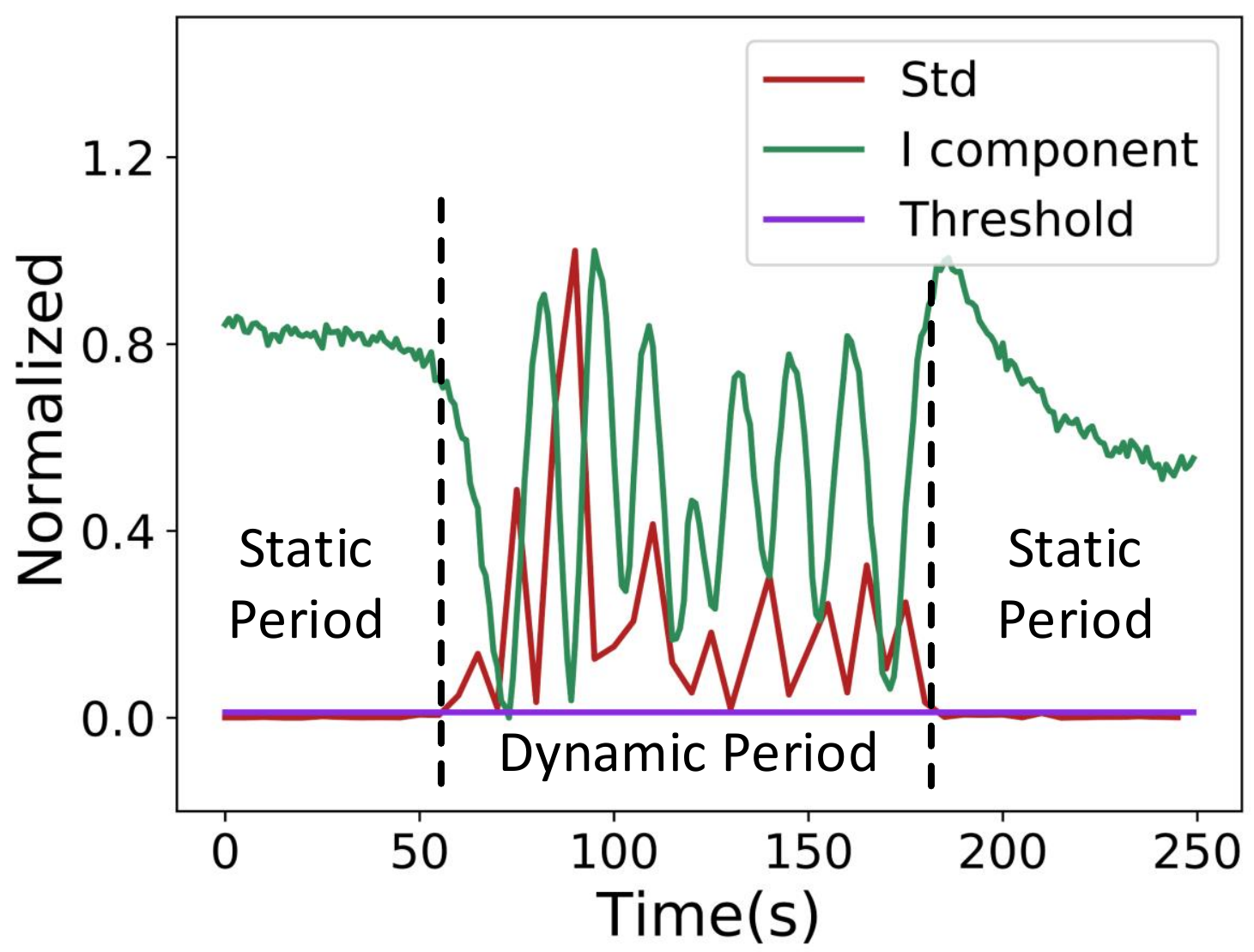}
\caption{Motion detection in static and dynamic periods.}
\label{fig10}
\end{center}
\end{figure}

For gesture tracking, we need to detect the start and end location of each hand trace over time. When hand keeps static, the amplitude of $H_{ratio}(f,t)$ is considerably stable. Once hand begins to move, the amplitude will undergo large fluctuations. As shown in Fig. \ref{fig10}, Std represents the standard deviation of $H_{ratio}(f,t)$ I component for each short period. We adopt a 0.1s sliding window with a 50\% overlay to detect the real-time signal variance. For simplicity, both Std and I component are normalized to $[0,1]$. It can be seen that the variance of static periods is much smaller than that of dynamic periods. Therefore, the motion start and end point can be easily detected by an empirical threshold, which is typically ten times of Std of the static periods. We also note that the Motion Detection module works on both pairs of transceivers simultaneously to complement each other.

\subsection{Initial Position Identification}
\subsubsection{Empirical Study}
\begin{figure}[htbp]
\begin{center}
\subfigure[1D space.]{
\includegraphics[width=4.15cm,height=3.0cm]{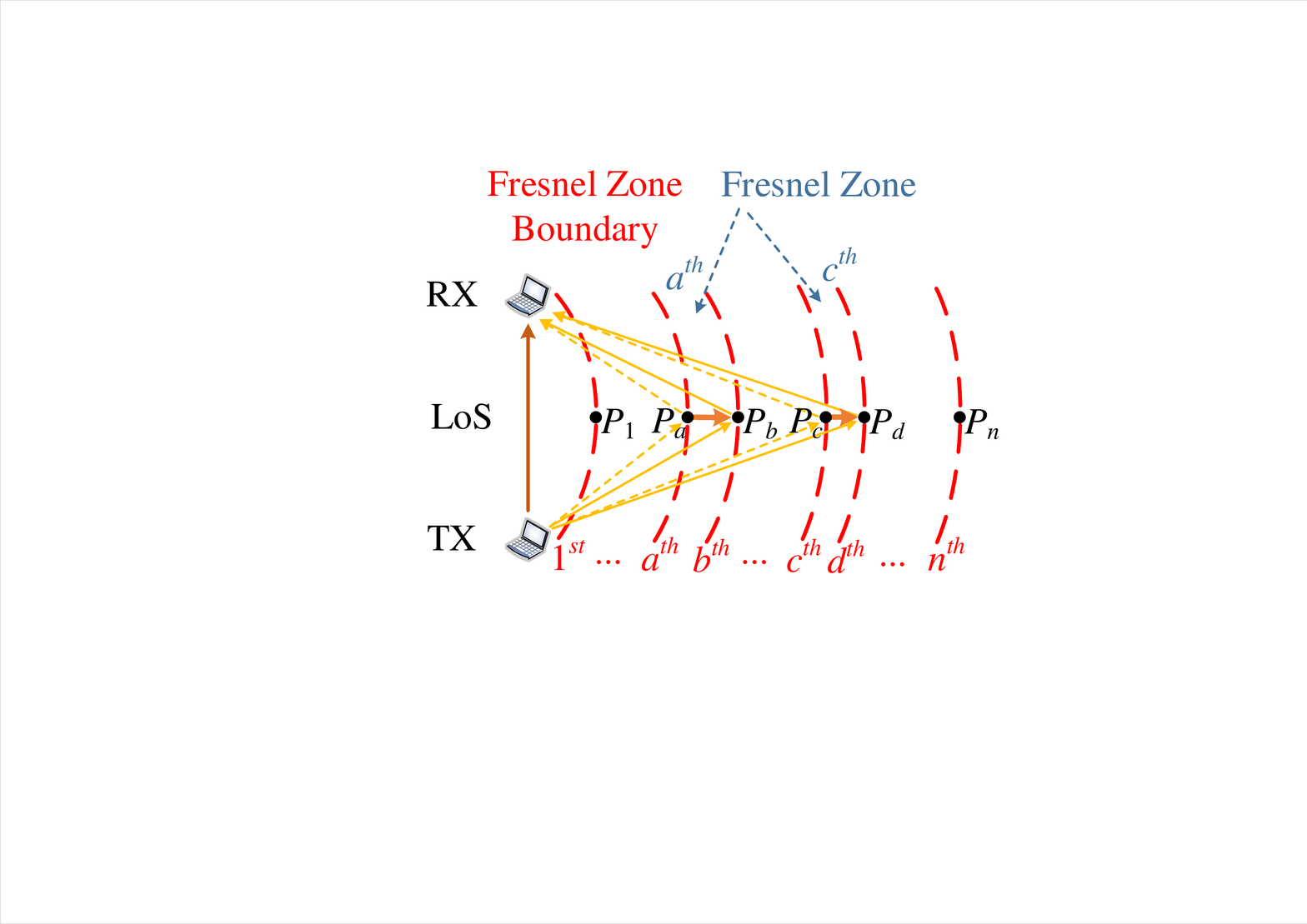}
}
\subfigure[2D space.]{
\includegraphics[width=4.1cm,height=3.0cm]{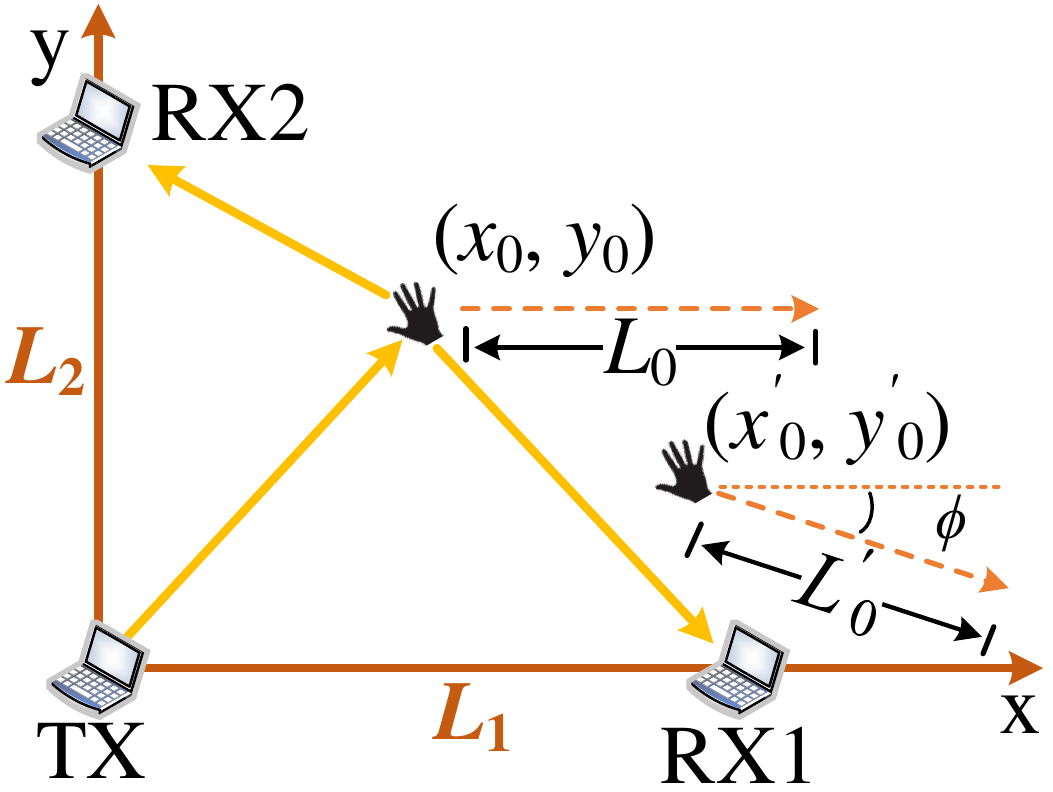}
}
\caption{Empirical study of initial position estimation in 1D and 2D space.}
\label{fig11}
\end{center}
\end{figure}

The trace of hand motion is determined not only by the changes of path length but also by its initial position. The biased initial position will severely degrade the tracking accuracy. Take the case that users push hands along the perpendicular bisector of a pair of transceiver with a constant LoS. We record the ground truth of the start position $P_s$ and end position $P_e$. Following simple geometry, the length change of the reflection path is $\Delta d = 2(\sqrt{(\frac{LoS}{2})^2 + P_e^2} - \sqrt{(\frac{LoS}{2})^2 + P_s^2})$. Given a fixed length change, the geometry distance $|P_e - P_s|$ will vary as $P_s$. We study these variations with the Fresnel Zone model \cite{fresnel}. Consider a free space, the transceiver TX and RX transmit WiFi signals with a wavelength $\lambda$, as illustrated in Fig. \ref{fig11} (a). The Fresnel Zones are concentric ellipses, and the $n^{th}$ Fresnel Zone corresponds to the region surrounded by the $(n-1)^{th}$ and the $n^{th}$ ellipse. The point $P_n$ on the outer boundary of the $n^{th}$ Fresnel Zone satisfies the formula:
\begin{equation}
\begin{aligned}
\begin{split}
|TXP_n| + |P_nRX| - LoS = n\lambda / 2,
\end{split}
\end{aligned}
\end{equation}
where $|TXP_n|$ and $|P_nRX|$ are the distances from TX and RX to point $P_n$, respectively. 
\begin{figure}[htbp]
\begin{center}
\includegraphics[width=5.3cm,height=3.0cm]{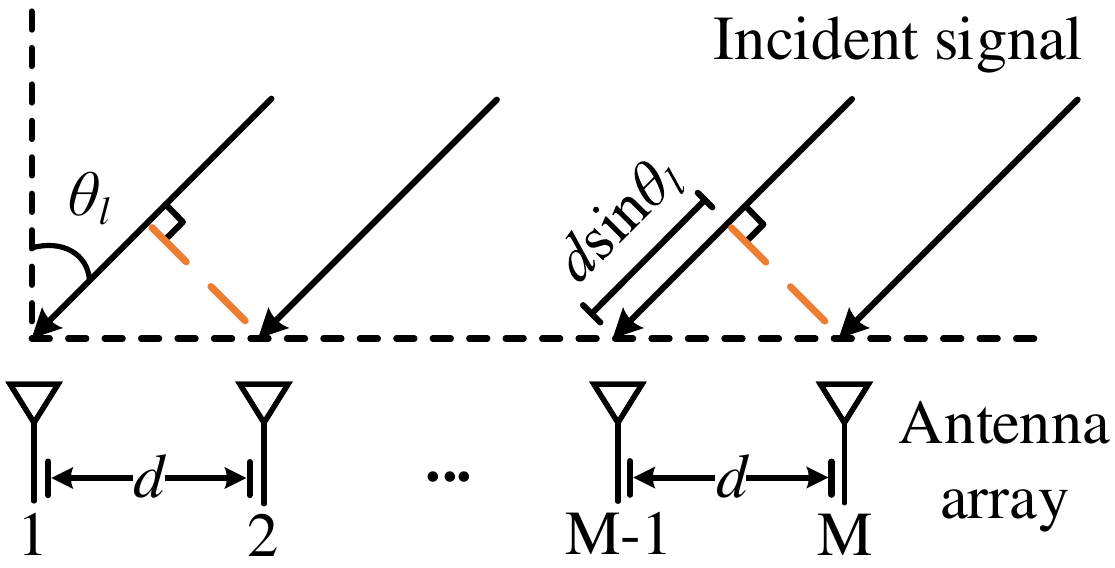}
\caption{Uniform linear M-antenna array: For AoA of $\theta_l$, the incident signal travels an additional distance of $d\sin(\theta_l)$ to the first antenna compared to the second, resulting in an additional phase of $-2\pi d \sin(\theta_l)/\lambda$ at the first antenna.}
\label{fig12}
\end{center}
\end{figure}Thus, the length difference $\Delta d$ of the reflection path between any two adjacent boundaries is always $\lambda/2$, e.g., boundary $a$ and $b$, or boundary $c$ and $d$. With the length change of $\lambda/2$, we make two assumptions about the hand movement: 1) from point $P_a$ to $P_b$ (i.e., $P_s=P_a$, $P_e=P_b$); 2) from point $P_c$ to $P_d$ (i.e., $P_s=P_c$, $P_e=P_d$).  Without loss of generality, we set LoS as 1 m, $a=10$, $c=20$. In this way, at the carrier frequency of 5.32 GHz ($\lambda= 5.6$ cm), the geometry distances $|P_b - P_a| = 2.2$ cm, $|P_d - P_c| = 1.8$ cm. Likewise, in 2D space in Fig. \ref{fig11} (b), hand also moves along the perpendicular bisector of one pair of transceiver on the y-axis. The initial position is marked with $(x_0, y_0)$ and the geometric distance is $L_0$. For the same change of path length, if the initial position changes to $(x_0^{'}, y_0^{'})$, the geometric distance would become $L^{'}_0$, and there would be a deviation of $\phi$ in moving direction \cite{witrace}.

\subsubsection{Initial Position Estimation}
Mutiple Signal Classification (MUSIC) \cite{music} is widely adopted to calculate the AoAs of incoming signals for device-based tracking or localization tasks \cite{spotfi, wicapture}, where the direct path (i.e., LoS) points to the active target. Nevertheless, in this paper, the location estimation belongs to device-free (i.e., passive) task, and the direct path is independent of hand location. Our intuition is that hand serves like a moving ``TX'' which reflects the delayed signals to RX. With one ``TX'' and two RXs, we can locate the hand with triangulation.

The basic idea of MUSIC is that incident signals from different angles will introduce different phase changes across RX antennas \cite{dynamic}. Consider a uniform linear M-antenna array in the RX prototype and a free-space environment with $L$ paths. Let the spacing between successive antennas at RX be $d$, which is half-wavelength of the signal, $\theta_l$ denotes the AoA for the $l^{th}$ path. The signal along $l^{th}$ path travels different distances from TX antenna to different RX antennas and hence, the signal at different antennas accumulates different phases. As illustrated in Fig. \ref{fig12}, for the $l^{th}$ path with AoA $\theta_l$, the signal travels an extra distance of $d\sin(\theta_l)$ received at the first antenna compared to the signal at the second one. Therefore, a phase difference of $-2\pi d \sin(\theta_l)/\lambda$ is accumulated at successive antennas. For simplicity, the phase shifts can be expressed as a function of the AoA $\theta_l$:
\begin{equation}
\Phi(\theta_l) = e^{-2\pi d \sin(\theta_l)/\lambda}.
\end{equation}
Hence, the vector of signals received at RX along the $l^{th}$ path can be denoted by:
\begin{equation}
\vec{a}(\theta_l) = [1 \ \Phi(\theta_l)\ \dots \ \Phi(\theta_l)^{M-1}]^{\text{T}},
\end{equation}
where $\vec{a}(\theta_l)$ is known as the steering vector. Steering vectors will be as many as the number of propagation paths, and the overall streeing matrix $A$ can be defined as:
\begin{equation}
A = [\vec{a}(\theta_1) \ ... \ \vec{a}(\theta_l) \ ...\ \vec{a}(\theta_L)].
\end{equation}

The received signal $X(t)$ at the antenna array is the superposition of all incident signals, i.e.,
\begin{equation}
\begin{aligned}
\begin{split}
X(t) = &[x_1(t) \ ... \ x_M(t)]^{\text{T}} \\
= &\sum \limits_{l=1} \limits^{L} \vec{a}(\theta_l) s_l(t) + N(t)                         \\
= &AS(t) + N(t),
\end{split}
\end{aligned}
\end{equation}
where $s_l(t)$ is the $l^{th}$ path signal at the first antenna, $N(t)$ is the Gaussian noise with zero mean and $\sigma^2$ variance. The key idea behind the MUSIC is the eigenstructure
analysis of correlation matrix $R_X$ of the received signal $X$:
\begin{equation}
\begin{aligned}
\begin{split}
R_X = &\mathbb{E}[XX^\text{H}] \\
= &A\mathbb{E}[SS^\text{H}]A^\text{H}+  \mathbb{E}[NN^\text{H}]   \\
= &AR_SA^\text{H} + \sigma^2I,
\end{split}
\end{aligned}
\end{equation}
where $X^\text{H}$ indicates the conjugate transpose of $X$, $R_S$ is the source correlation matrix, and $I$ is an identity matrix. The correlation matrix $R_X$ has $M$ eigenvalues $\zeta_1 \ ...\ \zeta_M$ associated with $M$ eigenvectors $\textbf{E}=[e_1 \ e_2 \ ... \ e_M]$. The smallest $M-n$ eigenvalues correspond to the noise while the other $n$ eigenvalues correspond to the $n$ incident signals. In this way, a noise vector subspace $\textbf{E}_\textbf{N}=[e_1 \  ... \ e_{M-n}]$ and a signal subspace $\textbf{E}_\textbf{S}=[e_{M-n+1} \ ... \ e_M]$ can be constructed. The signal and the noise subspaces are orthogonal so that the spatial spectrum can be expressed as:
\begin{figure}[htbp]
\begin{center}
\subfigure[Dynamic and static paths.]{
\includegraphics[width=3.7cm,height=3.2cm]{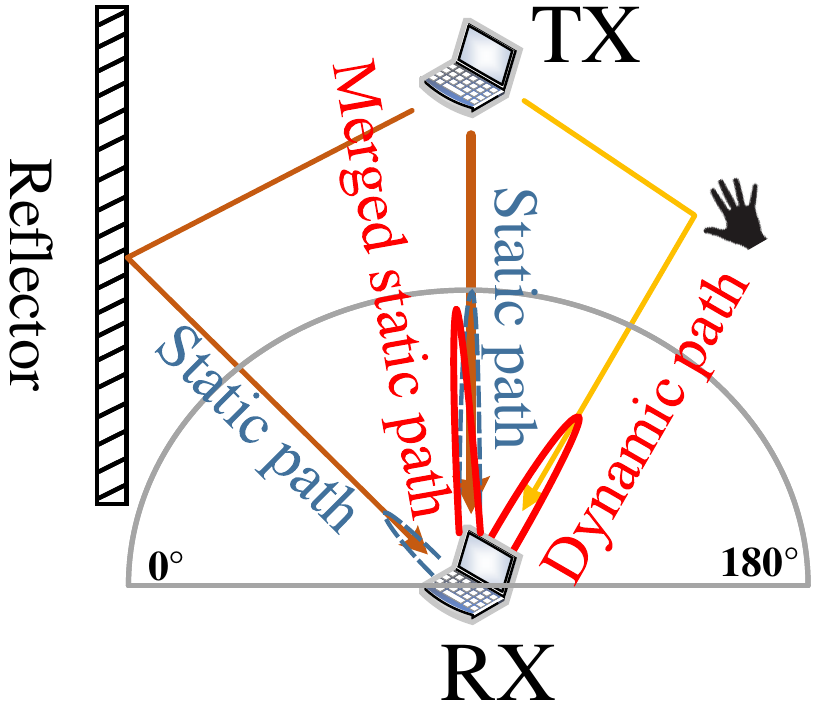}
}
\subfigure[Location with triangulation.]{
\includegraphics[width=4.4cm,height=3.2cm]{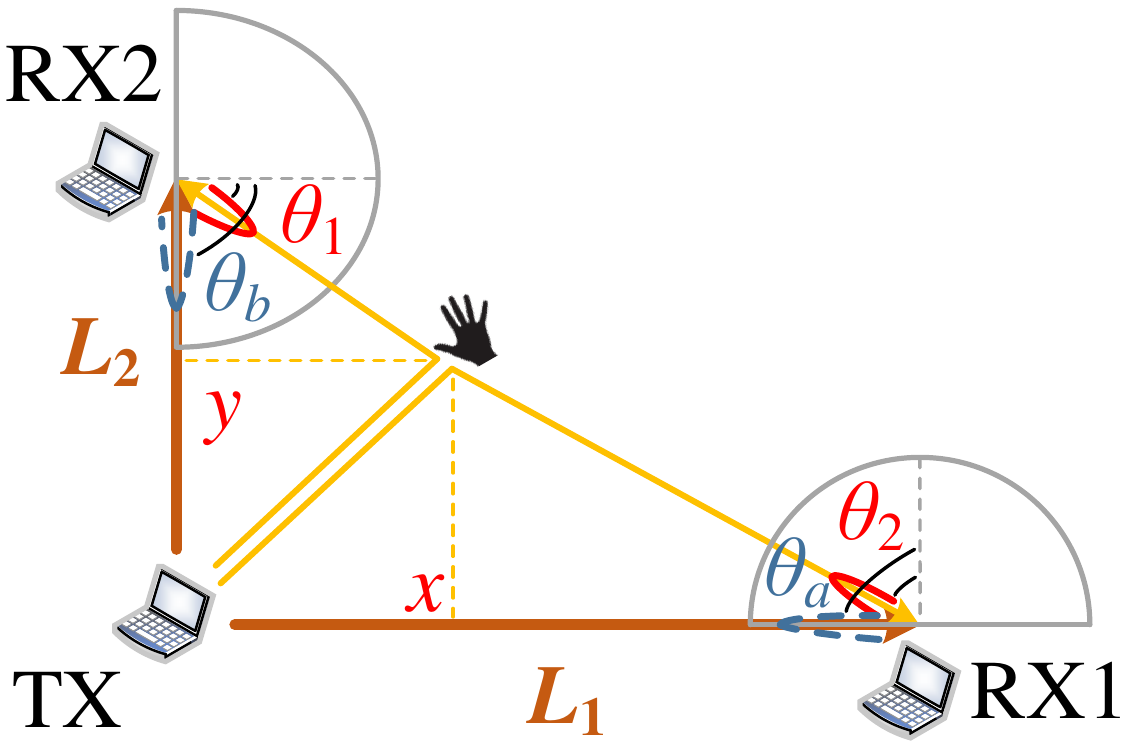}
}
\caption{Initial position estimation by AoAs.}
\label{fig13}
\end{center}
\end{figure}
\begin{equation}
P(\theta)_{MUSIC} = \frac{1}{\vec{a}^\text{H}(\theta)\ \textbf{E}_\textbf{N} \ \textbf{E}_\textbf{N}^\text{H} \ \vec{a}(\theta)},
\end{equation}
where peak locations in the pseudospectrum represent AoAs of incident signals. Techniques-wise, the AoAs of paths are relatively stationary and cannot change rapidly from packet to packet \cite{wicapture}. Inspried by this, CSI from $N$ packets are concatenated to further mitigate random noises following:
\begin{equation}
X = [X_1 \ \dots \ X_N]^{\text{T}}.
\end{equation}

In addition, CnetiTrack automatically evaluates the number of signals with MUSIC improved by the Minimum Description Length (MDL) principle \cite{mdl}. MDL can identify the division of signals into subspaces confronting low Signal-to-Noise Ratio (SNR) or non-adaptive eigenvalue threshold \cite{signaltrack}. The MDL is performed as:
\begin{equation}
\mathop{\arg\min}_{\omega} \frac{1}{2} \omega(2M-\omega)\lg{N}-N(M-\omega)\lg{\left( \frac{\prod \limits_{i=\omega+1} \limits^{M} \zeta_i^{1/(M-\omega)}}{\frac{1}{M-\omega}\sum \limits_{i=\omega+1} \limits^{M} \zeta_i} \right)},
\end{equation}
where $\omega \in [0, 1,..., M-1]$. The number of signals is determined as the value of $\omega$ for which MDL is minimized.

A typical indoor environment usually contains around 4$\thicksim$8 significant reflection paths \cite{pathnumber}. However, with three-antenna array of most commodity WiFi NICs (e.g., Intel 5300 NIC), only two paths can be captured with MUSIC. Therefore, how to estimate AoAs related to hands with limited antennas becomes the key to the problem. We have divided all the signal propogation paths into one dynamic path and multiple static paths. Note that signals of  static paths have the same central frequency and fixed phase differences, they are therefore coherent with each other. The motion of hands will not only introduce DFS on the dynamic path signal, but also change the path length. For signals to be coherent with each other, they must be with the same central frequency and fixed phase differences over time \cite{dynamic}. Thus, the signal of dynamic path is incoherent with those of static paths. MUSIC is designed to identify incoherent multiple signals, so all the coherent static paths which are out of the interest can be merged into one path \cite{arraytrack}, which significantly reduce the path number on the pseudospectrum. The signal of LoS is typically much stronger than those of static reflection paths, thus the merged path AoA searched on the spectrum is very close to the AoA of LoS. When hands begin to move, the dynamic path will be detected along with the merged static path, as shown in Fig. \ref{fig13} (a). 

Therefore, the very properties of whether paths are coherent or not make three-antenna array competent for initial position estimation. We also note that works \cite{spotfi, arraytrack} focus on eliminating the coherence by sacrificing the effective number of sensors and all incident signals will be detected. In this paper, all the static paths except LoS are useless for gesture tracking, so there is no need to distinguish them with extra efforts. Given one TX, and two RXs with LoS of $L_1$ and  $L_2$, we can solve AoAs of dynamic path  $[\theta_1, \theta_2]$ and LoS $[\theta_a, \theta_b]$ on RXs, and thus locate the hand $(x, y)$ with triangulation as shown in Fig. \ref{fig13} (b):
\begin{equation}
\begin{aligned}
\begin{split}
x = &\frac{L_2 - L_1 \cdot \tan{(\theta_a - \theta_2)}}{\tan{(\theta_b - \theta_1)}^{-1} - \tan{(\theta_a - \theta_2)}},\\
y = &\frac{L_1 - L_2 \cdot \tan{(\theta_b - \theta_1)}}{\tan{(\theta_a - \theta_2)}^{-1} - \tan{(\theta_b - \theta_1)}}.
\end{split}
\end{aligned}
\end{equation}

\subsection{Tracking Calibration}

Theoretically, given the initial positions and the derived length changes of dynamic path, the 2D hand motion traces can be estimated. However, the CSI deviates from the above theoretical model, which is mainly reflected in three aspects. Firstly, the signals of static paths mainly composed of LoS and static reflectors are orders of magnitude stronger than that of hand-reflected dynamic path, which will make CSI trace deviate from the origin. Secondly, the static components are not absolutely constant, where the circular traces are not strictly concentric. This is mainly because the hand blocks the static reflectors, as well as the erratic motions of other body parts during tracking, such as posture changes or limb swings. Last, the magnitude of the dynamic path, i.e., circle radius, changes continuously, which is inversely proportional to the reflection path length. These aspects above will severely degrade the tracking accuracy. Thus, we propose the following so-called Static Component Elimination (SCE) algorithm for tracking calibration.

\begin{figure*}[htbp]
\begin{center}
\subfigure[Hand movement.]{
\includegraphics[width=3.5cm,height=3.37cm]{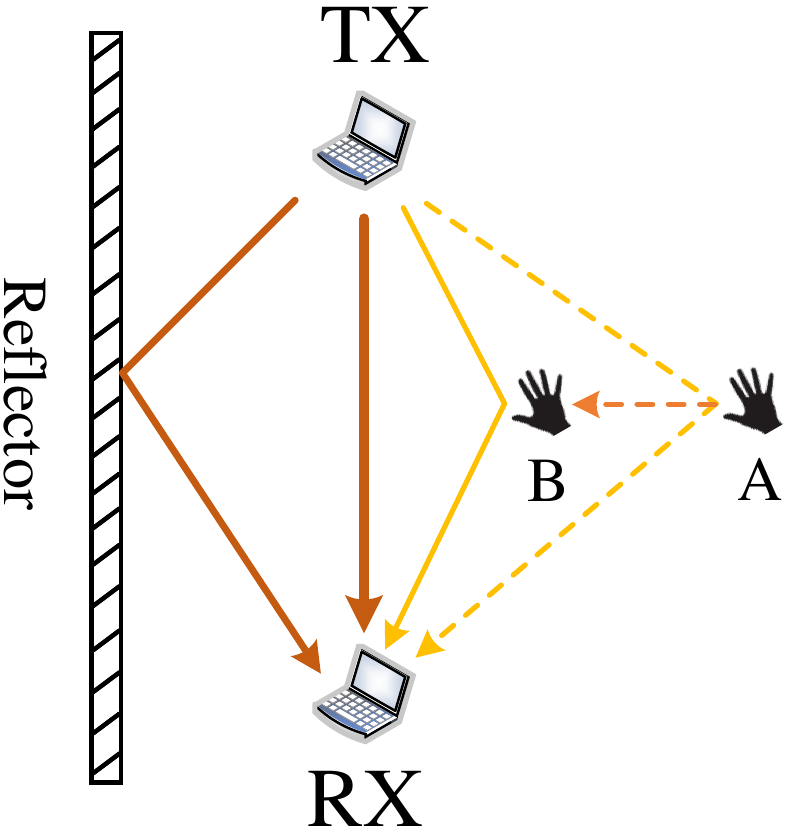}
}
\subfigure[CSI trace.]{
\includegraphics[width=3.8cm,height=3.37cm]{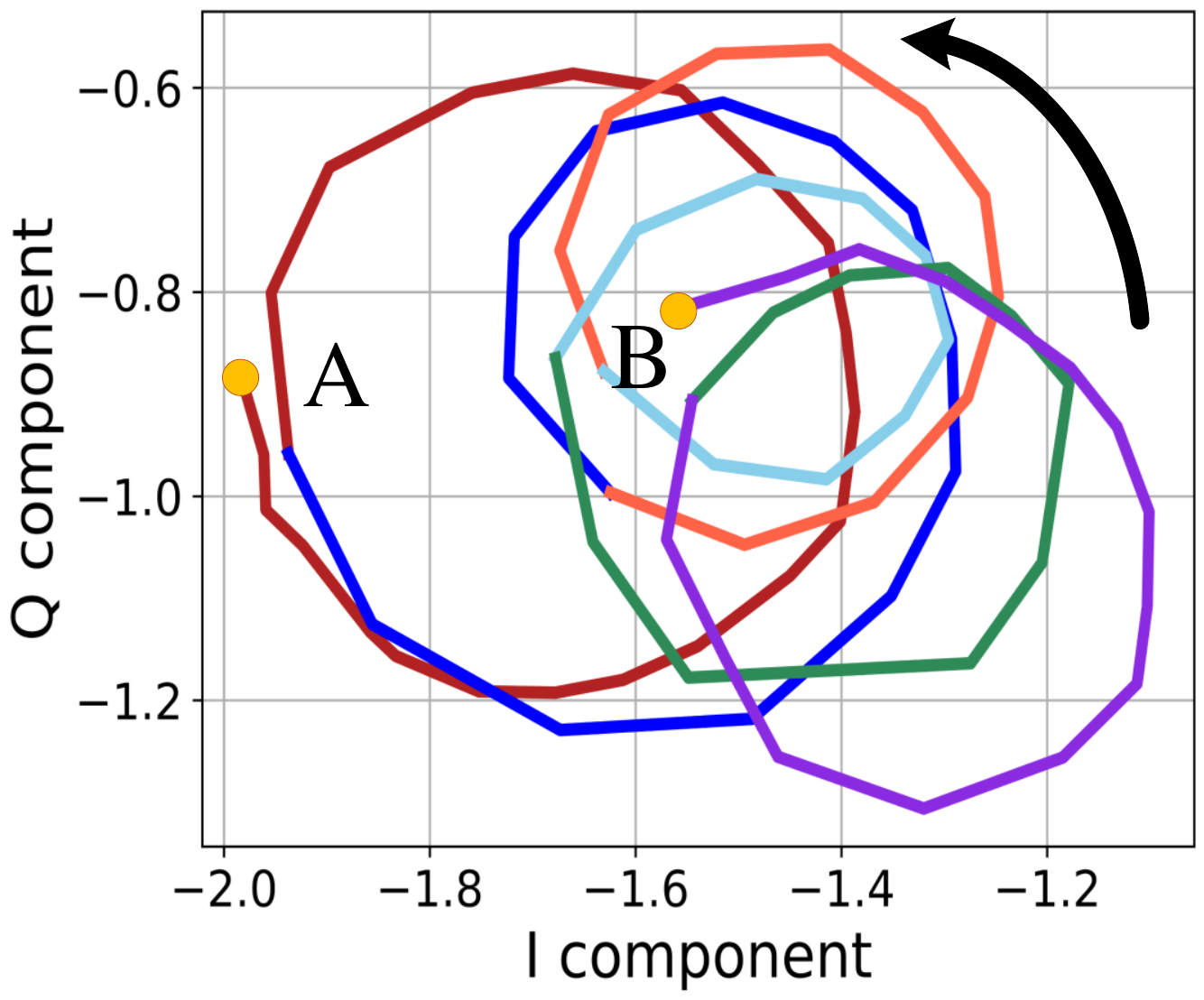}
}
\subfigure[Phase $\mathcal{A}$.]{
\includegraphics[width=3.8cm,height=3.37cm]{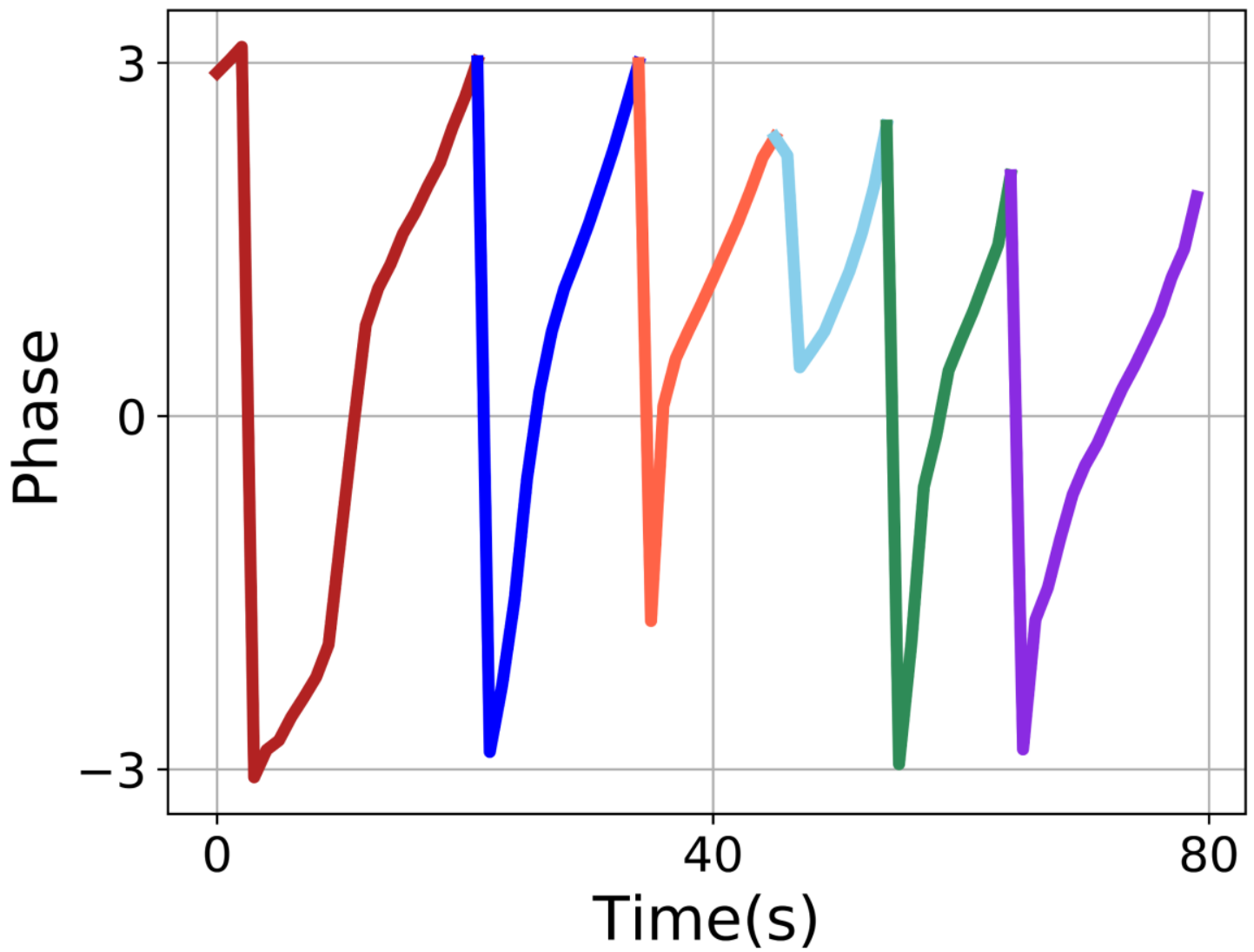}
}
\subfigure[Unwrapped phase $\mathcal{A}$.]{
\includegraphics[width=3.8cm,height=3.37cm]{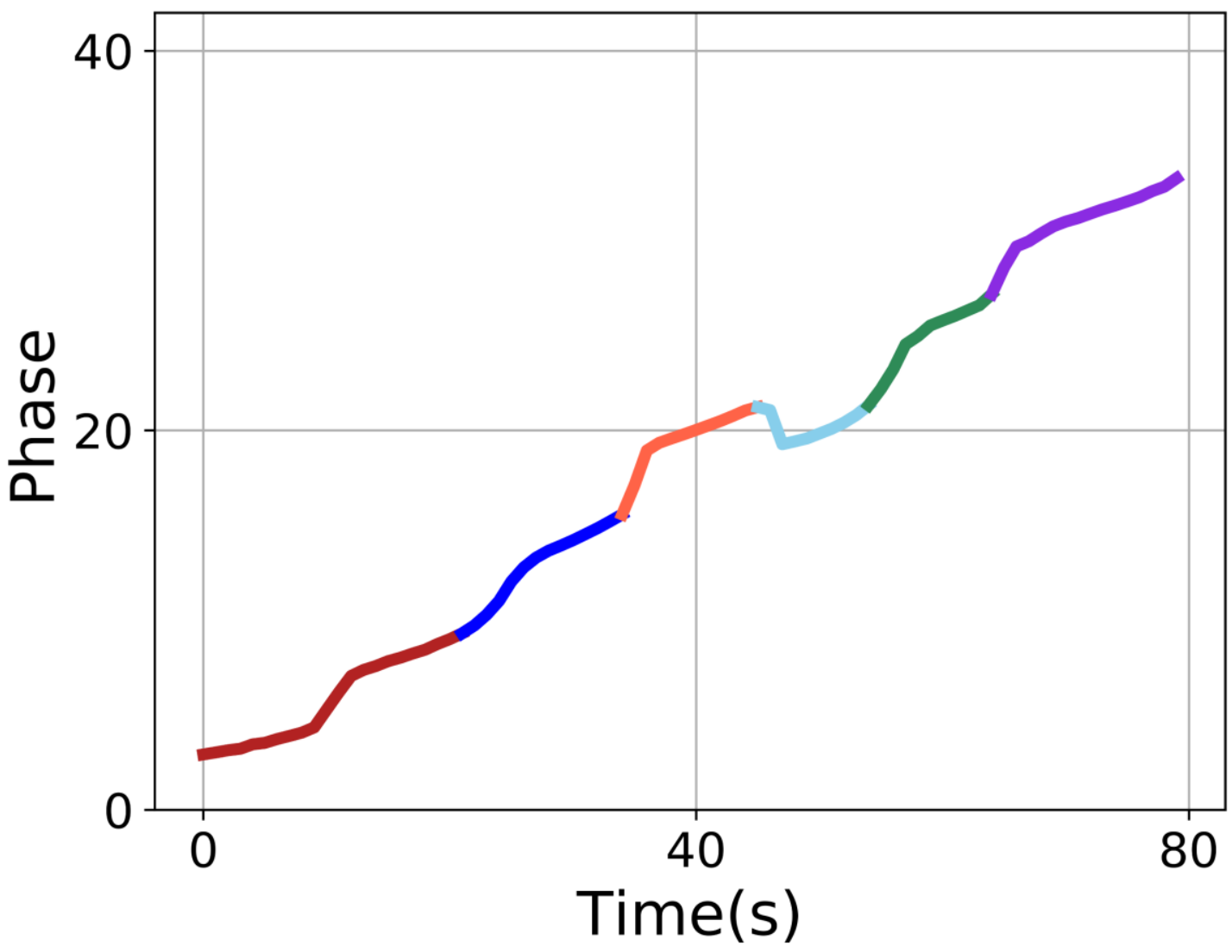}
}
\subfigure[CSI trace after SCE.]{
\includegraphics[width=3.8cm,height=3.37cm]{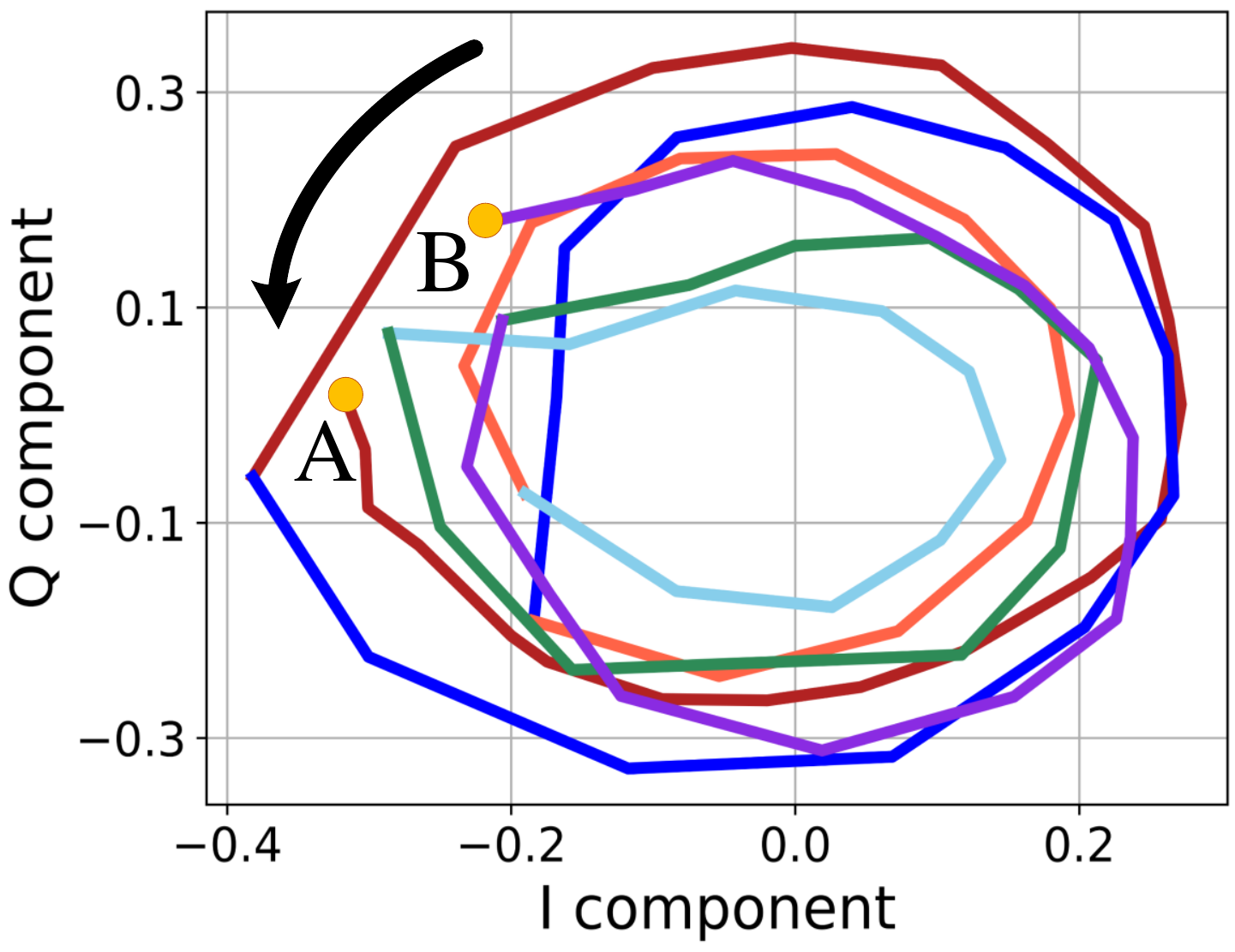}
}
\subfigure[Phase $\mathcal{A}^{'}$.]{
\includegraphics[width=3.8cm,height=3.37cm]{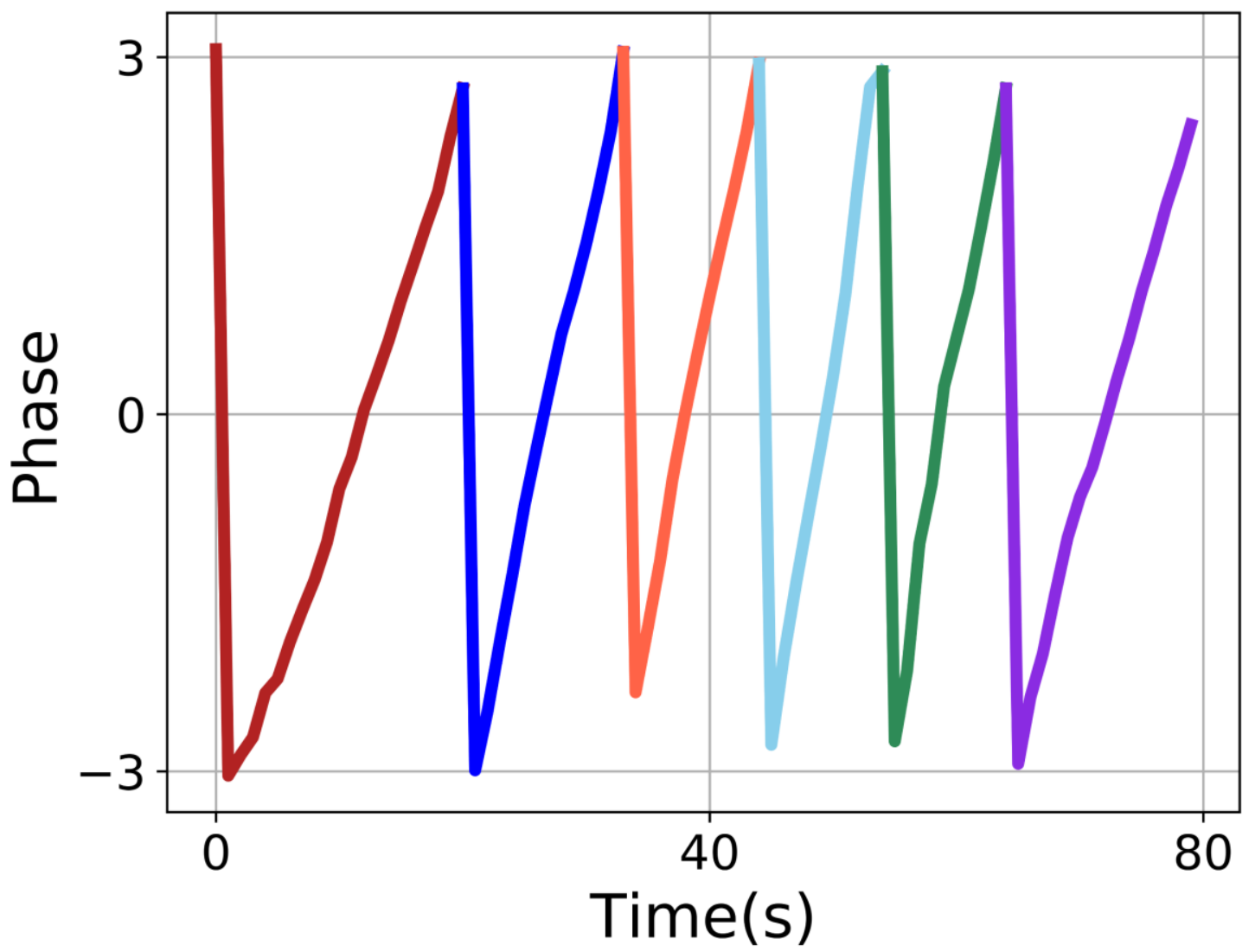}
}
\subfigure[Unwrapped phase $\mathcal{A}^{'}$.]{
\includegraphics[width=3.8cm,height=3.37cm]{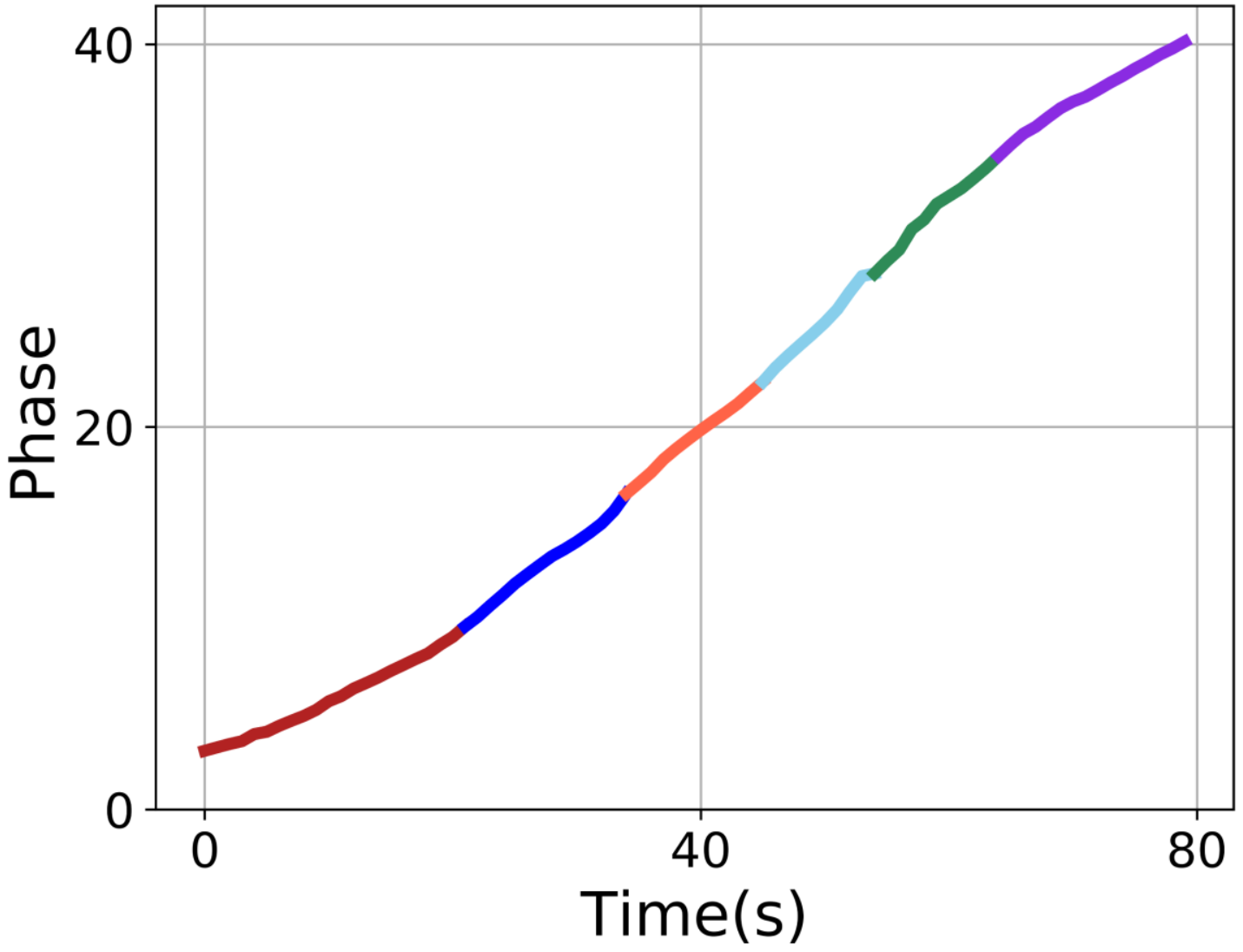}
}
\caption{Phase calibration when moving towards transceiver.}
\label{fig14}
\end{center}
\end{figure*}

We first take the case that hand moves towards transceiver, $H_{ratio}(f,t)$ should rotate counterclockwise on the complex plane with the decrease of path length. As illustrated in Fig. \ref{fig14}, when user moves hand from position A to B in a short distance, the rotation is a non-concentric spiral trace. The trace is first translated to the origin by subtracting its average. In this way, most of the strong static components can be roughly eliminated, and the result is denoted by $\mathcal{M}$. We can get the phase of each CSI sample based on $\mathcal{M}$, denoted by $\mathcal{A}$, as shown in Fig. \ref{fig14} (c). Due to phase flip, there are several peaks and troughs in $\mathcal{A}$, in which each group of adjacent peaks is mapped to a complete circle segment of the whole trace. We directly unwrap $\mathcal{A}$ by changing absolute jumps greater than $1.5\pi$ to a complement of $2\pi$ in Fig. \ref{fig14} (d). The unwrapped curve increases over time, which indicates user moves hand towards the transceiver, and the slope of curve is positively related to the moving speed. 

\begin{algorithm}
\caption{Static Component Elimination}
\label{alg1}
\KwIn {Denoised CSI trace, $H_{ratio}(f,t)$.}
\KwOut {Sequence of phase differential iterations, $\mathcal{A}^{'}$.}
\Begin
{
Translate $H_{ratio}(f,t)$ to origin\;
Split motion segments by phase flip directions\;
\For{each segment $s$}
{
Decompose $s$ into circles $[\mathcal{M}_1, \mathcal{M}_2,..., \mathcal{M}_n]$ and two arcs by multi-peak searches\;
Solve centers of circle $[\mathcal{O}_{\mathcal{M}_1}, \mathcal{O}_{\mathcal{M}_2},..., \mathcal{O}_{\mathcal{M}_n}]$\;
  \For{each group of adjacent circles $\mathcal{M}_u$ and $\mathcal{M}_v$}
  {
  Carry out linear interpolation between $[\mathcal{O}_{\mathcal{M}_u}, \mathcal{O}_{\mathcal{M}_v}]$\;
  Calibrate $\mathcal{M}_u$ and $\mathcal{M}_v$ by interpolation results\;
  }
  Calibrate the two arcs by $\mathcal{O}_{\mathcal{M}_1}$ and $\mathcal{O}_{\mathcal{M}_n}$\;
  Aggregate the calibrated results\;
  Achieve the phase of samples\;
} 
\Return Sequence of phase differential iterations $\mathcal{A}^{'}$\;
}
\end{algorithm}

The unwrapped result $\mathcal{A}$, however, is suffered from the slow changes of static paths, leading to the translations to certain parts of whole trace. Therefore, some radians corresponding to CSI trace in unwrapped $\mathcal{A}$ may be incomplete, and results in a large deviation of the estimated traces. Recall to $\mathcal{M}$, without loss of generality, it can be decomposed into several circles $[\mathcal{M}_1, \mathcal{M}_2,..., \mathcal{M}_n]$ and two independent arc segments by the multi-peak searching algorithm. The mean value of each circle, i.e., the coordinate of circle center, can be denoted by 
$[\mathcal{O}_{\mathcal{M}_1}, \mathcal{O}_{\mathcal{M}_2},..., \mathcal{O}_{\mathcal{M}_n}]$. For $H_{ratio}(f,t)$ samples belonging to two consecutive circles $\mathcal{M}_u$ and $\mathcal{M}_v$, we carry out linear interpolation in interval $[\mathcal{O}_{\mathcal{M}_u}, \mathcal{O}_{\mathcal{M}_v}]$, with the number of interpolations equal to that of CSI samples. In this case, each sample on $\mathcal{M}_u$ and $\mathcal{M}_v$ can be calibrated by subtracting interpolated result accordingly, and the results are marked with $[\mathcal{M}_1^{'}, \mathcal{M}_2^{'},..., \mathcal{M}_n^{'}]$. The arc segments are the first and last segments left after division, and they can be calibrated by $\mathcal{O}_{\mathcal{M}_1}$ and $\mathcal{O}_{\mathcal{M}_n}$, according to the proximity principle. The intuition of decomposition and interpolation is that slow changes of static paths will have a translation of CSI trace in part over time, so all circles are not completely concentric. These translations applied to the CSI trace are actually the translations of the trace segments to diverse degrees, so we decompose the trace and adopt interpolation between adjacent circle centers to cope with it. The trace after SCE is in Fig. \ref{fig14} (e), and the phase marked with $\mathcal{A}^{'}$ is in Fig. \ref{fig14} (f) and (g), where most of the static components have been removed, with more complete and accurate radians.

\begin{figure*}[htbp]
\begin{center}
\subfigure[Hand movement.]{
\includegraphics[width=3.5cm,height=3.37cm]{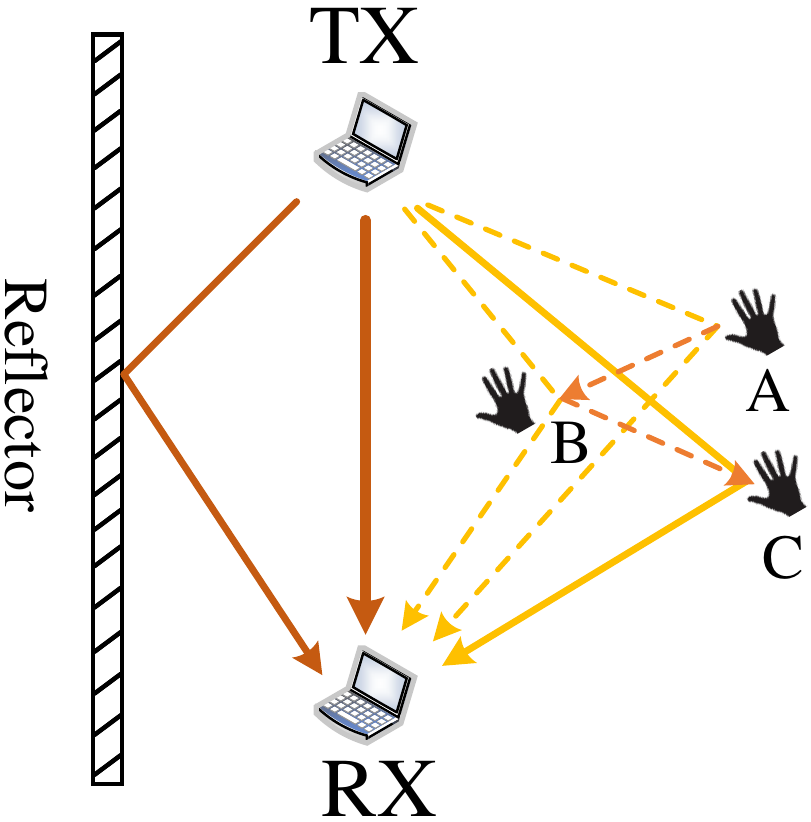}
}
\subfigure[CSI trace.]{
\includegraphics[width=3.8cm,height=3.37cm]{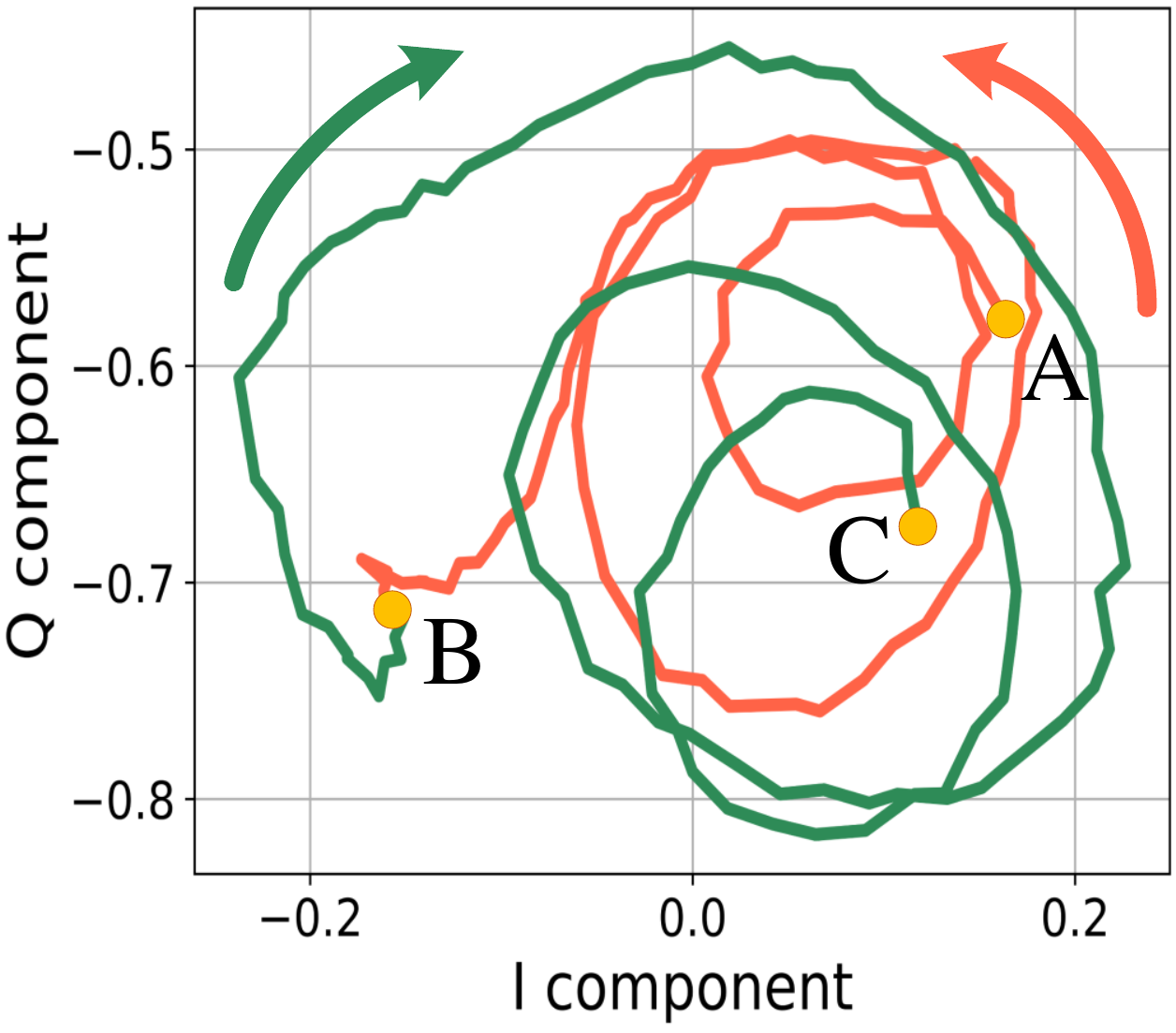}
}
\subfigure[The whole calibration.]{
\includegraphics[width=3.8cm,height=3.37cm]{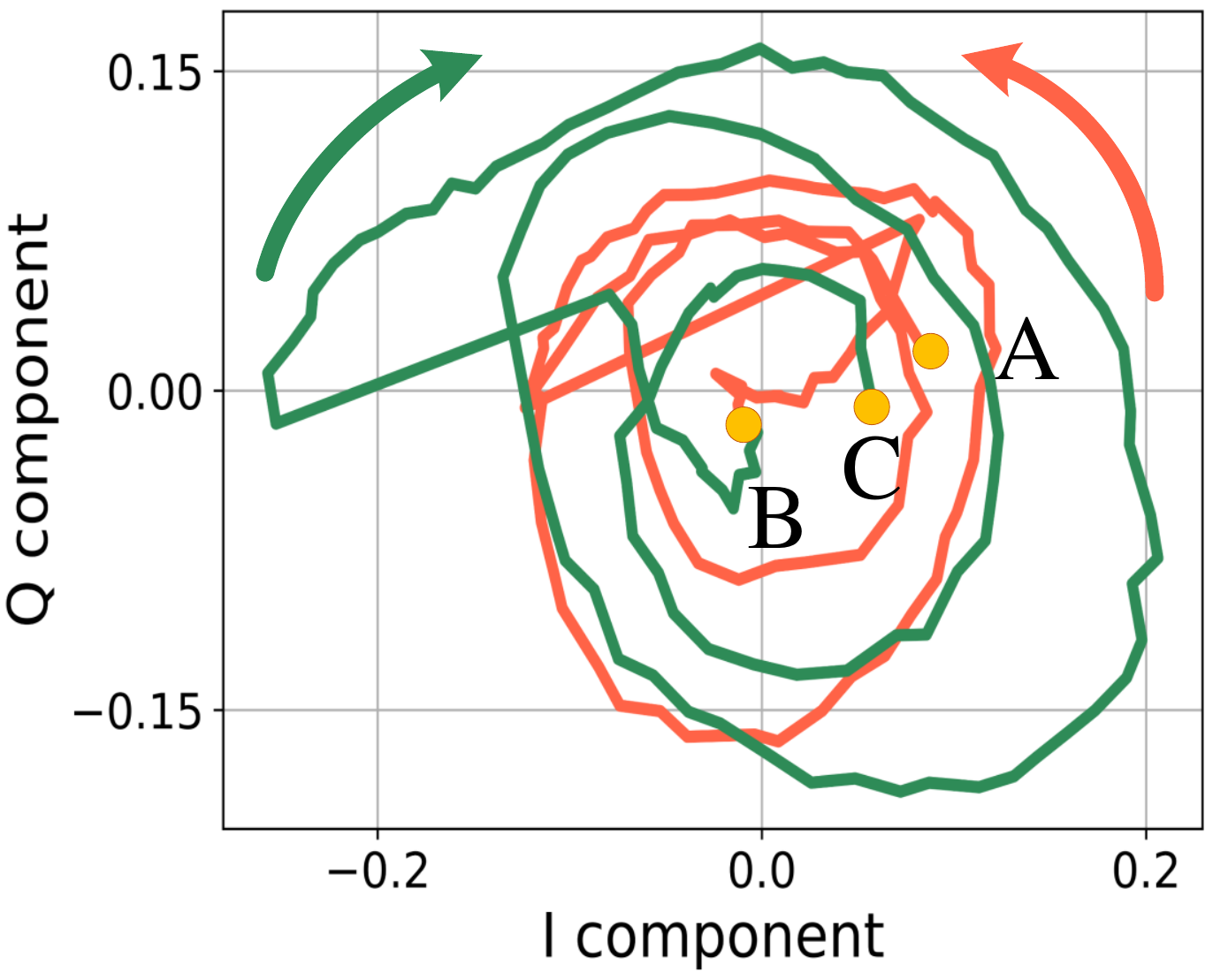}
}
\subfigure[Unwrapped phase of (c).]{
\includegraphics[width=3.8cm,height=3.37cm]{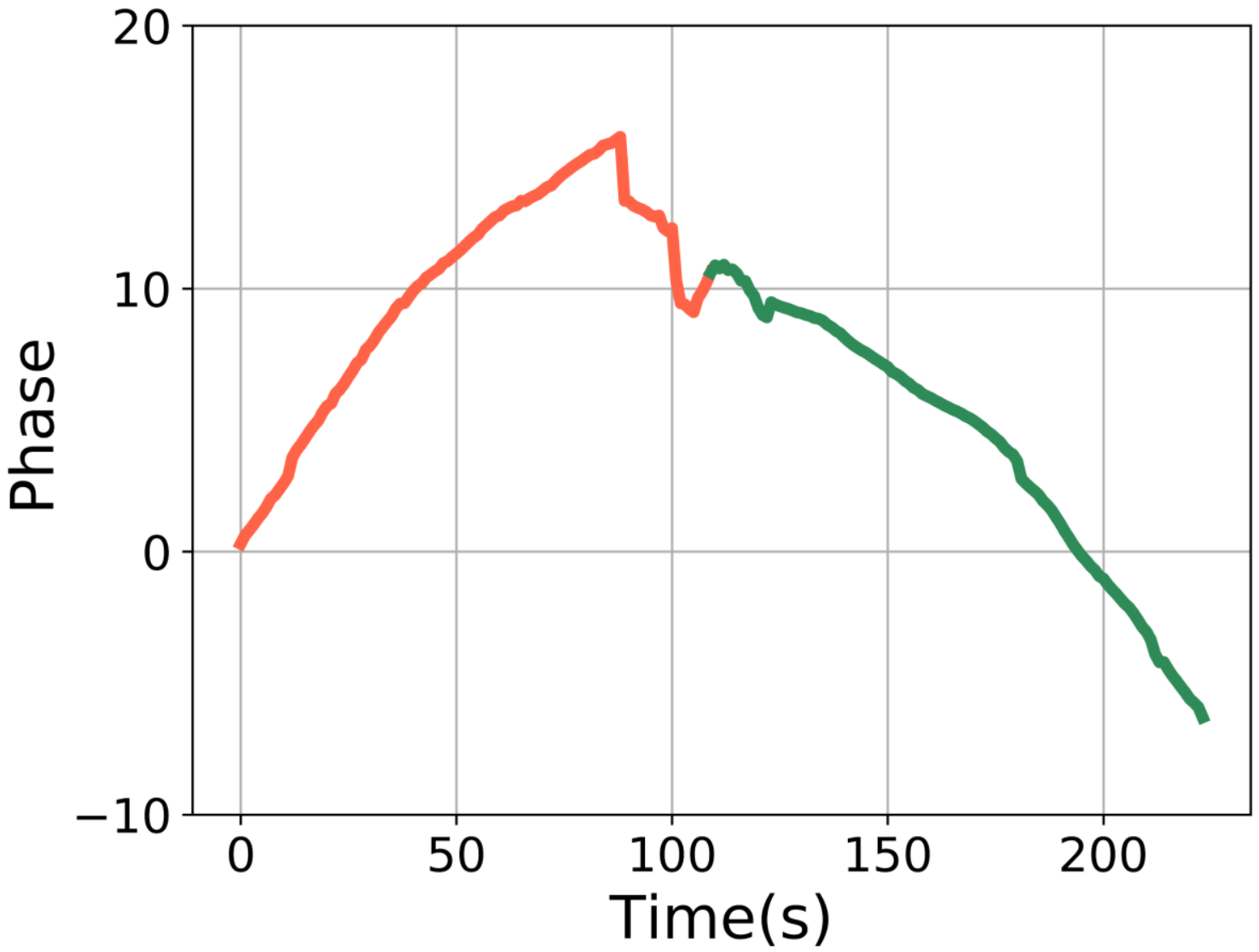}
}
\subfigure[The segment calibration.]{
\includegraphics[width=3.8cm,height=3.37cm]{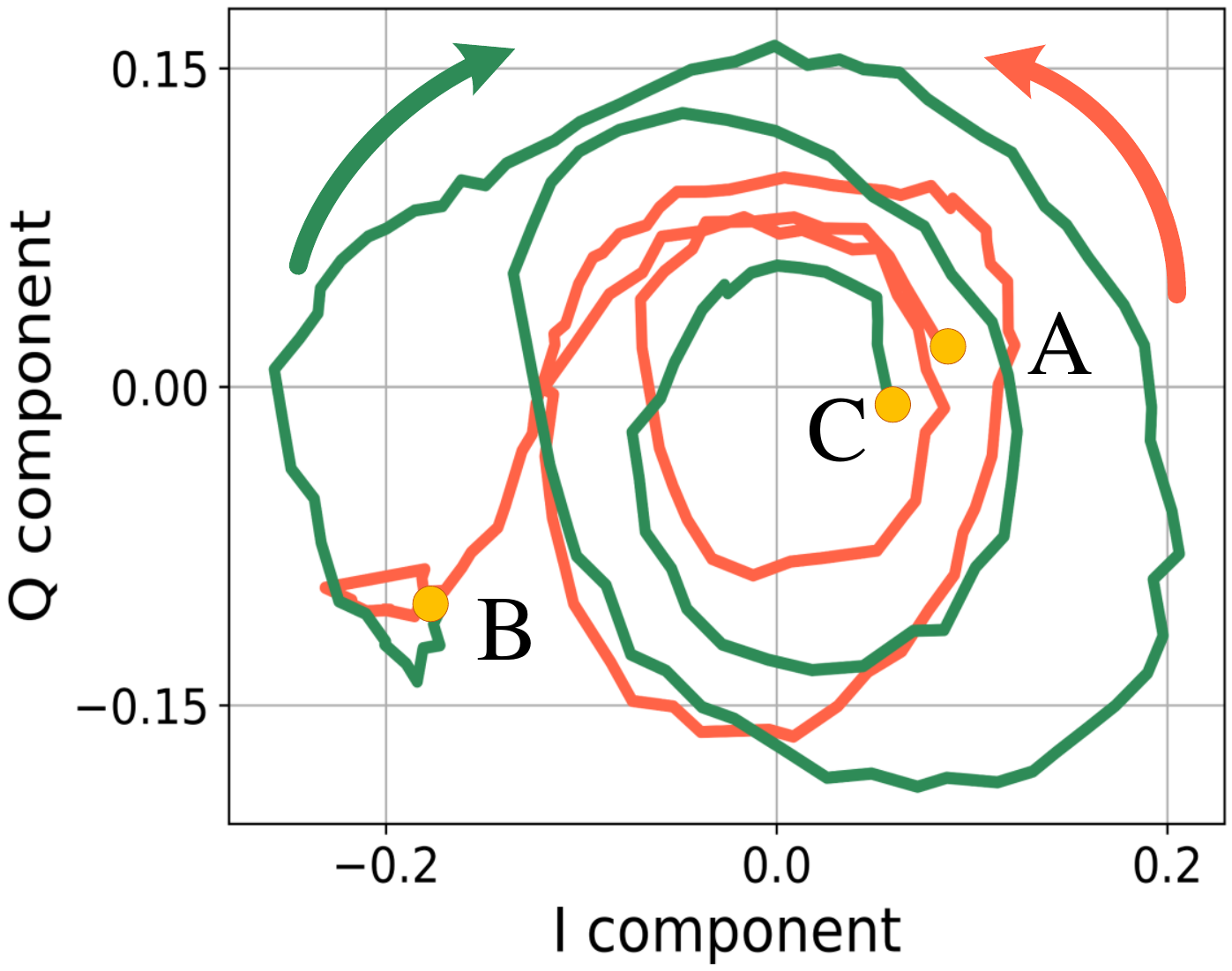}
}
\subfigure[Phase of (e).]{
\includegraphics[width=3.8cm,height=3.37cm]{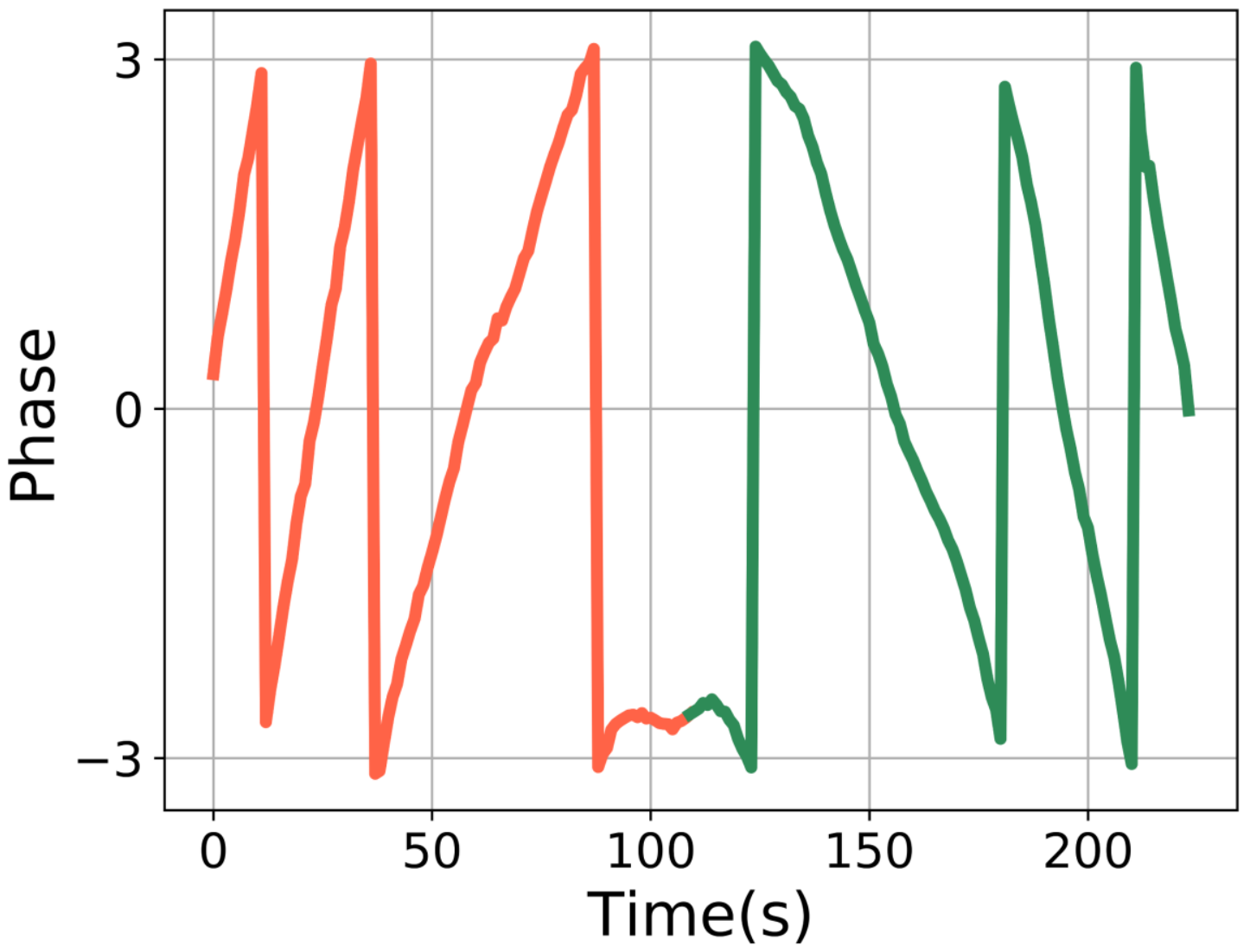}
}
\subfigure[Unwrapped (f).]{
\includegraphics[width=3.8cm,height=3.37cm]{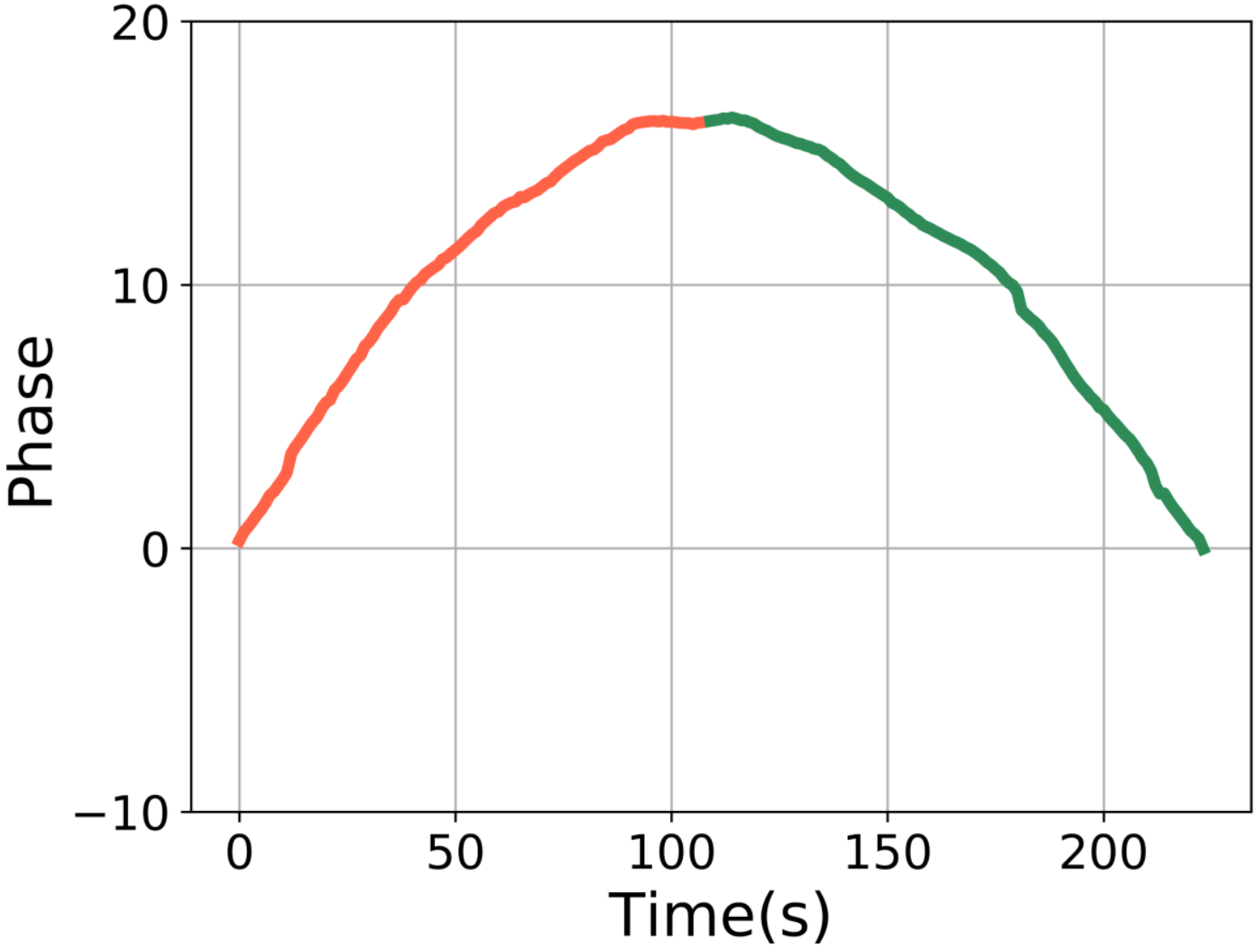}
}
\caption{Phase calibration when reciprocating transceiver.}
\label{fig15}
\end{center}
\end{figure*}

Besides, user may move hand back and forth around the transceiver while performing gestures. As illustrated in Fig. \ref{fig15}, user moves hand first towards (from position A to B) and then away from the transceiver (from position B to C). As aforementioned, $H_{ratio}(f,t)$ should rotate first counterclockwise (red curve) and then clockwise (green curve). In this way, the entire CSI trace is interrupted by the two opposite rotations (i.e., AB and BC). If the CSI trace is calibrated as an entirety, a large range of errors may be introduced, thus degrading the tracking accuracy, as illustrated in Fig. \ref{fig15} (c) and (d). Hence, CSI trace segments corresponding to towards or away from the transceiver need to be separated based on phase flip directions. The above static components elimination operations can be conducted for each segment thereafter, and calibrated results are as shown in Fig. \ref{fig15} (e) to (g). With $\mathcal{A}^{'}$ of both pairs of transceivers, the precise 2D length changes of dynamic path over time can be derivated, i.e., $\frac{\mathcal{A}^{'}}{2\pi}\lambda$. The pseudocode of this module is shown in Algorithm \ref{alg1}.
\subsection{Tracking Smoothing}
Given the initial position in the Cartesian coordinate system, the reflection path length can be initialized. According to the changes of path length $\frac{\mathcal{A}^{'}}{2\pi}\lambda$, the absolute length can be updated, and the motion traces of hands can be estimated thereafter.
Due to persistent system noise, the estimated fine-grained trace still carries errors. Here we turn to the Savitzky-Golay filter \cite{sg}, which leverages the least square to fit a particular polynomial to a windowed part of the signal, and further replaces the central point of the window with the polynomial value of the point to generate smoothed output. Due to its low computational cost and ability to effectively preserve the envelope of the signal waveform, Savitzky-Golay filter is adopted for tracking smoothing.

\begin{figure*}[htbp]
\begin{center}
\subfigure[Basement.]{
\includegraphics[width=5.5cm,height=5.2cm]{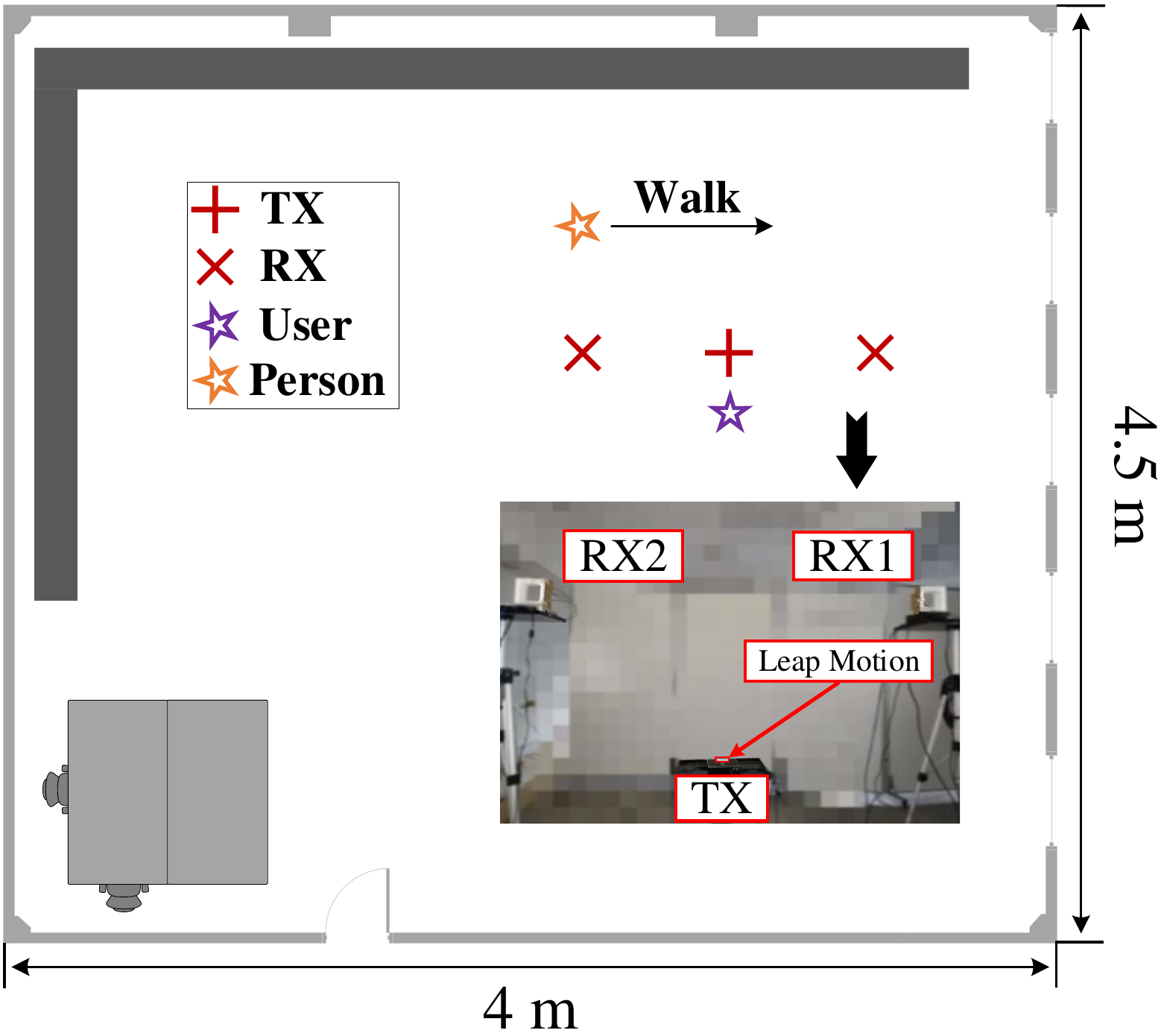}
}
\subfigure[Office.]{
\includegraphics[width=5.5cm,height=5.2cm]{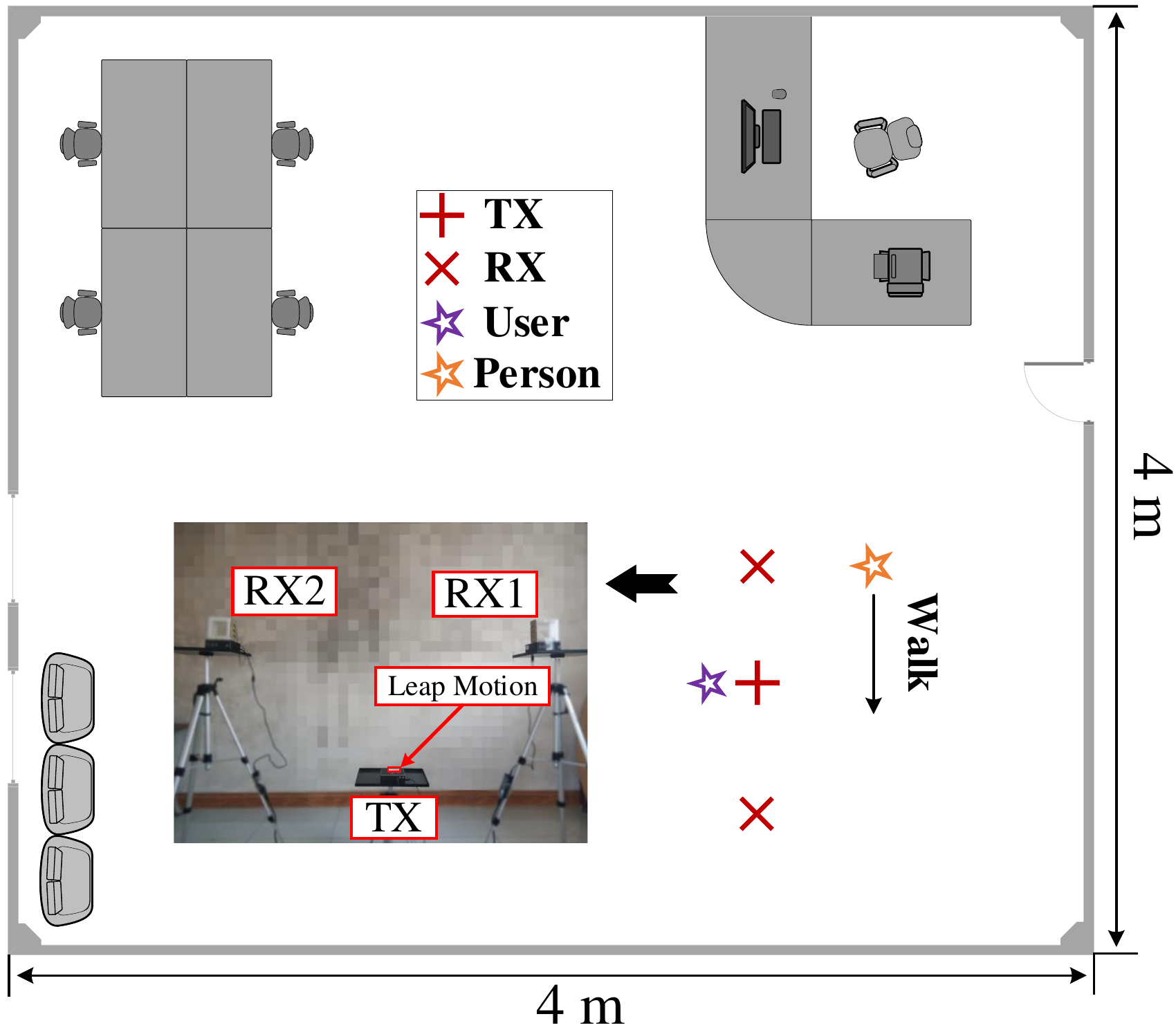}
}
\subfigure[Laboratory.]{
\includegraphics[width=5.5cm,height=5.2cm]{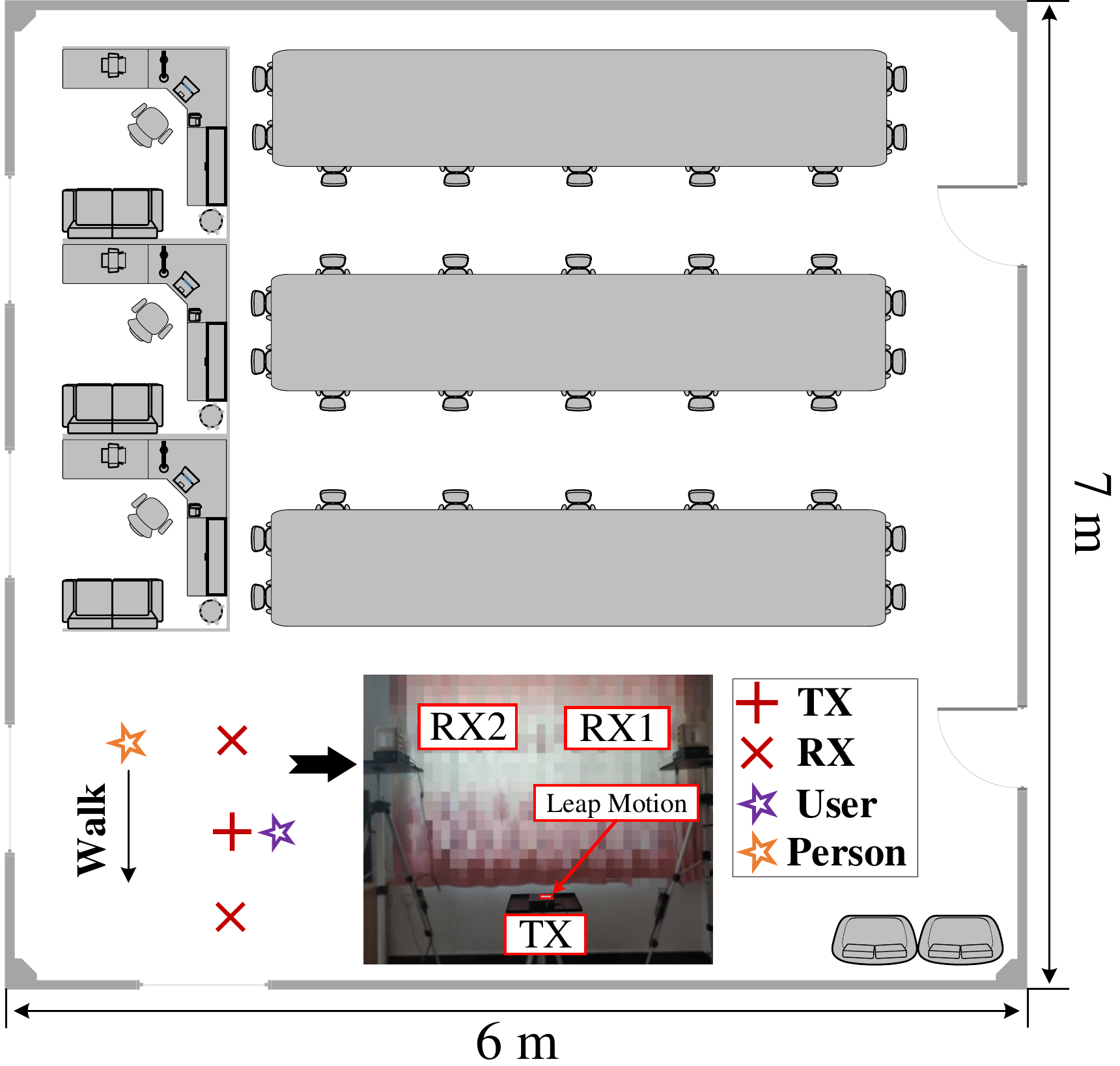}
}
\caption{Vertical view of experimental scenarios: (a) a basement of size 4.5 m $\times$ 4 m; (b) an office of size 4 m $\times$ 4 m; (c) a laboratory of size 7 m $\times$ 6 m.}
\label{fig16}
\end{center}
\end{figure*}

\section{Implementation and Evaluation}
In this section, we implement CentiTrack and evaluate its 2D gesture tracking accuracy, as well as related parameter study. Experiments are conducted in three indoor scenarios, as shown in Fig. \ref{fig16}. The sensing area is set perpendicular to the ground, and the LoS of transceivers is 1.5 m. Unless otherwise stated, the users stand in front of our system and make gestures on the sensing plane. The moving speed is less than or equal to 0.5 m/s, which is in line with the real handwriting speed range of users. The ground truth is supervised by Leap Motion \cite{camera2}, which can compute hands' 2D coordinates in millimeters with respect to the Leap Motion controller as the origin. The reason why we do not adopt the template as the ground truth is that hands will shake slightly during the movement, which will lead to additional estimation errors. The tracking error of each trace is reported by the Euclidean distance between the ground truth and estimated trace point by point, and then averaged by all the points.

\subsection{Implementation}
We implement CentiTrack on totally three mini-PCs equipped with Intel 5300 WiFi NICs, where the TX is with one antenna, while three for each RX to form a uniform linear array. All the antennas are commonly-seen horizontally-polarized omni-directional antennas. Here, we make WiFi NICs run at the same configuration as section \ref{111}. The Intel 5300 NIC reports 30 out of 56 subcarriers for each of its antennas. The sampling rate of CSI defaults to 300 Hz. Our system can run either offline or online via User Datagram Protocol (UDP) protocol.

\subsection{Evaluation Metrics}
We evaluate the gesture tracking performance with up to four users from eight metrics: (i) Initial position error. (ii) 2D tracking error. (iii) Impact of different scenarios. (iv) Impact of different moving directions. (v) Impact of sampling rate. (vi) Impact of user and trace length diversity. (vii) Impact of other person walking around. (viii) Benefits of individual modules.

\subsection{Experimental Results}

\begin{figure*}[htbp]
\begin{center}
\subfigure[Initial position error.]{
\includegraphics[width=4.1cm,height=3.5cm]{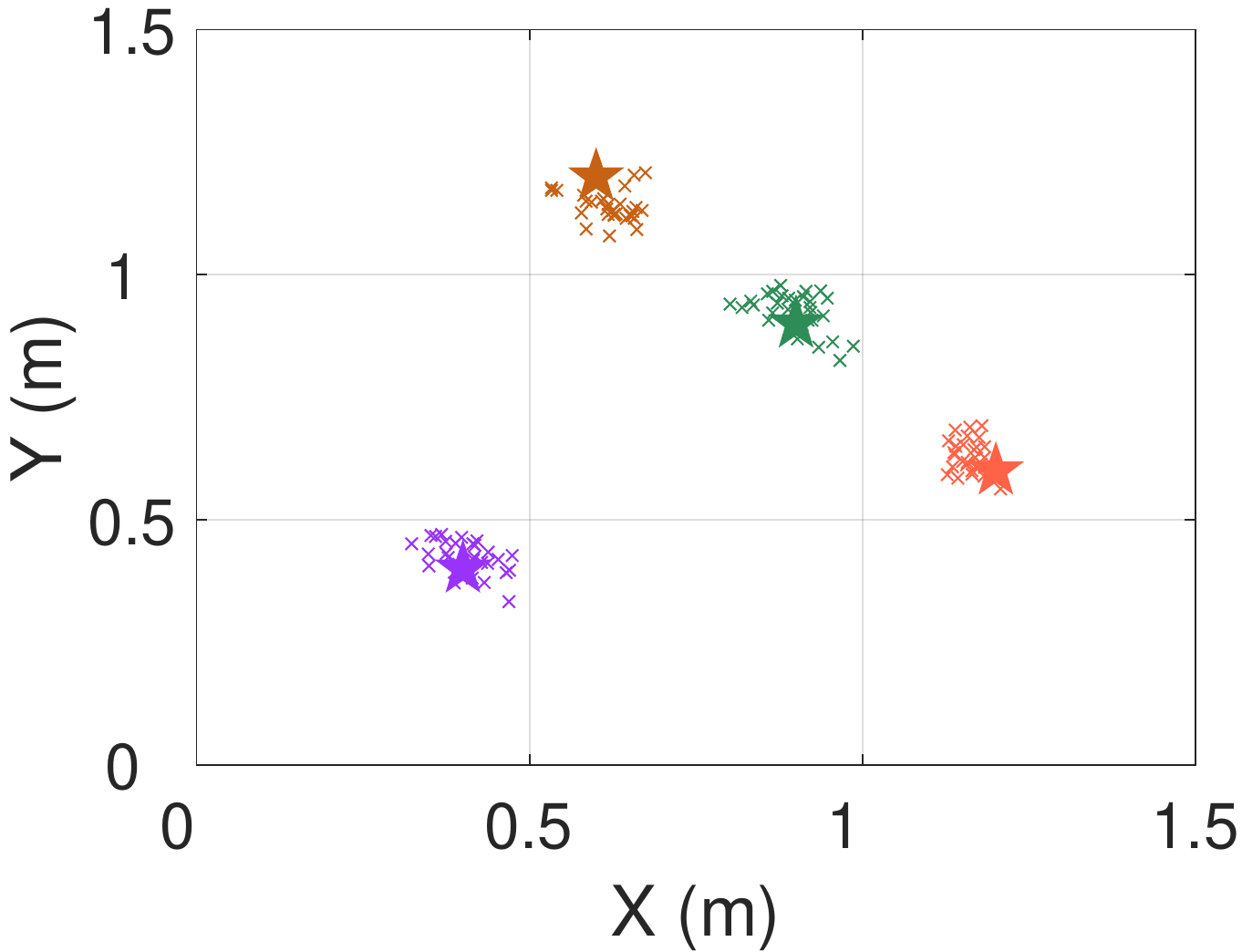}
}
\subfigure[CDF of initial position error.]{
\includegraphics[width=4.1cm,height=3.5cm]{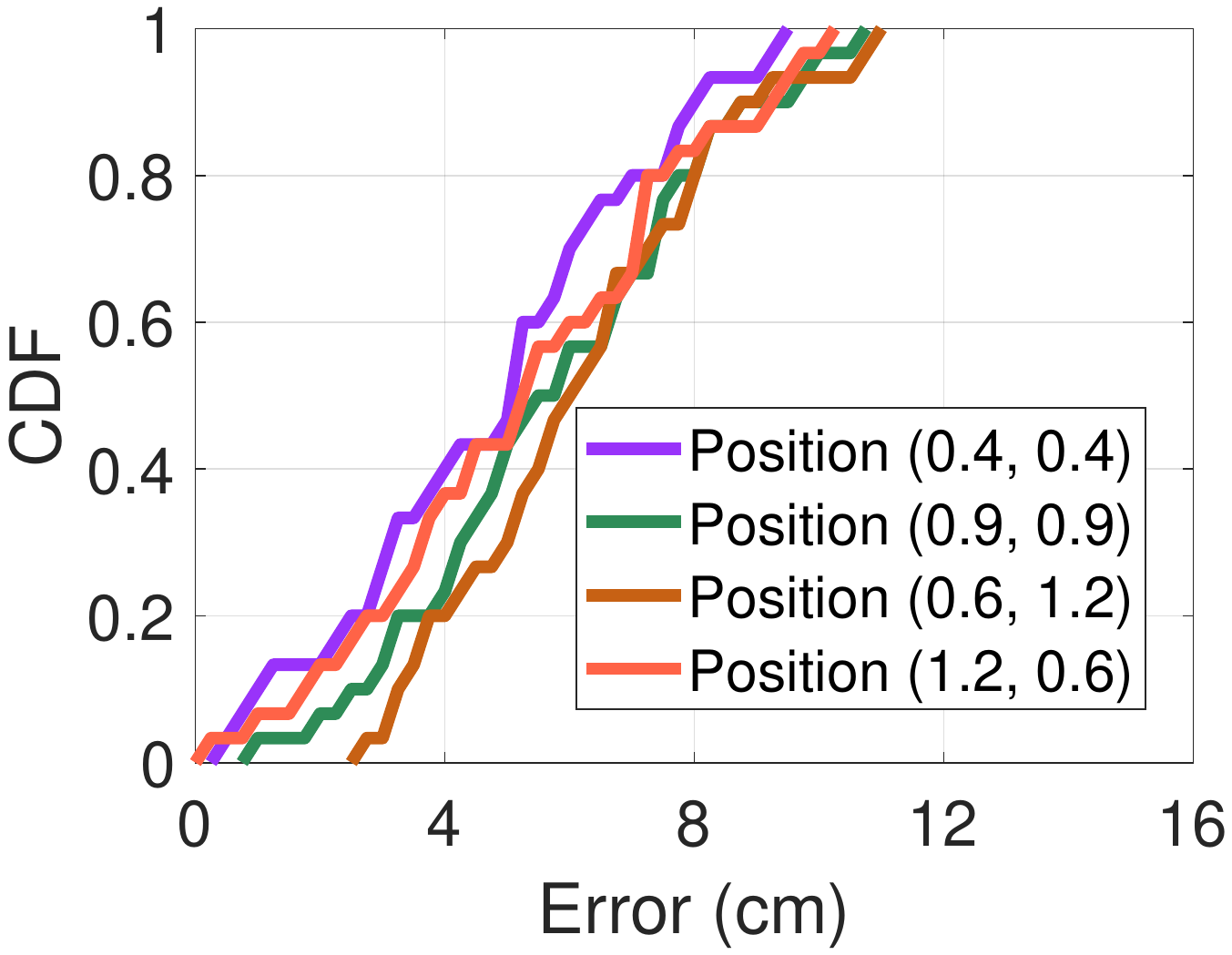}
}
\subfigure[CDF of 2D tracking error.]{
\includegraphics[width=4.1cm,height=3.5cm]{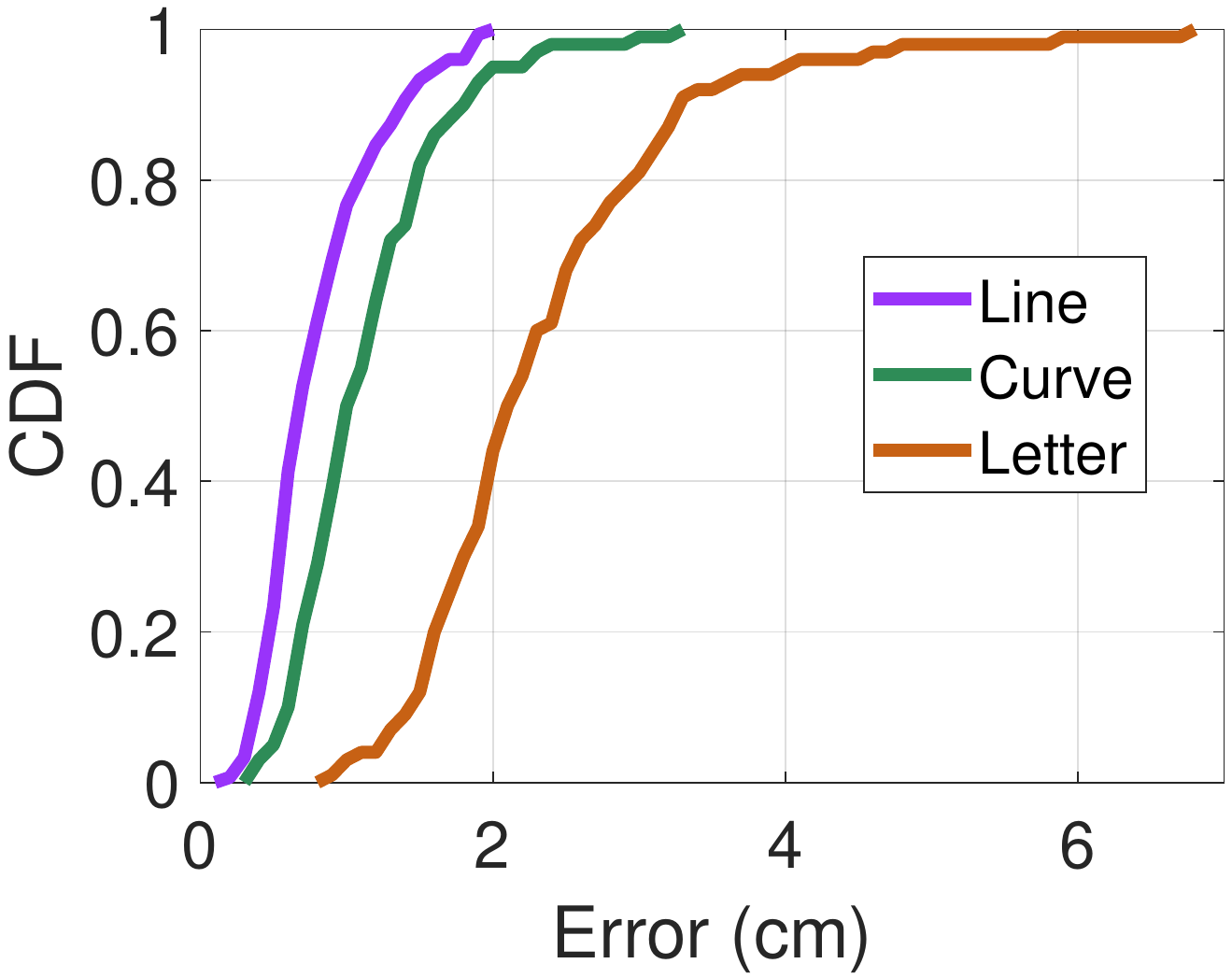}
}
\subfigure[Tracking W/O initial position.]{
\includegraphics[width=4.1cm,height=3.5cm]{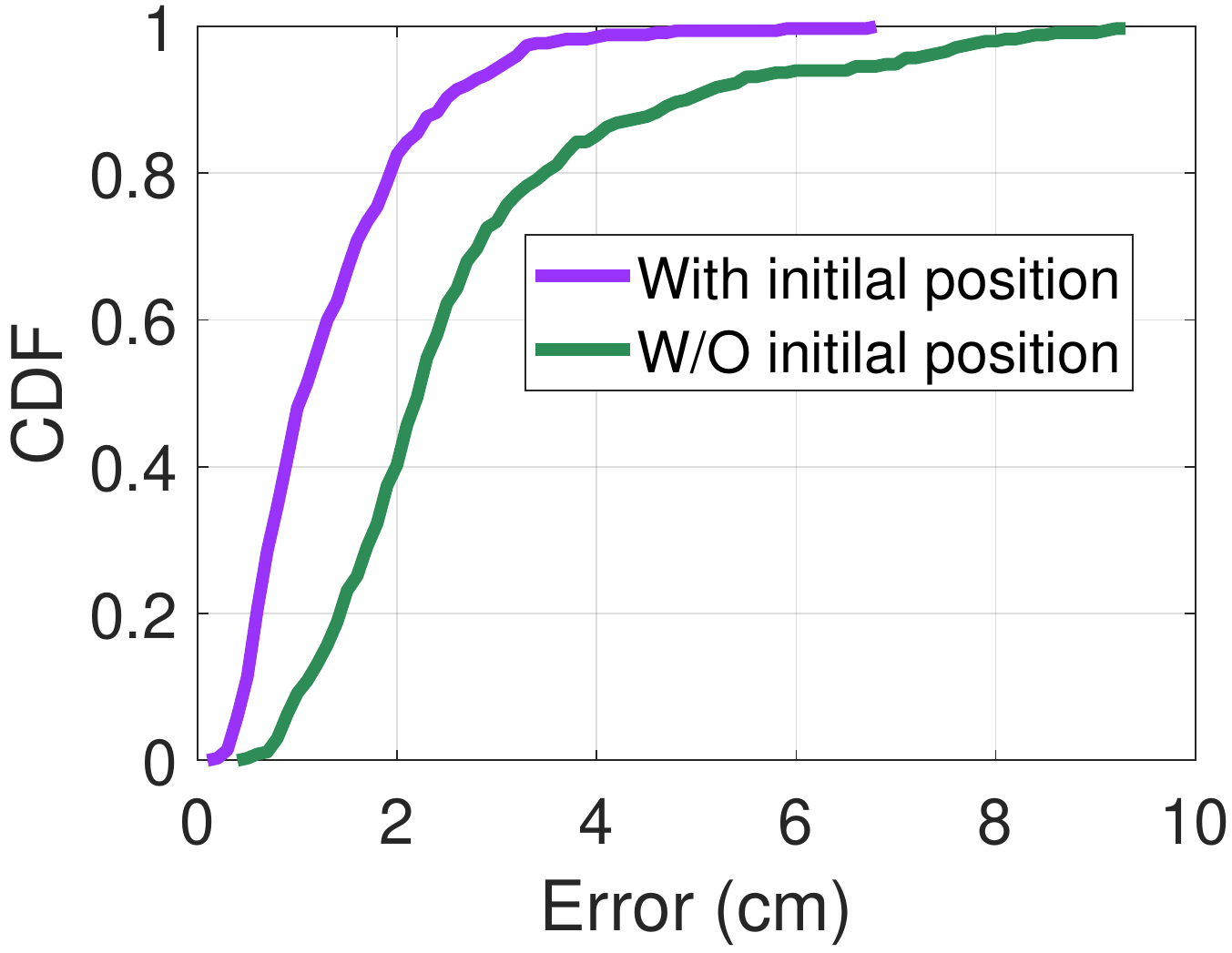}
}
\caption{2D measurement including initial position estimation and 2D traces tracking.}
\label{fig18}
\end{center}
\end{figure*}

\begin{figure*}[htbp]
\begin{minipage}{0.23\linewidth}
\centering
\includegraphics[width=4.1cm,height=3.5cm]{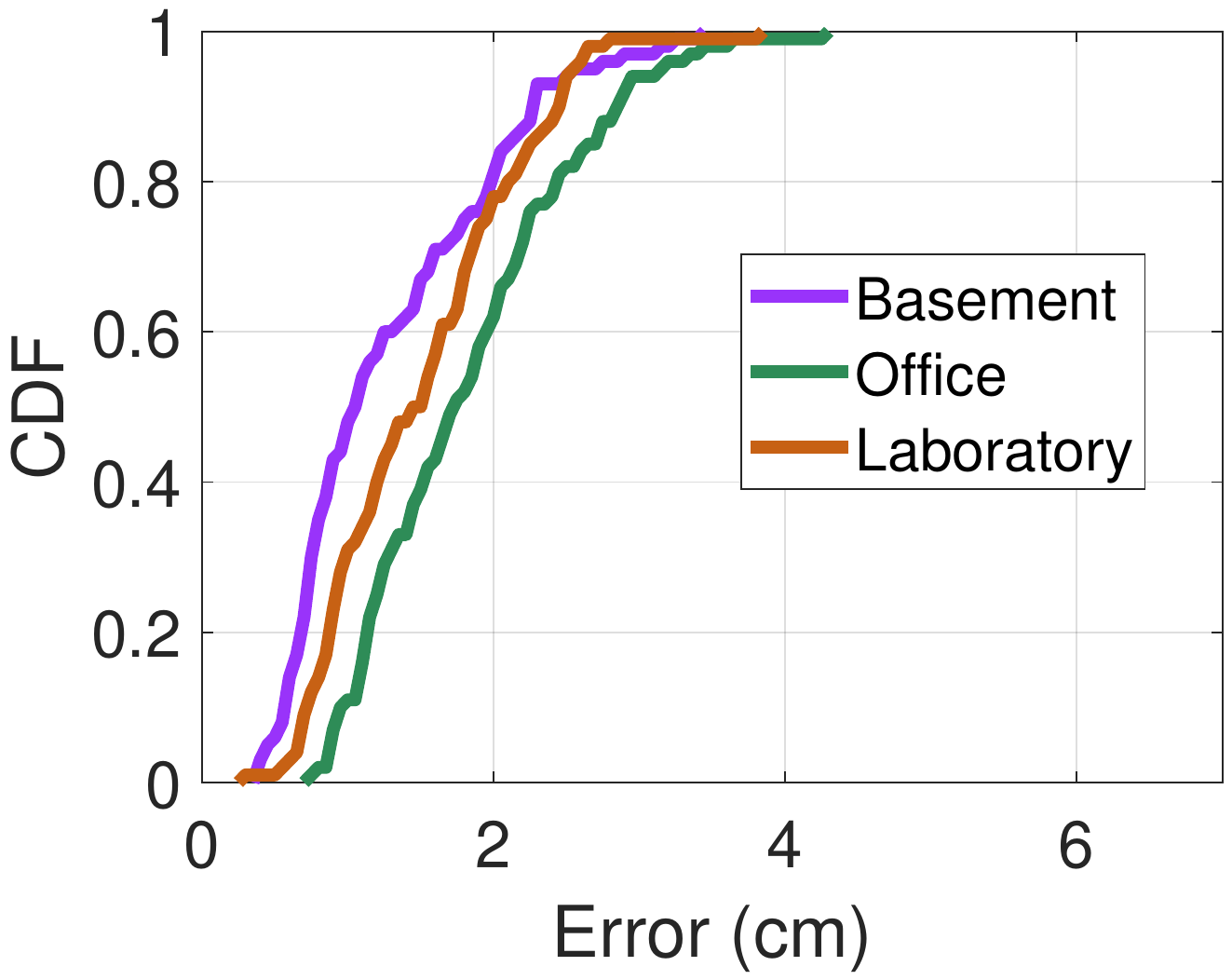}
\caption{Impact of environments.}
\label{fig19}
\end{minipage}
\hfill
\begin{minipage}{0.5\linewidth}
\centering
\subfigure[CDF of distance error.]{
\includegraphics[width=4.1cm,height=3.5cm]{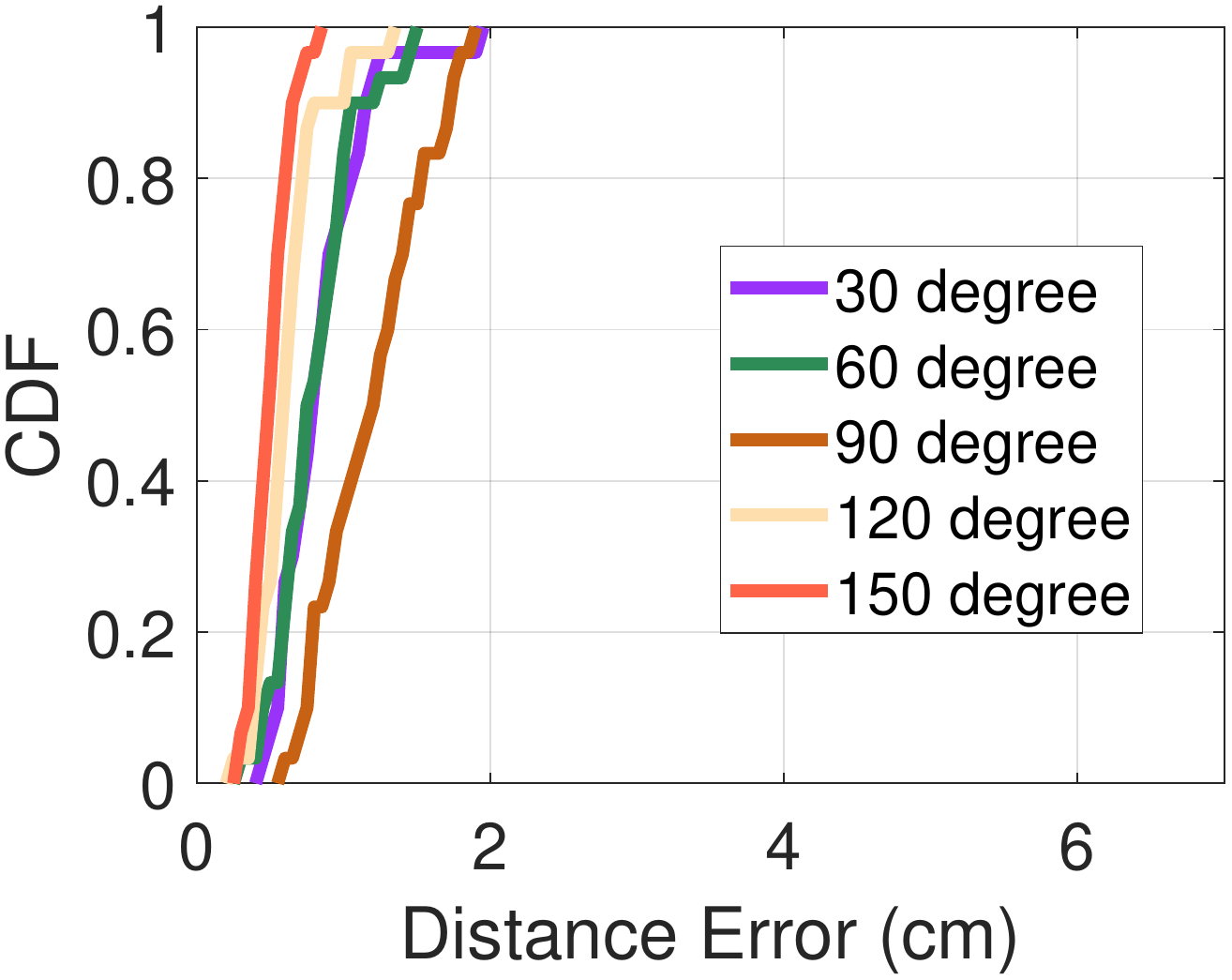}
}
\subfigure[CDF of direction error.]{
\includegraphics[width=4.1cm,height=3.5cm]{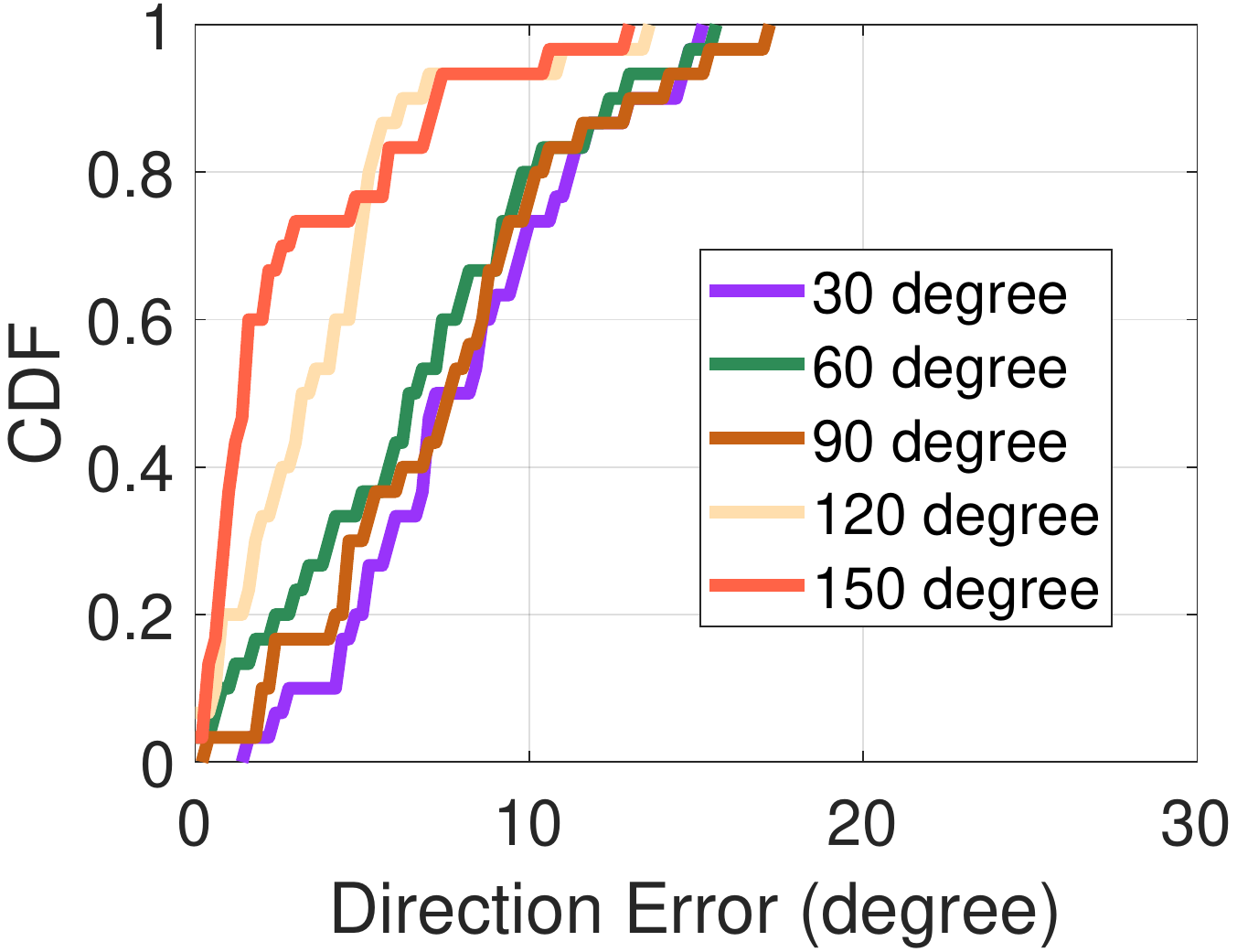}
}
\caption{Impact of moving directions.}
\label{fig20}
\end{minipage}
\hfill
\begin{minipage}{0.23\linewidth}
\centering
\includegraphics[width=4.1cm,height=3.5cm]{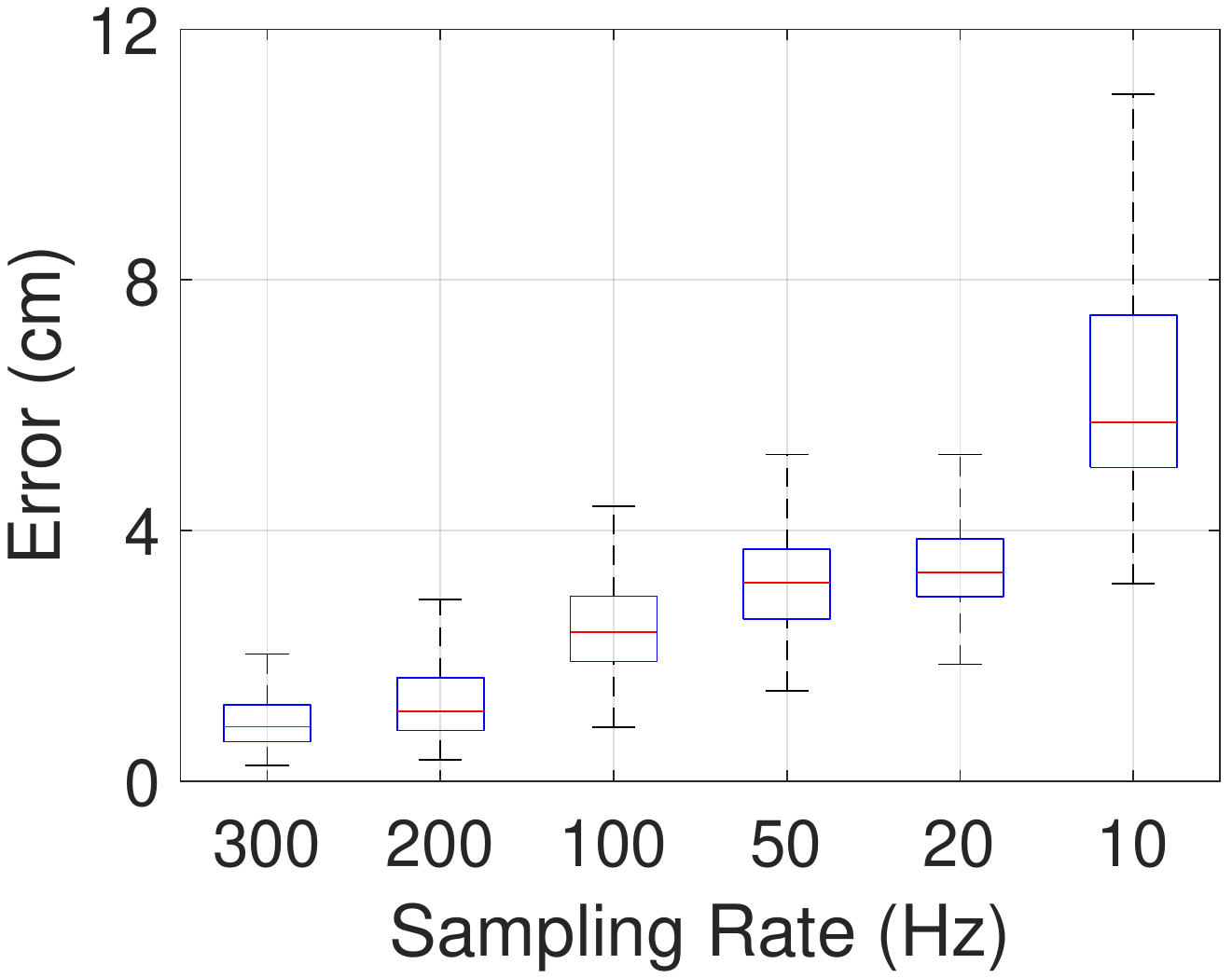}
\caption{Impact of sampling rate.}
\label{fig21}
\end{minipage}
\end{figure*}

\begin{figure*}[htbp]
\begin{minipage}{0.23\linewidth}
\centering
\includegraphics[width=4.1cm,height=3.5cm]{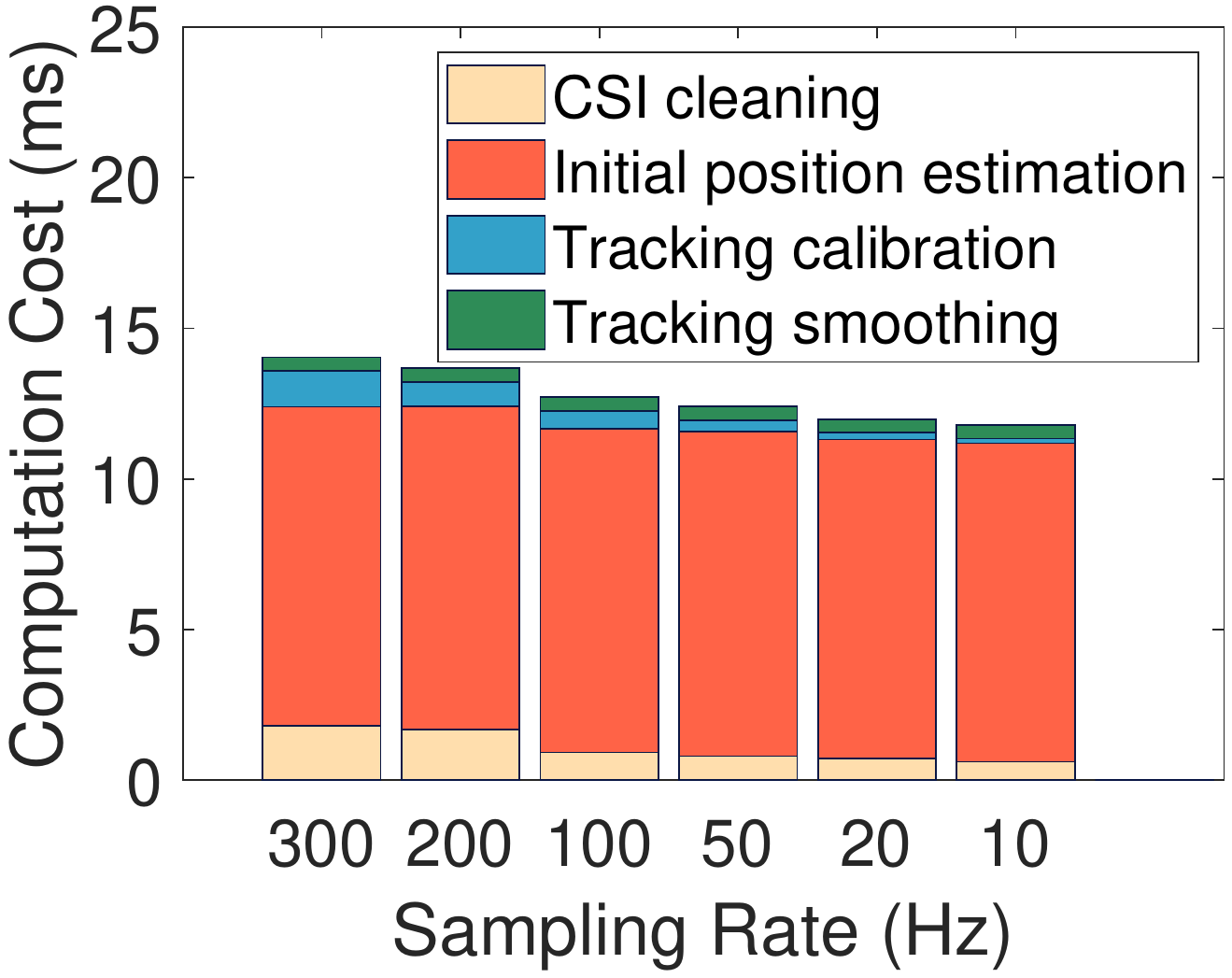}
\caption{Per-second computation cost.}
\label{fig22}
\end{minipage}
\hfill
\begin{minipage}{0.23\linewidth}
\centering
\includegraphics[width=4.1cm,height=3.5cm]{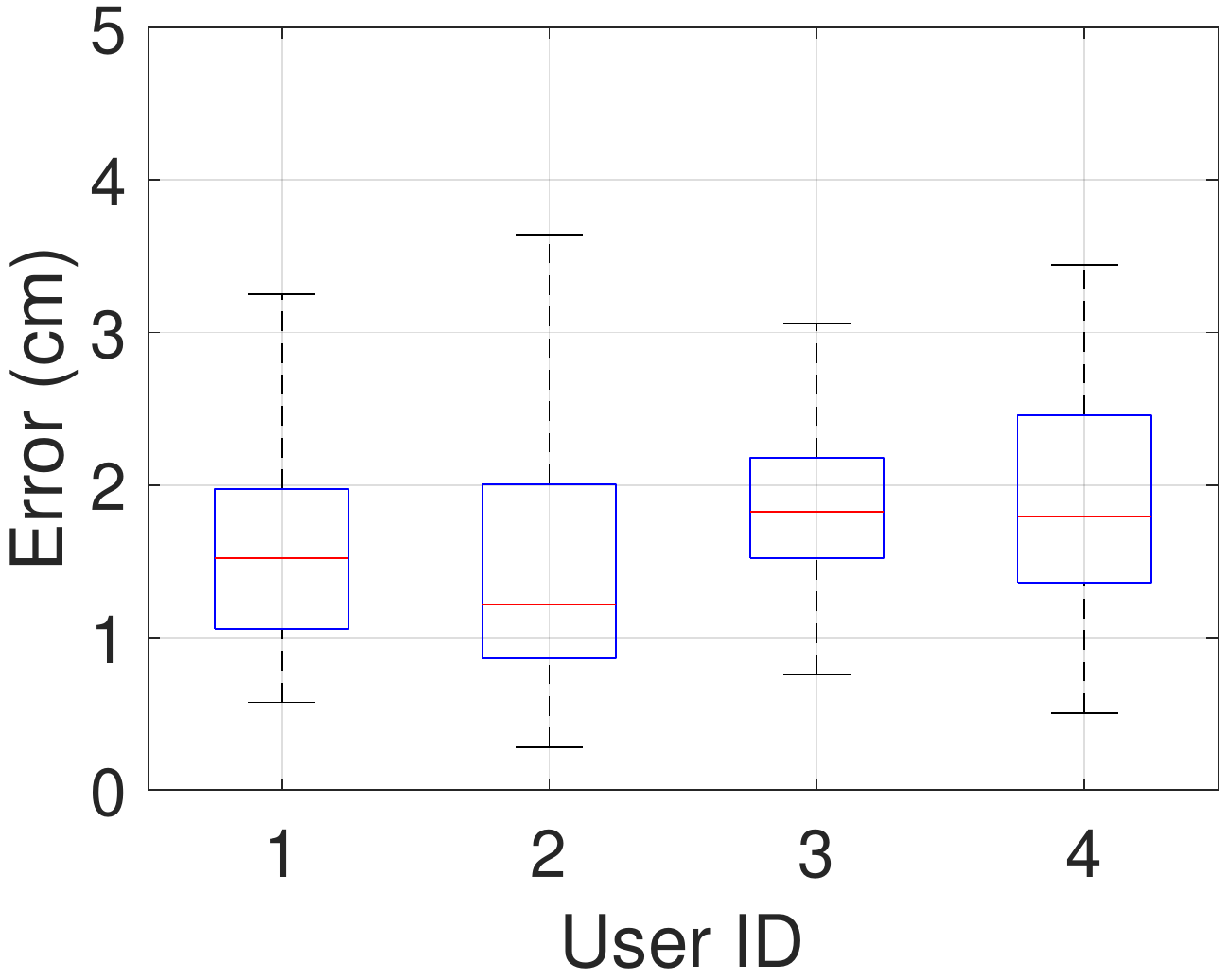}
\caption{Impact of user diversity.}
\label{fig23}
\end{minipage}
\hfill
\begin{minipage}{0.23\linewidth}
\centering
\includegraphics[width=4.1cm,height=3.5cm]{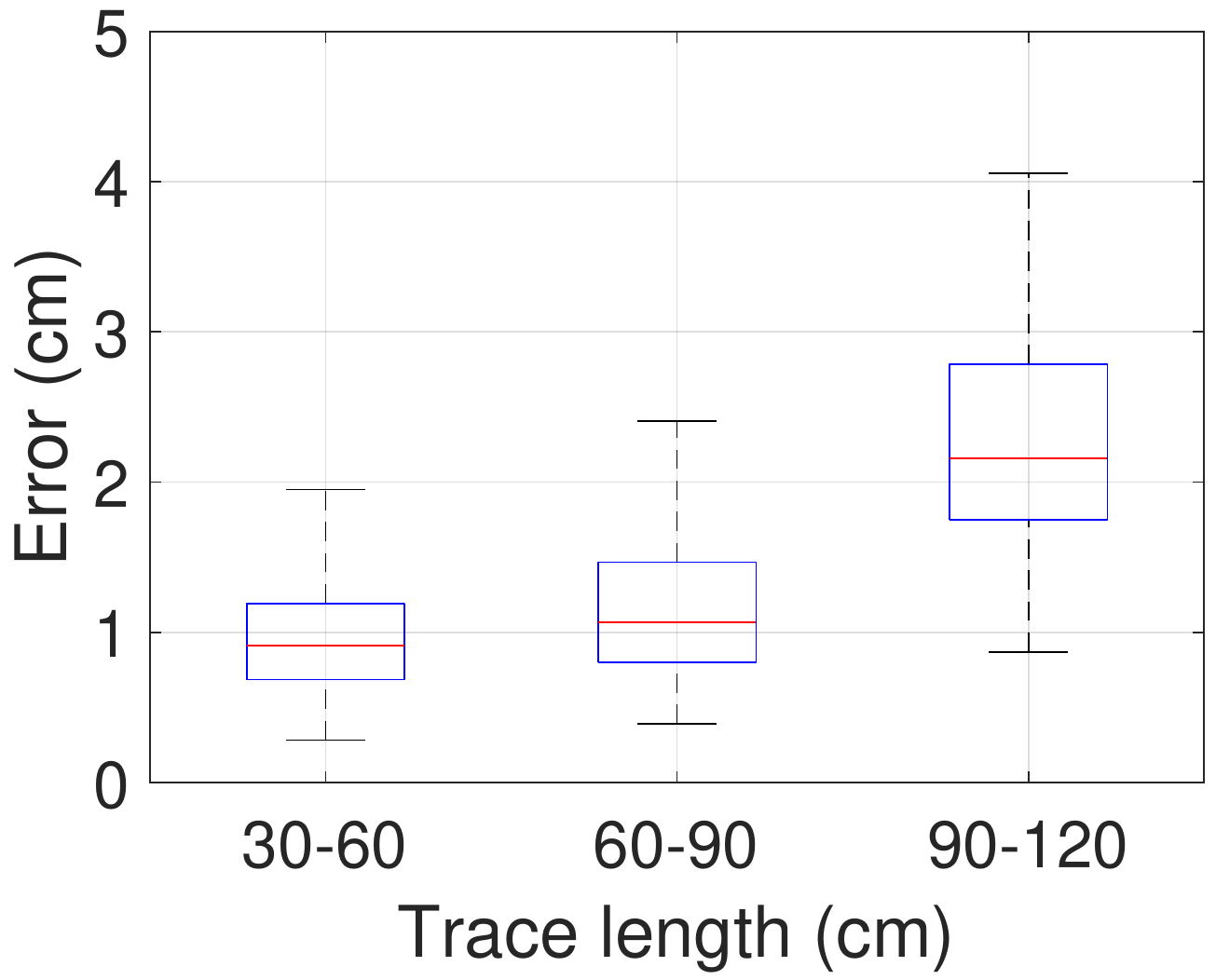}
\caption{Impact of length diversity.}
\label{fig24}
\end{minipage}
\hfill
\begin{minipage}{0.23\linewidth}
\centering
\includegraphics[width=4.1cm,height=3.5cm]{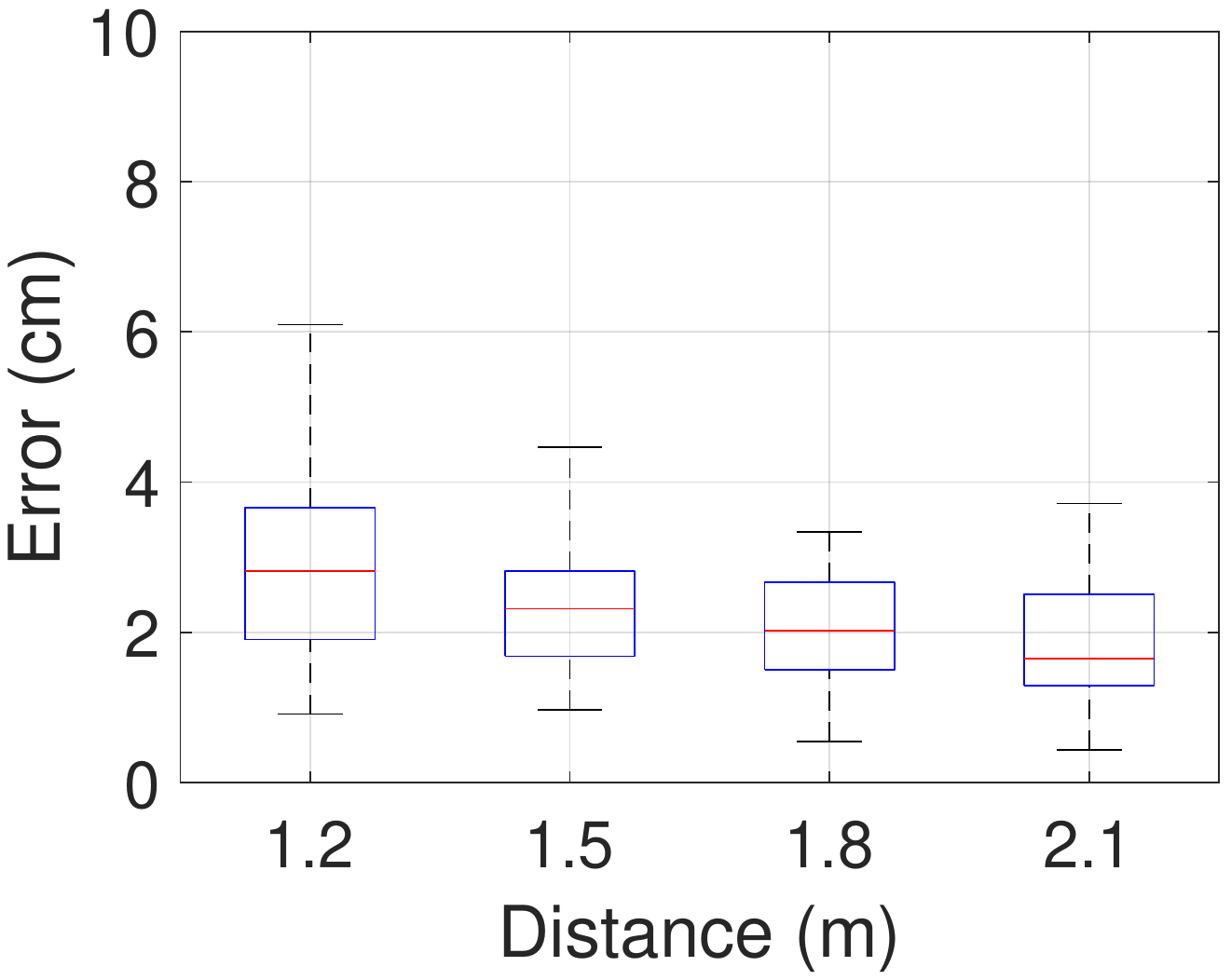}
\caption{Impact of other person walking around.}
\label{fig25}
\end{minipage}
\end{figure*}

$\textbf{Initial position error}$: We first evaluate the initial position estimation when users perform 120 pairs of slight random heading movements, with four evenly distributed ground truth initial positions as (0.4, 0.4), (0.6, 1.2), (1.2, 0.6) and (0.9, 0.9). As shown in Fig. \ref{fig18} (a), the estimated results have an excellent clustering, and the errors are independent of the spatial position. Fig. \ref{fig18} (b) shows that the 80$^{th}$ percentile estimated position error is within 8.0 cm. This is mainly due to the inherent search error of MUSIC algorithm and some weak dynamic path interference. All in all, the results indicate that CentiTrack is competent for the task of initial position estimation.

$\textbf{Tracking error}$: We next evaluate the 2D tracking accuracy when drawing lines, random curves, and English letters in the air. All the movements range from 30 cm to 120 cm. We strat to evaluate the tracking accuracy of a total number of 250 movements including lines and circular curves on the 2D sensing plane. Fig. \ref{fig18} (c) depicts the distributions of 2D tracking errors. It can be seen that for lines and curves, their median tracking errors are 0.6 cm and 1.0 cm, and their 80$^{th}$ percentile errors are both within 1.5 cm. Then, we evaluate if CentiTrack can effectively track the trace of English letters, which is more challenging due to sharp turns or segments. In our study, 100 letters are evaluated in total. As illustrated, the median and mean errors both remain below 2.3 cm, and 80$^{th}$ percentile errors are within 3.0 cm. Due to sharp turns and larger length, the errors of letters are slightly larger than those of lines and curves. Note that without the initial position, the above tracking performance will be seriously degraded. As shown in Fig. \ref{fig18} (d), we take the case that the initial position is 30 cm horizontally away from the ground truth, the median error of total 350 traces will increase by 144\%, i.e., from 0.9 to 2.2 cm. In addition, the distribution of errors will have a worse tailing.

$\textbf{Impact of different scenarios}$: We conduct experiments in different scenarios in order to demonstrate the influence of environment diversity. As shown in Fig. \ref{fig16}, there are three scenarios including basement, office and laboratory, and we collected 100 2D traces in each of them. Fig. \ref{fig19} shows the distributions of tracking errors at different spots. As shown, it achieves considerably low mean tracking errors of 1.5 cm, 2.2 cm and 2.0 cm, respectively. We note that CentiTrack still maintains high tracking performance confronting diverse environments. However, this stability is inversely related to the size of sensing area. With a larger sensing area, the signal reflected by hands tends to be more vulnerable to noises, resulting in degradation of accuracy. Besides, given the fixed moving speed and distance, the longer the LoS is, the smaller the DFS will be, which further degrades the sensitivity of system. Therefore, we configure the LoS by 1.5 m, which not only is sufficient for gesture tracking, but also does not cause the reflected signal to be weak or DFS to be insensitive.

$\textbf{Impact of different moving directions}$: To evaluate how moving direction impacts the tracking performance, users are asked to move hands in straight lines with various degrees related to y-axis. Fig. \ref{fig20} shows the distribution of direction and distance errors for five directions. CentiTrack remains high performance for diverse moving directions, indicating it has slight impact on the tracking accuracy.

\begin{figure*}[htbp]
\begin{minipage}{0.5\linewidth}
\subfigure[Benefits of individual modules.]{
\includegraphics[width=4.1cm,height=3.5cm]{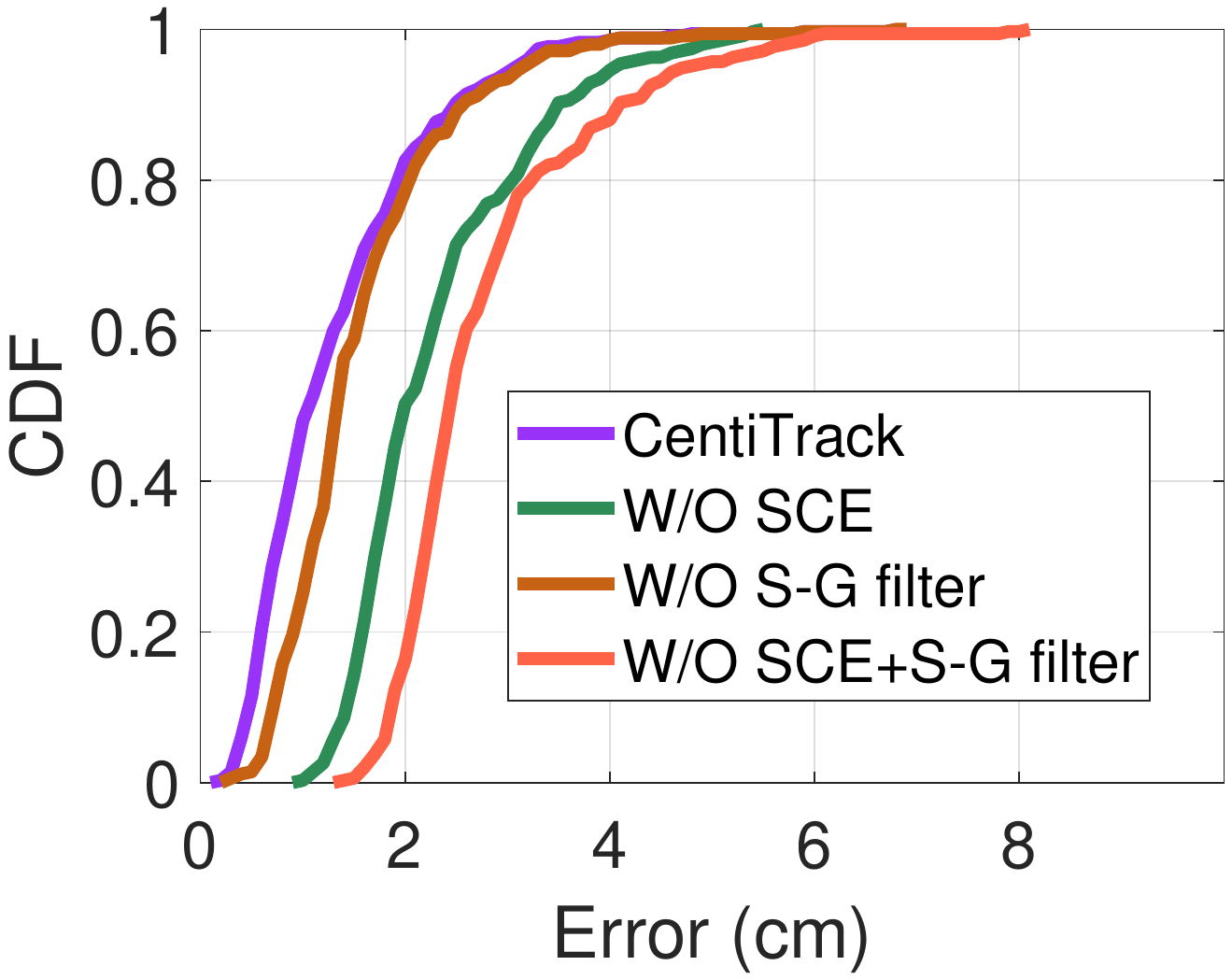}
}
\subfigure[Algorithm comparison.]{
\includegraphics[width=4.1cm,height=3.5cm]{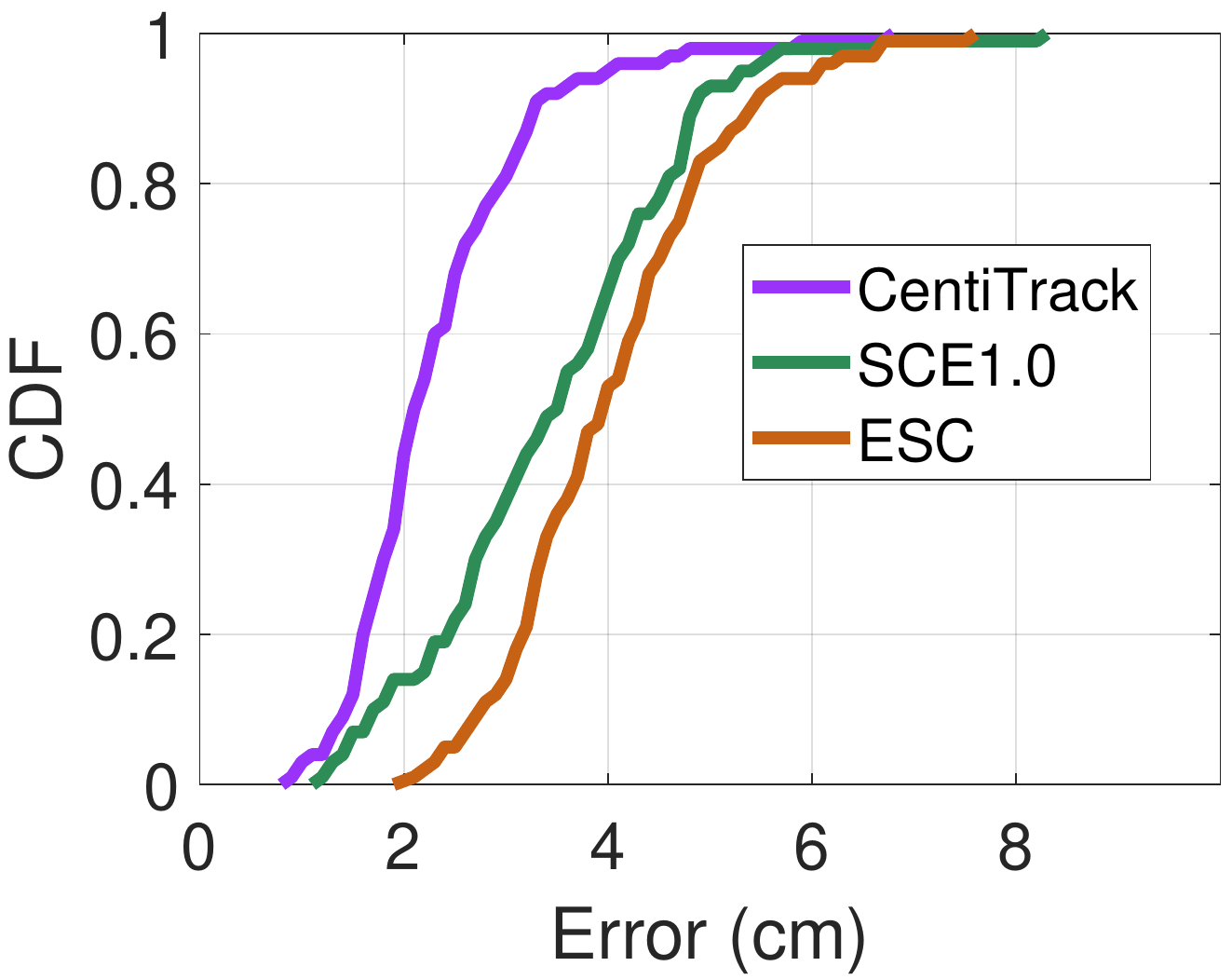}
}
\caption{Benefits of individual modules.}
\label{fig26}
\end{minipage}
\hfill
\begin{minipage}{0.5\linewidth}
\centering
\subfigure[10 digits and 26 letters.]{
\includegraphics[width=4.8cm,height=3.2cm]{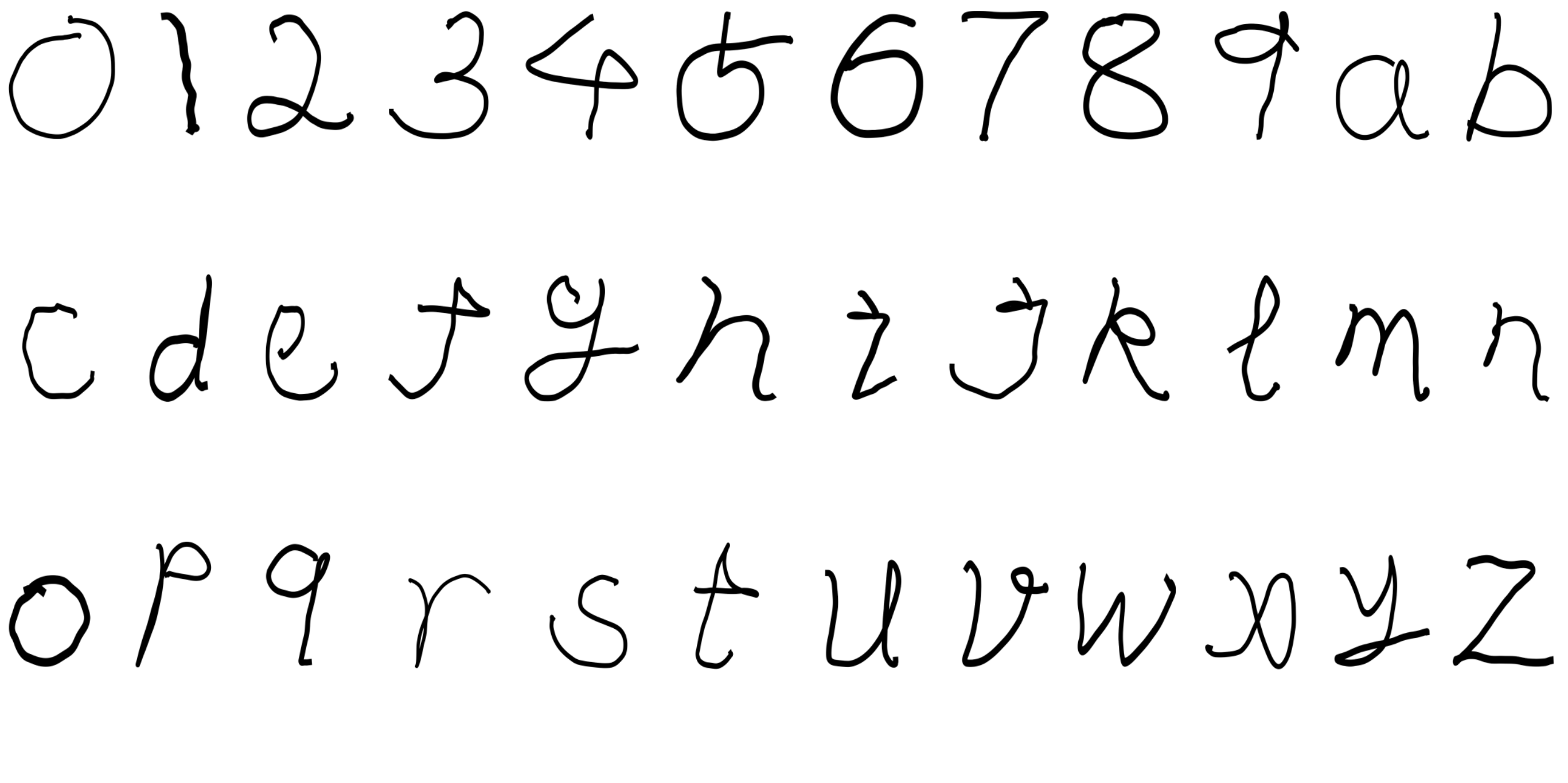}
}
\subfigure[Confusion matrix of text recognition.]{
\includegraphics[width=3.6cm,height=3.5cm]{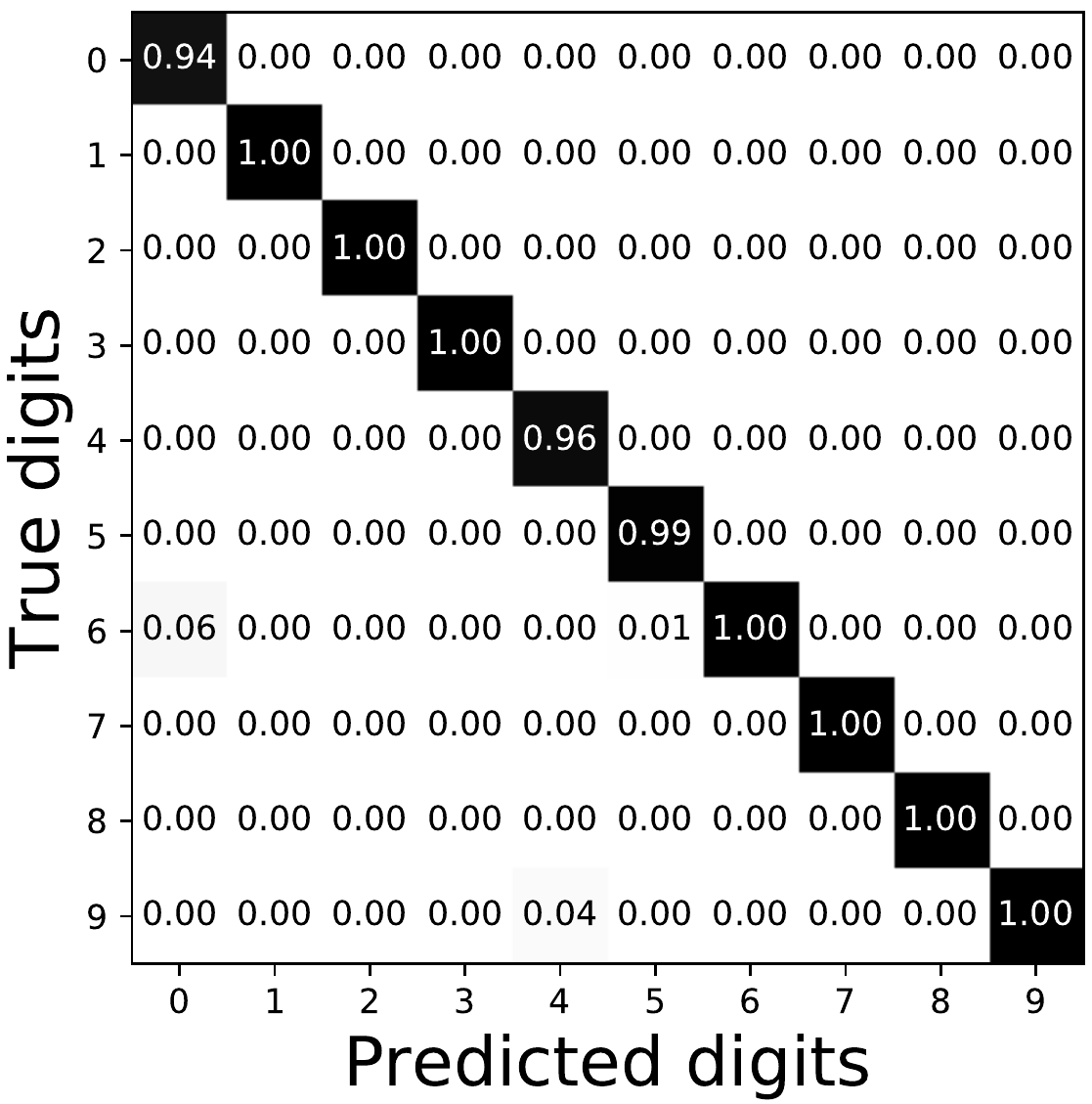}
}
\caption{Handwriting and text recognition.}
\label{fig27}
\end{minipage}
\end{figure*}

$\textbf{Impact of sampling rate}$: To find out the minimum CSI sampling rate required to ensure high performance, we initialize the sampling rate at TX as 300 Hz, and gradually drop CSI samples at regular intervals to achieve rates of 200 Hz, 100 Hz, 50 Hz, 20 Hz and 10 Hz. Fig. \ref{fig21} shows the tracking errors with different sampling rates, we can find that a higher sampling rate does lead to a higher accuracy, but the performance degradation with lower rate is not significant, when the rate exponentially decreases from 300 Hz to 20 Hz. It demonstrates that CentiTrack can be compatible with low packet rate, which is practical with real WiFi transmissions. However, further reducing the sampling rate, from 20 Hz to 10 Hz, may result in aliasing of DFS estimation. Given 0.5 m/s moving speed, accroding to DFS definiton caused by hand motion, $f_{D} = \frac{v_{hand}}{\lambda}$, the range of DFS is therefore within [-10 Hz, 10 Hz]. Thus, the minimum sampling rate required is at least ${2f_D}_{max}$, i.e. 20 Hz, due to the Nyquist Theorem. If users draw at a higher speed, a higher minimum sampling rate is required accordingly.

Next, we also evaluate the per-second computation cost. The processing computer is with the Intel i7-7700 CPU. As shown in Fig. \ref{fig22}, the major time cost comes from initial position estimation by MUSIC. By decreasing the sampling rate, the computation cost is gradually reduced. Given the rate by 20 Hz, the per-second cost is only 12 ms, enabling a real-time gesture tracking.

$\textbf{Impact of user and trace length diversity}$: To determine whether CentiTrack consistently works for diverse users, we in total recruit four volunteers to participate in experiments. The volunteers are with different genders, heights and body shapes, and without training in advance. Given the differences in moving speeds, ranges and styles, the tracking accuracy varies among users. Fig. \ref{fig23} plots the boxplot of the mean tracking errors for different users. As shown, our system tracks all users' hands with high accuracy, without knowing any hand motion features of users. The results indicate that CentiTrack is robust to user diversity and can achieve a high and stable tracking performance. Fig. \ref{fig24} also shows that CentiTrack can track reasonably long traces (up to 120 cm) with only a slight increase in estimation error.

$\textbf{Impact of other person walking around}$: To evaluate the system robustness, we conduct control groups where other volunteers were working or walking around during tracking. As shown in Fig. \ref{fig25}, the closer the distance between the walking volunteer and the transceiver, the larger the tracking error will be. The signal variations induced by other volunteers will make it more difficult to detect hand motion accurately. The result above shows that CentiTrack has high precision even if another person walking 1.5 m away from users. Therefore, our system is robust to interference caused by indoor daily activities.

$\textbf{Benefits of individual modules}$: The accurate tracking performance stems from both the proposed Tracking Calibration and Tracking Smoothing processes. Fig. \ref{fig26} (a) shows the effects of these two modules on total 350 traces. It can be seen that, without the SCE, the mean tracking error increases from 1.3 to 2.3 cm. The necessity of the static components elimination is underlined here. The tracking accuracy is also improved by 0.5 cm with the Savitzky-Golay (S-G) filter, which benefits from the smooth envelope. Besides, we compare the performance of the proposed SCE with the state-of-the-arts, i.e., Extracting Static Component algorithm (ESC) \cite{witrace} and our previous work Static Components Elimination algorithm (SCE1.0) \cite{airdraw}. As shown in Fig. \ref{fig26} (b), we compute the tracking errors on 100 letter traces, because the algorithms will have more obvious correction effect on the more challenging letters tracking. ESC isolates the static components by detecting whether the gap between alternate local maximum and minimum points is larger than an empirical threshold. For CSI signals, the static components are always affected by surrounding noise and interference, therefore it is difficult to reliably detect the local maximum and minimum points by a pre-configured threshold. Different from ESC, SCE1.0 combines the I/Q parts in CSI trace and decomposes the whole trace to calibrate the signals, which has better static components elimination results. As a subsequent work of SCE1.0, we decompose CSI traces, and adopt linear interpolation between the circle centers to further eliminate slow changing static components in CentiTrack. 

\subsection{Use Case: Handwriting and Text Recognition}
In this section, we evaluate CentiTrack by allowing the users to input digits or letters to devices by writing in the air. An in-air handwriting prototype is thus implemented, where RXs report CSI and sends to a central server via UDP protocol real time. Users can write a digit, letter or cursive word once a time to splice them into a word or sentence. When the signal variance caused by user pauses is below the empirical threshold for $\tau_a$, we regard it as a pause and wait for next writing for splicing. When the pause exceeds $\tau_b$, the splicing is considered complete. We interface CentiTrack with
Google Input Method Editors (GIME). As users write in the air, we feed the coordinates of estimated traces into GIME and use the recognition functionality to interpret the texts.

\begin{figure*}[htbp]
\centering
\subfigure[Circle.]{
\includegraphics[width=3.95cm,height=3.7cm]{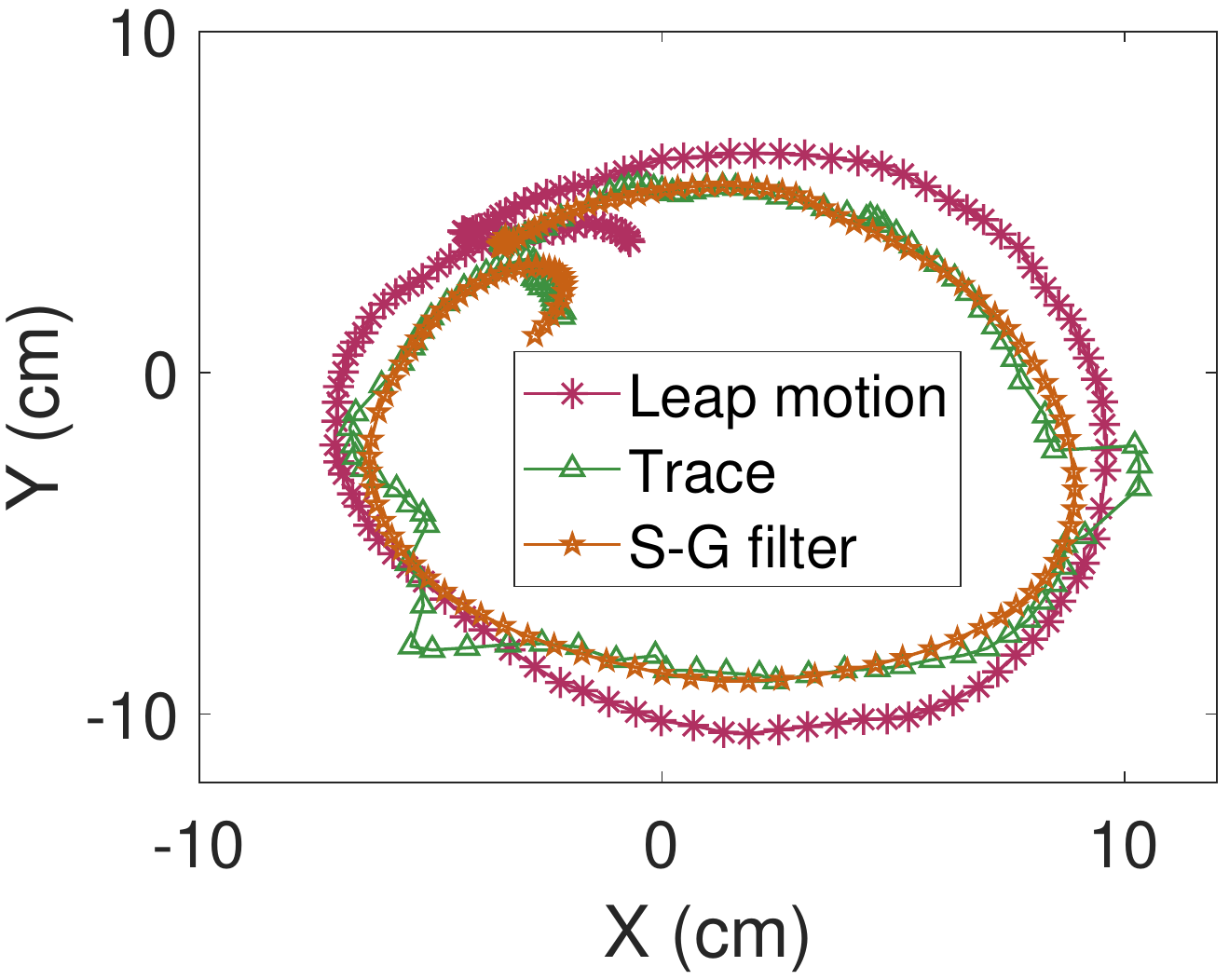}
}
\subfigure[Triangle.]{
\includegraphics[width=3.95cm,height=3.7cm]{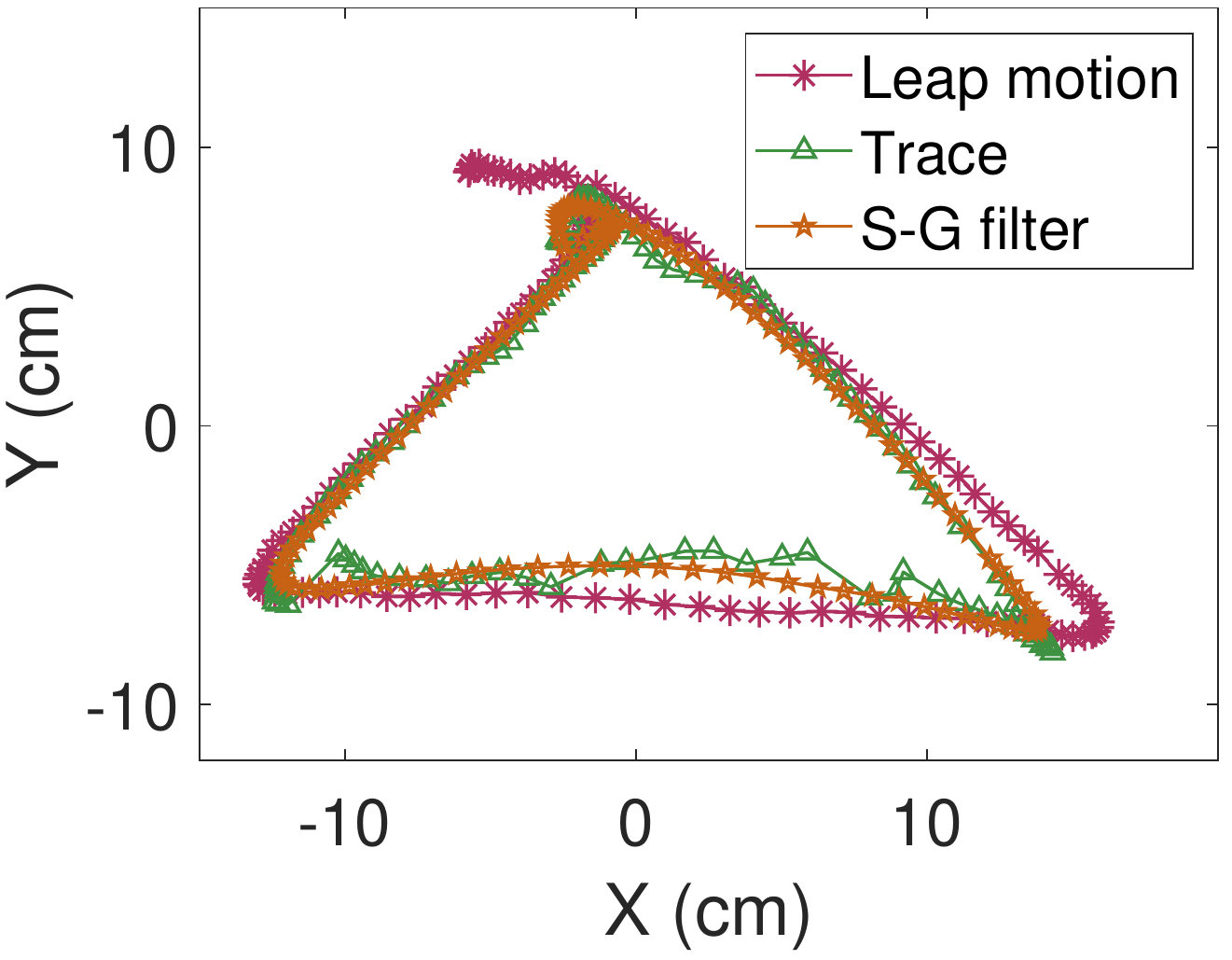}
}
\subfigure[Five-pointed star.]{
\includegraphics[width=3.95cm,height=3.7cm]{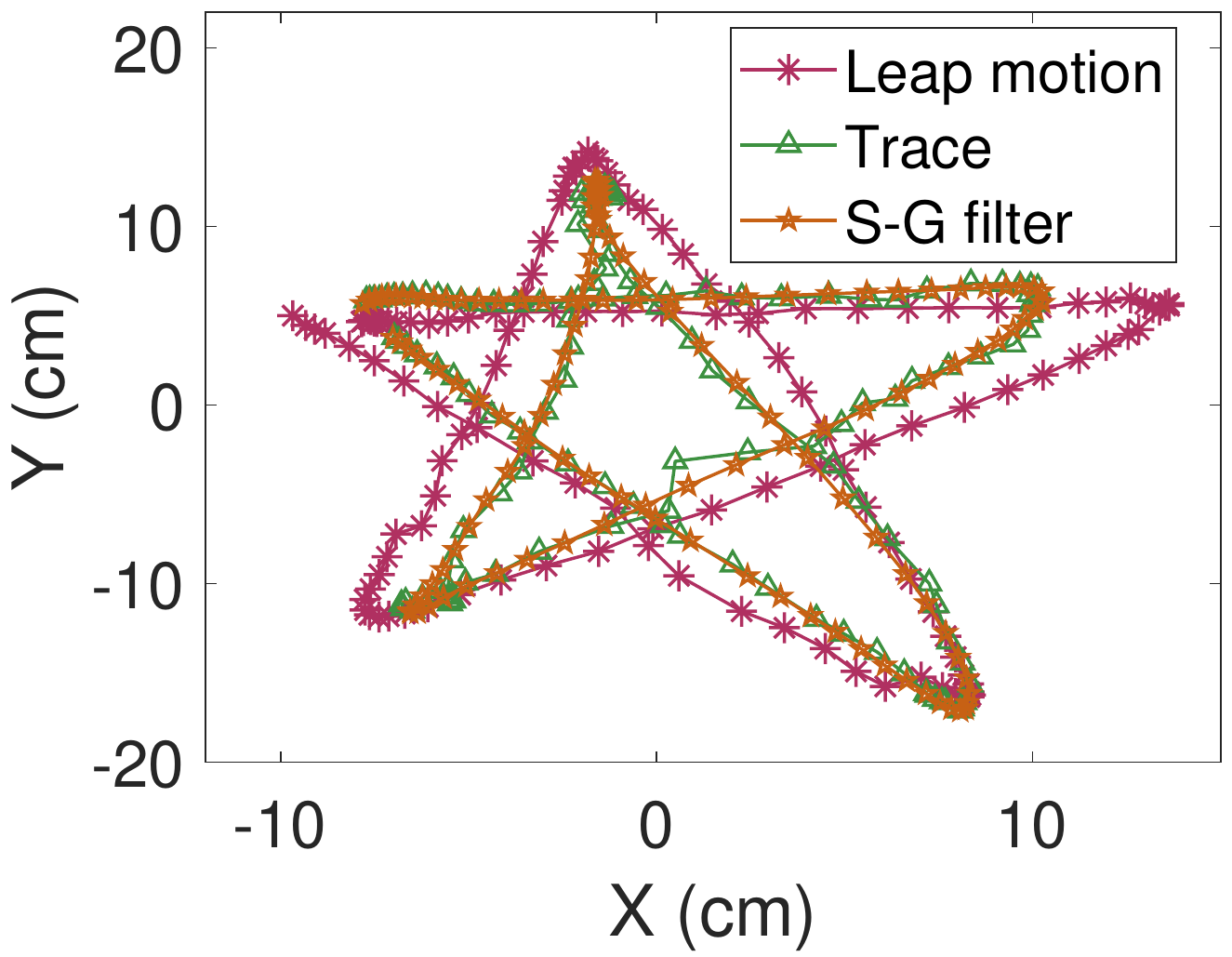}
}
\subfigure[Cursive ``CentiTrack''.]{
\includegraphics[width=4.4cm,height=3.9cm]{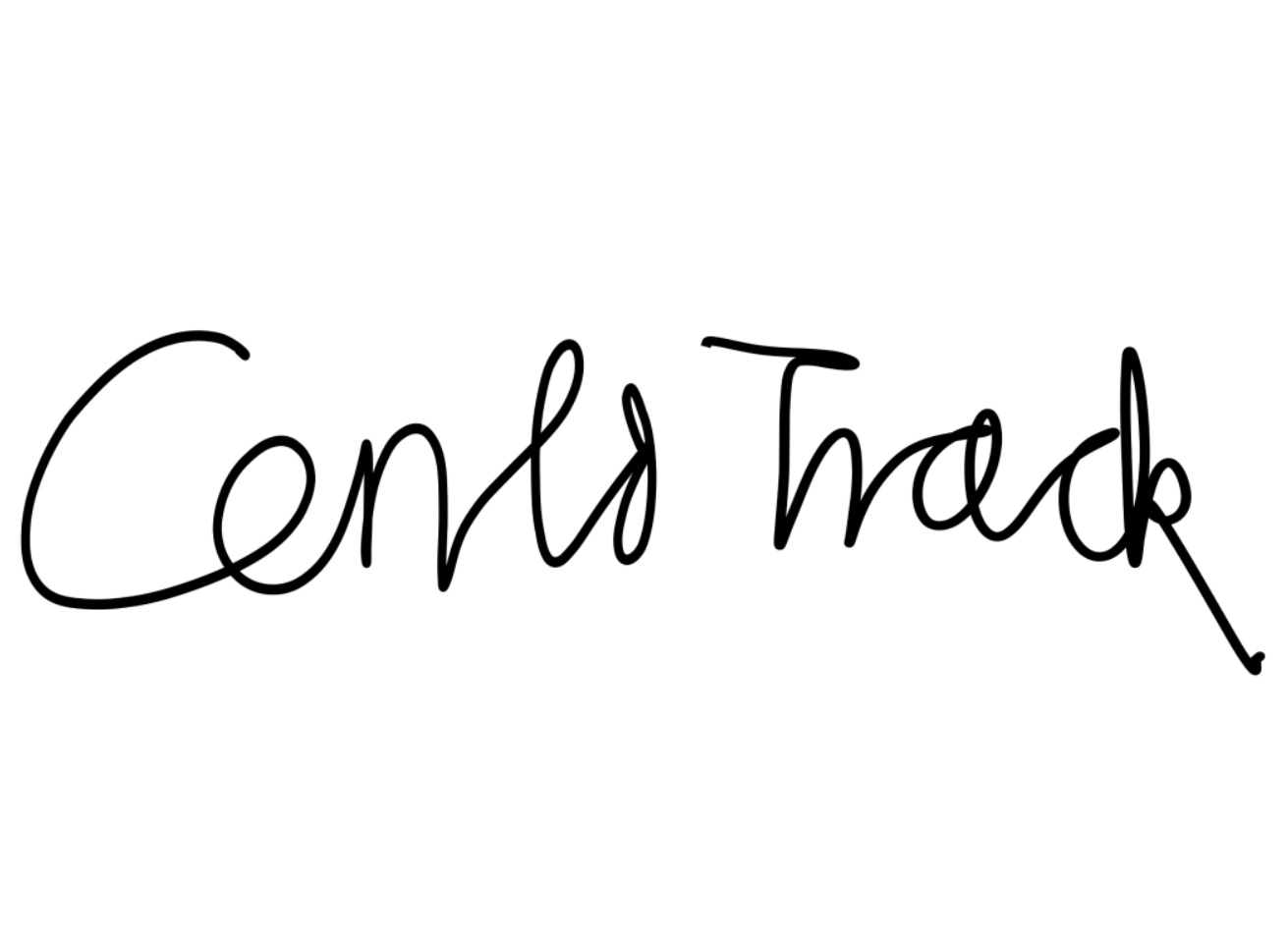}
}
\caption{Samples of estimated traces.}
\label{fig28}
\end{figure*}

We evaluate the recognition accuracy of digits for text input. The experimental settings are the same as the one in 2D tracking scenario. The users can draw digits or letters in different sizes and initial positions. We collect 20 samples for each digit from 4 users, i.e., $20\times10\times4 = 800$ data in total. An example of the estimated digits is illustrated in Fig. \ref{fig27} (a).  Fig. \ref{fig27} (b) shows the detail of recognition accuracy. It can be seen that most of the digits achieve an accuracy of greater than 94\%. Fig. \ref{fig28} also shows some samples of the estimated trace, including several arbitrary movements and the cursive form of ``CentiTrack'', indicating that our system is as fine-grained as the actual handwriting.

\section{Discussion and Limitations}
All the major commodity WiFi chip families (Atheros, Intel, and Marvell) expose CSI \cite{tool1, tool2}. With low computation cost, robustness and wireless transmission compatibility, we believe CentiTrack can easily apply to any commodity WiFi infrastructure. In CentiTrack, two paris of transceiver are placed vertically to establish a Cartesian coordinate system. However, this can also be achieved by non-vertical placement, except that the transceiver pairs are not on the axis. To be specific, based on the AoAs [$\theta_1$, $\theta_2$] of two LoS path from TX to RXs, and the vertical arrangement of RX antenna array, we can determine a unique triangle with three WiFi devices as its vertices. In this way, the coordinate system can be established, and the gesture tracking task can be realized.

Our current implementation has following limitations. We currently estimate the initial position of hand by AoAs. Nevertheless, this method cannot work when user starts to move hand on the connection of two RXs. This is because all points on this connection have the same pair of AoAs. To cope with it, we delay several CSI samples to derive the initial position when the estimated AoAs are on the connection line. Besides, our scheme is not based on range information, it will have the accumulative tracking errors. While, the tracking error is quite acceptable (2.2 cm) even the trace length is up to 120 cm.

\section{Conclusion}
In this paper, we propose CentiTrack, the first passive gesture tracking system with only three commodity WiFi devices, which achieves centimeter-level accuracy without any dedicated devices or extra hardware modifications. We fully implement CentiTrack and carry out comprehensive evaluations as well as related parameter study. The extensive experiments yield that it can track gestures with mean error lower than 1.5 cm, which is comparable to camera-based solutions. Benefits provided by its accurate performance can track users' writings/drawings in practical applications, be it handwriting or intelligent household appliance controlling in smart homes. In the future, we will prove that CentiTrack can achieve 3D hand motion tracking.

\section*{Acknowledgment}
This work was supported by the National Natural Science Foundation of China under grant No.61671073.

\end{document}